\documentclass[twocolumn]{aastex63}
\usepackage{amsmath}
\usepackage{empheq}
\usepackage{mathrsfs}
\usepackage{textcomp}
\usepackage{enumitem}   
\usepackage{gensymb}
\usepackage{hyperref}
\usepackage{graphicx}
\usepackage[caption=false]{subfig}
\usepackage{multirow}
\usepackage{longtable}
\usepackage{booktabs} % To thicken table lines
\hypersetup{colorlinks, linkcolor={blue}, citecolor={blue}, urlcolor={blue}} 

\newcommand{\name}{SN2019dge}

\definecolor{DarkOrange}{RGB}{204, 85, 0}
\definecolor{LincolnGreen}{RGB}{17, 102, 0}
\def\ion#1#2{#1$\;${\footnotesize\rm{#2}}\relax}

\newcommand{\swift}{\textit{Swift}}

\shorttitle{\name: an Ultra-Stripped SN}
\shortauthors{Yao et al.}

\begin{document}
\pagenumbering{arabic}

\title{\name: a Helium-rich Ultra-Stripped Envelope Supernova}

\author[0000-0001-6747-8509]{Yuhan Yao}\email{yyao@astro.caltech.edu}
\affiliation{Cahill Center for Astrophysics, 
             California Institute of Technology, MC 249-17, 
             1200 E California Boulevard, Pasadena, CA 91125, USA}

\author[0000-0002-8989-0542]{Kishalay De}
\affiliation{Cahill Center for Astrophysics, 
	California Institute of Technology, MC 249-17, 
	1200 E California Boulevard, Pasadena, CA 91125, USA}

\author[0000-0002-5619-4938]{Mansi M. Kasliwal}
\affiliation{Cahill Center for Astrophysics, 
	California Institute of Technology, MC 249-17, 
	1200 E California Boulevard, Pasadena, CA 91125, USA}

\author[0000-0002-9017-3567]{Anna Y. Q.~Ho}
\affiliation{Cahill Center for Astrophysics, 
	California Institute of Technology, MC 249-17, 
	1200 E California Boulevard, Pasadena, CA 91125, USA}

\author[0000-0001-6797-1889]{Steve Schulze}
\affiliation{Department of Particle Physics and Astrophysics, 
	Weizmann Institute of Science, 234 Herzl Street, 76100 Rehovot, Israel}

\author[0000-0001-5113-7558]{Zhihui Li}
\affiliation{Cahill Center for Astrophysics, 
	California Institute of Technology, MC 249-17, 
	1200 E California Boulevard, Pasadena, CA 91125, USA}

\author[0000-0001-5390-8563]{S. R. Kulkarni}
\affiliation{Cahill Center for Astrophysics, 
	California Institute of Technology, MC 249-17, 
	1200 E California Boulevard, Pasadena, CA 91125, USA}

\author[0000-0002-6652-9279]{Andrew Fruchter}
\affiliation{Space Telescope Science Institute, 
	3700 San Martin Drive, Baltimore, MD 21218, USA}

\author[0000-0001-5402-4647]{David Rubin}
\affiliation{Space Telescope Science Institute, 
	3700 San Martin Drive, Baltimore, MD 21218, USA}
\affiliation{Department of Physics and Astronomy, 
	University of Hawai'i, 2680 Woodlawn Drive, Honolulu, HI 96822, USA}

\author[0000-0001-8472-1996]{Daniel A. Perley}
\affiliation{Astrophysics Research Institute, 
	Liverpool John Moores University,\\ 
	IC2, Liverpool Science Park, 146 Brownlow Hill, Liverpool L3 5RF, UK}

\author[0000-0002-4544-0750]{Jim Fuller}
\affiliation{TAPIR, 
	California Institute of Technology, MC 350-17, 
	1200 E California Boulevard, Pasadena, CA 91125, USA}

\author[0000-0002-4223-103X]{C. Fremling}
\affiliation{Division of Physics, Mathematics, and Astronomy, 
	California Institute of Technology, 
	Pasadena, CA 91125, USA}

\author[0000-0001-8018-5348]{Eric C. Bellm}
\affiliation{DIRAC Institute, Department of Astronomy, University of Washington, 
	3910 15th Avenue NE, Seattle, WA 98195, USA}

\author{Rick Burruss}
\affiliation{Caltech Optical Observatories, California Institute of Technology, Pasadena, CA  91125}

\author[0000-0001-5060-8733]{Dmitry A. Duev}
\affiliation{Division of Physics, Mathematics, and Astronomy, 
	California Institute of Technology, 
	Pasadena, CA 91125, USA}

\author{Michael Feeney}
\affiliation{Caltech Optical Observatories, California Institute of Technology, Pasadena, CA  91125}

\author[0000-0002-3653-5598]{Avishay Gal-Yam}
\affiliation{Department of Particle Physics and Astrophysics, 
	Weizmann Institute of Science, 234 Herzl Street, 76100 Rehovot, Israel}

\author[0000-0001-8205-2506]{V. Zach Golkhou}
\affiliation{DIRAC Institute, Department of Astronomy, University of Washington, 
	3910 15th Avenue NE, Seattle, WA 98195, USA} 
\affiliation{The eScience Institute, University of Washington, Seattle, WA 98195, USA}

\author[0000-0002-3168-0139]{Matthew J. Graham}
\affiliation{Division of Physics, Mathematics, and Astronomy, 
	California Institute of Technology, 
	Pasadena, CA 91125, USA}

\author[0000-0003-3367-3415]{George Helou}
\affiliation{IPAC, California Institute of Technology, 
	1200 E. California Blvd, Pasadena, CA 91125, USA}

\author[0000-0002-6540-1484]{Thomas Kupfer}
\affiliation{Kavli Institute for Theoretical Physics, 
	University of California, Santa Barbara, CA 93106, USA}

\author[0000-0003-2451-5482]{Russ R. Laher}
\affiliation{IPAC, California Institute of Technology, 
	1200 E. California Blvd, Pasadena, CA 91125, USA}

\author[0000-0002-8532-9395]{Frank J. Masci}
\affiliation{IPAC, California Institute of Technology, 
	1200 E. California Blvd, Pasadena, CA 91125, USA}

\author[0000-0001-9515-478X]{Adam A.Miller}
\affiliation{Center for Interdisciplinary Exploration and Research in 
	Astrophysics (CIERA) and Department of Physics and Astronomy, 
	Northwestern University, 1800 Sherman Road, Evanston, IL 60201, USA}
\affiliation{The Adler Planetarium, Chicago, IL 60605, USA}

\author[0000-0001-6806-0673]{Anthony L. Piro}
\affiliation{The Observatories of the Carnegie Institution for Science, 
	813 Santa Barbara St., Pasadena, CA 91101, USA}

\author[0000-0001-7648-4142]{Ben Rusholme}
\affiliation{IPAC, California Institute of Technology, 
	1200 E. California Blvd, Pasadena, CA 91125, USA}

\author[0000-0003-4401-0430]{David L. Shupe}
\affiliation{IPAC, California Institute of Technology, 
	1200 E. California Blvd, Pasadena, CA 91125, USA}

\author[0000-0001-7062-9726]{Roger Smith}
\affiliation{Caltech Optical Observatories, California Institute of Technology, Pasadena, CA  91125}

\author[0000-0003-1546-6615]{Jesper Sollerman}
\affiliation{The Oskar Klein Centre, Department of Astronomy, 
	Stockholm University, AlbaNova, SE-10691 Stockholm, Sweden}

\author[0000-0001-6753-1488]{Maayane T. Soumagnac}
\affiliation{Lawrence Berkeley National Laboratory, 
	1 Cyclotron Road, Berkeley, CA 94720, USA}
\affiliation{Department of Particle Physics and Astrophysics, 
	Weizmann Institute of Science, 234 Herzl Street, 76100 Rehovot, Israel}

\author{Jeffry Zolkower}
\affiliation{Caltech Optical Observatories, California Institute of Technology, Pasadena, CA  91125}

\begin{abstract}

We present observations of ZTF18abfcmjw (\name), a helium-rich supernova with a fast-evolving 
light curve indicating an extremely low ejecta mass ($\approx 0.3\,M_\odot$) 
and low kinetic energy ($\approx 1.2\times 10^{50}\,{\rm erg}$). Early-time ($<$4\,d after explosion) 
photometry reveal evidence of shock cooling from an extended helium-rich envelope of 
$\sim0.1\,M_\odot$ located at $\sim 3\times 10^{12}\,{\rm cm}$ from the progenitor. Early-time 
\ion{He}{II} line emission and subsequent spectra show signatures of interaction with helium-rich 
circumstellar material, which extends from $\gtrsim 5\times 10^{13}\,{\rm cm}$ to $\gtrsim 2\times 
10^{16}\,{\rm cm}$. We interpret \name\ as a helium-rich supernova from an ultra-stripped 
progenitor, which originates from a close binary system consisting of a mass-losing helium star and a 
low-mass main sequence star or a compact object (i.e., a white dwarf, a neutron star, or a black hole). 
We infer that the local volumetric birth rate of 19dge-like ultra-stripped SNe is in the range 
of 1400--8200$\,{\rm Gpc^{-3}\, yr^{-1}}$ (i.e., 2--12\% of core-collapse supernova rate). This can be 
compared to the observed coalescence rate of compact neutron star binaries that are not formed by 
dynamical capture.

\end{abstract}

%% Keywords should appear after the \end{abstract} command. 
%% See the online documentation for the full list of available subject
%% keywords and the rules for their use.
\keywords{supernovae: general -- supernovae: individual (SN2019dge, iPTF14gqr) -- stars: neutron}

%% From the front matter, we move on to the body of the paper.
%% Sections are demarcated by \section and \subsection, respectively.
%% Observe the use of the LaTeX \label
%% command after the \subsection to give a symbolic KEY to the
%% subsection for cross-referencing in a \ref command.
%% You can use LaTeX's \ref and \label commands to keep track of
%% cross-references to sections, equations, tables, and figures.
%% That way, if you change the order of any elements, LaTeX will
%% automatically renumber them.
%%
%% We recommend that authors also use the natbib \citep
%% and \citet commands to identify citations.  The citations are
%% tied to the reference list via symbolic KEYs. The KEY corresponds
%% to the KEY in the \bibitem in the reference list below. 

\vspace{1em}

\section{Introduction}
Type Ibc supernovae (SNe Ibc) are believed to be explosions of massive stars that have lost their 
hydrogen envelopes \citep{Filippenko1997, GalYam2017}. Their typical rise times ($t_{\rm rise}$ in the 
range of 10--25\,d) and peak luminosities ($M_{R\rm , peak}$ between $-17$ and $-19$\,mag) 
suggest ejecta masses ($M_{\rm ej}$) of 1--5\,$M_\odot$ and $^{56}$Ni masses ($M_{\rm Ni}$) of 
0.1--0.4\,$M_\odot$ \citep{Drout2011, Taddia2018, Prentice2019}. The relatively low $M_{\rm ej}$ and 
high rates of SNe Ibc are not compatible with prediction from the evolution of single massive stars, 
whose mass-loss rates are not high enough to strip most of the outer layers \citep{Smith2011, 
Lyman2016}. In contrast, Wolf-Rayet (WR) or helium star descendants of massive stars in close binary 
systems are thought to be the dominant progenitors for the SN Ibc population \citep{Dessart2012, 
Eldridge2013}. 
%The pre-SN star sheds its envelope by mass transfer to the companion, leaving a final 
%envelope mass of $1\,M_\odot$ or more prior to explosion \citep{Yoon2010}.

SNe Ibc with the lowest $M_{\rm ej}$ arise from core-collapse of a stellar core with a small envelope. 
This can occur in tight binaries where a helium star transfers 
mass to a companion that is small in size. Such a scenario was invoked by \citet{Nomoto1994} as one
way to explain the fast evolution of the Type Ic SN1994I with a carbon-oxygen progenitor star of $\sim 
2\,M_\odot$ and $M_{\rm ej }\sim 0.9\, M_\odot$. Should the degree of stripping be even more extreme, 
we may expect the so-called \textit{ultra-stripped} envelope SNe where $M_{\rm ej}$ and $M_{\rm Ni} $ 
are on the order of $0.1 \, M_\odot$ and $0.01 \, M_\odot$, respectively \citep{Tauris2013,Tauris2015, 
Suwa2015}. These weak explosions are one of the two channels to form double neutron 
star (DNS) binaries that are compact enough to merge within a Hubble 
time due to gravitational wave (GW) radiation \citep{Tauris2017}\footnote{The other channel to 
	form compact DNSs is dynamical capture in a dense stellar environment 
	such as a globular cluster \citep{East2012, Andrews2019}.}. Ultra-stripped SNe are therefore 
a promising progenitor channel of multi-messenger sources that can be  jointly studied 
by the LIGO/VIRGO network and electromagnetic efforts \citep{GW170817, MMA, Goldstein2017, 
Coulter17, Hallinan17, Kasliwal2017}.

Compared with canonical SNe Ibc, we expect light curves of ultra-stripped SNe to be rapidly evolving 
and subluminous due to the small amount of $M_{\rm ej}$ and $M_{\rm Ni}$ produced. Among the 
group of faint and fast objects, SN2005ek \citep{Drout2013}, SN2010X \citep{Kasliwal2010}, as well as 
some of the calcium-rich gap transients such as PTF10iuv \citep{Kasliwal2012}, iPTF16hgs 
\citep{DeKC2018}, and SN2019ehk \citep{Nakaoka2020} have been suggested to be good candidates 
for ultra-stripped SNe \citep{Moriya2017}. However, properties of these objects are also consistent 
with alternative interpretations, including core-collapse of stars with extended hydrogen-free envelopes
\citep{Kleiser2014, KleiserFuller2018, KleiserKasen2018}, and explosive detonation of a helium shell on 
the surface of white dwarfs \citep{Shen2010, Sim2012, Polin2019, De2020b, Jacobson-Galan2020}.
%and mergers of white dwarfs with neutron stars \citep{Margalit2016}.

The most convincing ultra-stripped event to date is the Type Ic SN iPTF14gqr \citep{De2018}. Its 
radioactivity-powered emission reveals $M_{\rm ej}\sim 0.2\, M_\odot$ and $M_{\rm Ni}\sim 
0.05\,M_\odot$, whereas the detection of early-time shock cooling signatures shows that the 
progenitor is an extended massive star instead of a white dwarf, and therefore pins down its 
core-collapse origin. Discovered within one day of explosion, iPTF14gqr also demonstrated the 
importance of early-time observations in securely identifying ultra-stripped SNe. 

Here we report the discovery, observations and modeling of the rapidly rising ($t_{\rm rise}\lesssim 
3$\,d) subluminous ($M_{r\rm ,\, peak} \sim -16.3$\,mag) helium-rich event ZTF18abfcmjw (SN2019dge) 
discovered by the Zwicky Transient Facility (ZTF; \citealt{Bellm2019b};  \citealt{Graham2019}). \name\ 
is consistent with being a helium-rich ultra-stripped SN. Section \ref{sec:obs} describes 
the discovery and follow up observations. Section 
\ref{sec:properties} outlines the basic properties of the explosion and its host galaxy. Section 
\ref{sec:modelling} shows modeling of this transient. Section 
\ref{sec:interpretation} provides a discussion on the progenitor system, and Section \ref{sec:rates} 
presents the estimated volumetric rates of 19dge-like ultra-stripped SNe. Section \ref{sec:conclusion} 
gives a conclusion of this paper. Calculations in this paper assume a $\Lambda$CDM 
cosmology with $H_0= 70 \, \rm km \, s^{-1}\, 
Mpc^{-1}$, $\Omega_m = 0.27$ and $\Omega_{\Lambda} = 0.73$ \citep{Komatsu2011}. 
%Alongside this paper, we have released our open-source analysis and all of 
% the data utilized in this study at \url{https://github.com/yaoyuhan/\name,}. 
All spectra and photometry will be made available by the WISeREP repository \citep{Yaron2012} 
following publication.

\section{Observations}  \label{sec:obs}
\subsection{Discovery}

% In this section, I report values on Marshal
\name\ was discovered by ZTF, which runs on the Palomar Oschin Schmidt 48-inch (P48) 
telescope \citep{Dekany2020}. The first real-time alert \citep{Patterson2019} was generated on 2019 
Apr 7 10:18:46 (JD $=2458580.9297$) for a $g$-band detection at $20.66\pm0.34$ mag and J2000 
coordinates $\alpha = 17^{\mathrm{h}}36^{\mathrm{m}}46.75^{\mathrm{s}}$, $\delta = 
+50^{\mathrm{d}}32^{\mathrm{m}}52.2^{\mathrm{s}}$.
On Apr 8, a new alert was flagged by a science program filter on the 
GROWTH Marshal \citep{Kasliwal2019} that is designed to look for fast evolving transients. 

\subsection{Follow-up Observations}
\subsubsection{HST Observation}
\begin{figure}
	\centering
	\includegraphics[width=\columnwidth]{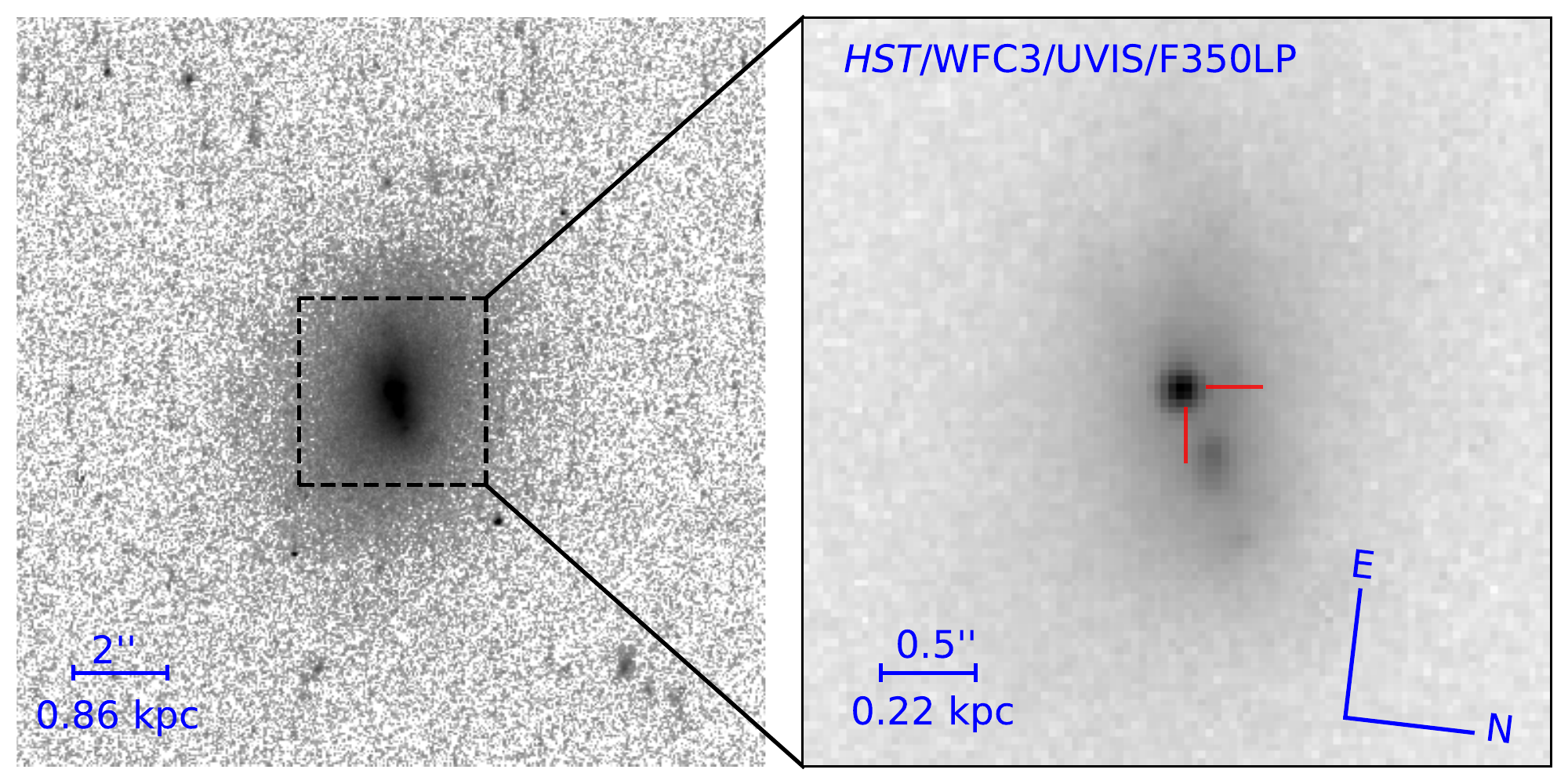}
	\caption{\textit{HST} image of the field on Apr 22 in the F350LP filter at two intensity scales. The 
	position of \name\ is marked by the red crosshairs in the right panel.
		\label{fig:offset}}
\end{figure}
\textit{Hubble Space Telescope} (\textit{HST}) observations were obtained as part of our \textit{HST}
``Rolling Snapshots'' pilot experiment (GO-15675, \citealt{Fruchter2018}). This new observational 
approach requires the PI to update a list of objects of interest each week before the schedule is built, 
giving the scheduler flexibility to choose a possible source for snapshots. Under this program, we 
obtained a NUV spectrum using the WFC3 G280 grism, a short (60\,s) direct image of this field in the 
F300X filter to set the wavelength scale of the spectrum, as well as a longer exposure (200\,s) in the 
F350LP filter. The image in the F350LP filter is shown in Figure~\ref{fig:offset}. It has very similar 
throughput to the zeroth order of the G280 grism. 

\name\ resides in a compact galaxy SDSS J173646.73+503252.3. From our follow up spectra (see 
Section \ref{subsec:spec_properties}), we measure a host redshift of $z=0.0213$ , corresponding to a 
luminosity distance of $D_L = 93$\,Mpc. Figure~\ref{fig:offset} shows that there is a surface brightness 
peak to the northwest of \name\ ($\sim0.2$\,kpc away), which might trace a star-forming region.

\begin{figure*}[htbp!]
	\centering
	\includegraphics[width=\textwidth]{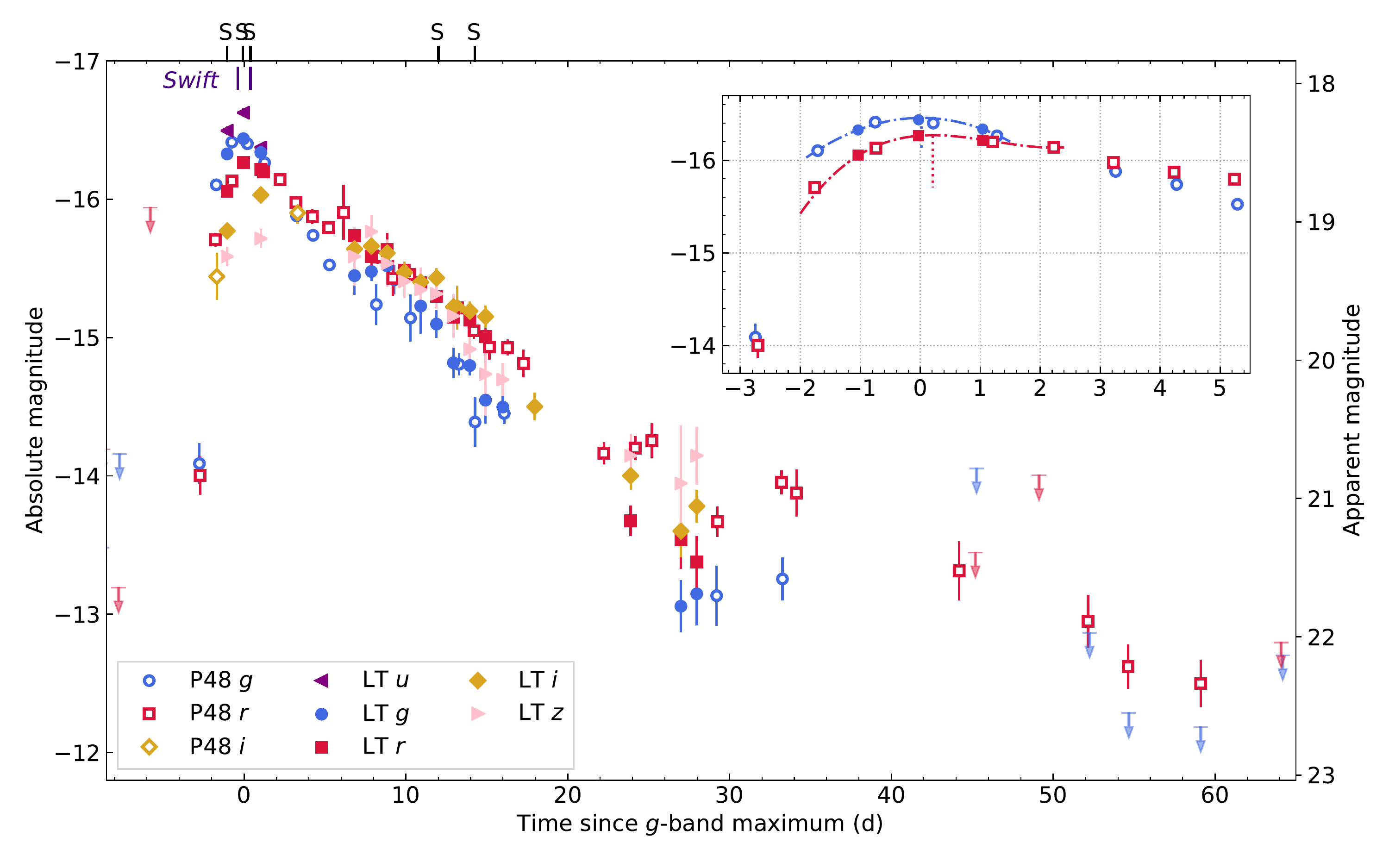}
	\caption{Galactic extinction corrected optical light curve of \name. The inset shows 
		the light curve in $g$ and $r$ bands zoomed around the region of maximum light. Along the 
		upper axis, epochs of spectroscopy are marked with the letter `S' above of the axis, while two 
		epochs of \swift/UVOT/XRT observations are marked below the axis.\label{fig:lightcurve}}
\end{figure*}
\subsubsection{Optical Photometry}
Following \citet{Yao2019}, we perform forced PSF photometry on ZTF difference 
images generated with the ZTF real-time reduction and image subtraction pipeline \citep{Masci2019}. 
ZTF image subtraction is based on the \citet{Zackay2016} image subtraction method. The sky region of 
\name\ is covered by two ZTF fields with ``fieldid'' (i.e., ZTF field 
identifier) 763 and 1799. We exclude all data in field 1799 since the reference image was constructed 
using images obtained between 2018 May 25 and 2019 Jul 12, which is after the explosion of the 
transient. Although the ZTF name of this object (ZTF18abfcmjw) may indicate that the transient was 
discovered in 2018, this is due to an alert generated on 2018 Jul 7 from a candidate detection in 
negative subtraction (reference minus science) in field 763. We note that the seeing during that night 
was $4.2^{\prime\prime}$, larger than 99\% of Palomar nights. The irregularly-shaped PSF might cause 
over-subtraction around the galaxy nucleus in the difference imaging process.

Since field 763 was included in both the Northern sky survey with two epochs (one epoch each in $g$ 
and $r$) per three nights and the extragalactic high-cadence survey with six epochs (three epochs 
each in $g$ and $r$) per night (see \citealt{Bellm2019a} for the design of ZTF experiments), \name\ was 
visited multiple times every night. Therefore, single-night flux measurements in the same 
filter are binned (by taking the inverse variance-weighted average). This gives a pre-explosion 
$r$-band limit of 18.95\,mag (5$\sigma$ limit computed at the expected position of the transient) on 
Apr 4 10:36:34. We convert 5$\sigma$ detections to AB magnitudes for further analysis.

Following the discovery of \name, we obtained follow-up photometry in $ugriz$ with an optical 
imager (IO:O) on the Liverpool Telescope (LT; \citealt{Steele2004}). Digital image subtraction and 
photometry for LT imaging was performed using the Fremling Automated Pipeline (\texttt{FPipe}; 
\citealt{Fremling2016}). \texttt{Fpipe} performs calibration and host subtraction against Sloan Digital 
Sky Survey reference images and catalogs (SDSS, \citealt{Alam2015}).

LT and P48 photometry are shown in Figure~\ref{fig:lightcurve}. Absolute and apparent magnitudes are 
corrected for Galactic extinction $E(B-V)=0.022$\,mag \citep{Schlafly2011}. We assume $R_V=3.1$, 
and adopt the reddening law from \citet{Cardelli1989}. We do not correct for host-galaxy 
contamination given the absence of \ion{Na}{I} D absorption in all spectra at the host redshift. To 
estimate the epoch of maximum light, we interpolated the $g$- and $r$-band photometry with 
three-order polynomial functions, as shown in the inset of Figure~\ref{fig:lightcurve}. The time 
window used in the fit is from ${\rm MJD}=58581.2$ to $58585.2$. \name\ was found to peak 
at $M_{g\rm , peak}=-16.45\pm0.03$\,mag on ${\rm MJD}=58583.19$, and $M_{r \rm ,peak} 
=-16.27\pm0.02$\,mag on ${\rm MJD}=58583.39$. Hereafter we use phase ($\Delta t$) to denote time
with respect to the $g$-band maximum light epoch, ${\rm MJD}=58583.2$.

We obtained one epoch of late-time imaging with the Wafer Scale Imager for Prime (WASP) mounted 
on the Palomar 200-inch telescope at $\Delta t \approx 85$\,d. The data were obtained 
in $r$ band with a total exposure time of 900\,s divided into dithered exposures of 300\,s each. The 
data were reduced using standard techniques as described in \citet{De2020a}. Image subtraction was 
performed using archival reference images from the Dark Energy Legacy Survey \citep{Dey2019}, using 
the method described in \citet{De2020b}. The median 5$\sigma$ limiting magnitude of the image is $r 
\approx 25$\,mag. However, the depth at the transient location is limited by the noise from the bright 
host galaxy, and the transient was not detected to a 5$\sigma$ limiting magnitude of $r = 22.1$\,mag. 

We also performed forced photometry on archival PTF/iPTF difference images spanning 2009 May 07 to 
2016 Jun 13 \citep{Law2009, Rau2009}. No historical detection was found. 

 \begin{deluxetable*}{llrccchh}
\tablecaption{Log of SN2019dge spectroscopy. \label{tab:spec}}
	\tablehead{
		\colhead{Start Time}   
		& \colhead{Instrument}
		& \colhead{Phase}  
		& \colhead{Exposure Time} 
		& \colhead{ Airmass }
		& \colhead{ Resolution (FWHM) }
		& \nocolhead{Observer}
		& \nocolhead{Reducer}\\
		\colhead{(UTC)}   
		& \colhead{  }
		& \colhead{ (day) }
		& \colhead{ (s) }
		& \colhead{  }
		& \colhead{  (\AA) }
		& \nocolhead{  }
		& \nocolhead{  }
	}
	\startdata
	% # DATE-OBS= '2019-04-09T03:30:28.157'             % # UTSTART = '03:30:28.157'
	% # MJD     =         58582.146159
	2019 Apr 09 03:30:28 & LT+SPRAT  & $-1.1$&  500   & 1.80 & 18& Perley & Perley\\
	% # DATE-OBS= '2019-04-10T03:06:09.604'
	% # UTSTART = '03:06:09.604'
	% # MJD     =         58583.129278
	2019 Apr 10 03:06:10 & LT+SPRAT & $-0.1$ &  500   & 1.80 & 18 & Perley & Perley\\
	% LRIS
	2019 Apr 10 14:21:44 & Keck1+LRIS & $+0.4$ & 300 & 1.17 & 6 &Graham & Ho\\
	% HST
	2019 Apr 22 05:08:00 & HST+WFC3+UVIS & $+12.0$& 2$\times$250 & --- & 43& --- & Rubin\\
	% DBSP
	2019 Apr 24 11:06:43 & P200+DBSP & $+14.3$& 1200 & 1.05 & 3--5& Yao & Yao\\
	% LRIS
	2019 Jul 04 11:49:18   & Keck1+LRIS & $+85.3$& 1740 & 1.42 &6  & De&De\\
	% LRIS
	2019 Aug 31 08:04:58   & Keck1+LRIS &$+143.1$ & 1150 & 1.41 & 6 & De&De\\
	% LRIS
	2019 Sep 28 08:14:27   & Keck1+LRIS & $+171.1$& 600 & 2.17 & 6& De&De\\
	% LRIS
	2020 Feb 18 15:23:40   & Keck1+LRIS & $+314.4$& 1450 & 1.38 &6& De&De\\
\enddata
\tablecomments{Phase is measured relative to $g$-band maximum (${\rm MJD}=58583.2$).}
\end{deluxetable*}

\subsubsection{Swift Observation}\label{subsubsec:swift}
Observations with the \textit{Neil Gehrels Swift Observatory} (\swift; 
\citealt{Gehrels2004}) was triggered on Apr 9 and Apr 10. Ultraviolet/Optical Telescope (UVOT; 
\citealt{Roming2005}) data were obtained in the $UVW1$, $UVM2$, $UVW2$, $U$, $B$, and $V$ 
filters. 

UVOT data are reduced using \texttt{HEAsoft} \citep{Heasarc} version 6.17 with a $3^{\prime\prime}$ 
circular aperture. To remove host-galaxy contribution at the location of the SN, we obtained a final 
epoch in all broad-band filters 
on 2019 Jun 23 and measured the photometry with the same aperture used for the transient. We 
present a table of our optical and UV photometry in \ref{sec:appphot_data}.

% /Users/yuhanyao/Documents/GitHub/AT2019dge/meet/SwiftBC.pdf

In parallel with the UVOT observations, \swift\ observed \name\ with its X-ray telescope
(XRT; \citealt{Burrows2005}) between 0.3 and 10\,keV in the photon counting mode. We note that no 
point sources were detected in the XRT event files with $\rm{SNR}>3$. The 3$\sigma$ limits (in 
count\,s$^{-1}$) in the Apr 9, Apr 10, and Jun 23 observations are $7.8\times 
10^{-3}$, $5.8\times 10^{-3}$, and $6.1\times 10^{-3}$, respectively. 
To convert the upper limit count-rate to flux, we adopted the Galactic neutral-hydrogen column 
density of $N_{\rm H} = 2.89\times 10^{20}\,{\rm cm^{-2}}$ towards \name\ \citep{Willingale2013} and a 
power-law spectrum in the form of $N(E) \propto E^{-1}$, where $N(E)$ has the unit of $\rm 
photons\,cm^{-2}\,s^{-1}\,keV^{-1}$. Using the PIMMS web 
tool\footnote{\url{https://heasarc.gsfc.nasa.gov/cgi-bin/Tools/w3pimms/w3pimms.pl}}, we obtained 
unabsorbed flux upper limits in the 0.3--10\,keV band of 5.22, 3.88, and 4.08 $\times 
10^{-13}\,{\rm erg\,s^{-1}\,cm^{-2}}$, corresponding to luminosities of 5.37, 4.00, and 4.20 $\times 
10^{41}\,{\rm erg\,s^{-1}}$. We note that these limits are shallower than X-ray luminosities expected to 
be seen in SNe \citep{OfekXray2013}.

\subsubsection{Radio Follow-up}
Shortly after the discovery of \name, we initiated radio follow-up observations in order to constrain the 
presence of a radio counterpart, as potentially expected in some rapid-rising transients with 
circumstellar interaction \citep{Weiler2007, Horesh2013, HoPhinney2019}. We observed at high 
frequency radio bands using the 
Submillimeter Array (SMA, \citealt{Ho2004}) on UT 2019 Apr 09 between 15:49:17 and 19:51:26 
under its target-of-opportunity program. The 
project ID is 2018B-S047 (PI: Anna Ho). We did not detect \name\ in the resulting image, and the 
3$\sigma$ upper limits are 2.25\,mJy at 230\,GHz and 8.4\,mJy at 
345\,GHz. 
%The actual rms are 0.75 mJy for the 230 GHz image and 2.8 mJy for the 345 GHz image.

\subsubsection{Spectroscopy}
We obtained eight optical spectroscopic follow-up observations of \name\ from $-1.1$\,d to 
$+314.4$\,d relative 
to $g$-band peak using the Spectrograph for Rapid Acquisition of Transients (SPRAT; 
\citealt{Piascik2014}) on the Liverpool Telescope (LT), the Double Spectrograph (DBSP) on the 
200-inch Hale telescope \citep{Oke1982}, and the Low Resolution Imaging Spectrograph (LRIS) on the 
Keck-I telescope \citep{Oke1995}. To extract the LT spectra, we use the automated SPRAT reduction 
pipeline, which is a modification of the pipeline for FrodoSpec \citep{Barnsley2012}. The DBSP spectrum 
was reduced using a \texttt{PyRAF}-based reduction pipeline \citep{Bellm2016}. LRIS 
spectra were reduced and extracted using \texttt{Lpipe} \citep{Perley2019lpipe}. 

A log of our spectroscopic observations is given in Table \ref{tab:spec}. We 
present our sequence of spectra in Figure~\ref{fig:spectra_early}, Figure~\ref{fig:spectra} and 
Figure~\ref{fig:spectra_late}.

\subsection{Host Galaxy Photometry}
To obtain archival photometry of the host galaxy, we retrieved images from the Sloan 
Digital Sky Survey data release (DR9) (SDSS; \citealt{Ahn2012a}), the Panoramic Survey Telescope and 
Rapid Response System (Pan-STARRS, PS1) 
DR1 \citep{Flewelling2016a}, the Two Micron All Sky Survey \citep[2MASS;][]{Skrutskie2006a}, and the 
unWISE images \citep{Lang2014a} from the NEOWISE Reactivation Year-3 \citep{Meisner2017a}. We 
augmented this data set with \swift/UVOT observations that extend our wavelength coverage to 
the UV. The photometry was extracted with the software package \texttt{LAMBDAR} \citep[Lambda 
Adaptive Multi-Band Deblending Algorithm in R;][]{Wright2016a}, to perform consistent photometry on 
images that are neither pixel nor seeing matched, and tools presented in Schulze et al. (in prep). The 
UVOT data were reduced in \texttt{HEAsoft} as described in Section \ref{subsubsec:swift}. The 
measured host photometry is given in \ref{sec:appphot_data}. 

\section{Properties of the Explosion and its Host Galaxy} \label{sec:properties}
\subsection{Light Curve Properties}\label{subsec:lc_properties}

\subsubsection{Peak Luminosity, Rise and Decline Timescale}\label{subsubsec:compare_mag}

\begin{figure}[htbp!]
	\centering
	\includegraphics[width=\columnwidth]{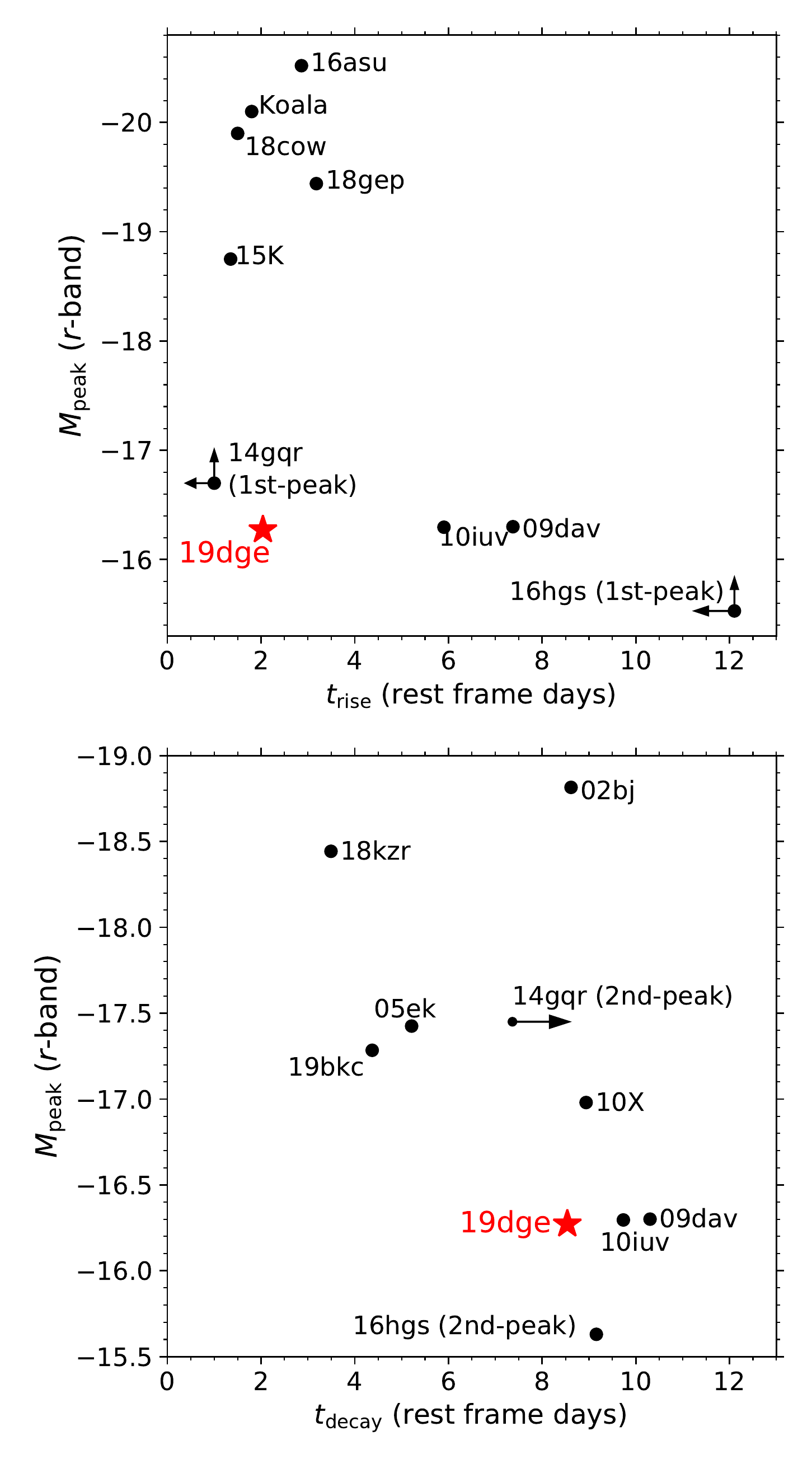}
	\caption{Comparison of the photometric evolution timescales ($t_{\rm rise}$ and $t_{\rm  
			decay}$) and $r$-band peak absolute magnitudes of \name\ (red asterisks) to other 
			fast-evolving transients (black dots). See the text for details.
		\label{fig:compare_mag}}
\end{figure}

The $g$- and $r$-band peak luminosity of \name\ ($\approx -16.3$\,mag) is around the lower limit 
of stripped envelope SNe \citep{Drout2011, Taddia2018, Prentice2019}, and akin to those of the Ca-rich 
gap transients, which occupy the luminosity `gap' between novae and SNe (peak absolute magnitude 
$M_R \approx -15.5$ to $-16.5$\,mag, \citealt{Kasliwal2012}).

To characterize the rise and decline timescales of \name, following \citet{Ho2020}, we calculate rise 
time ($t_{\rm rise}$) defined by how long it takes the $r$-band light curve to rise from 0.75\,mag below 
peak to peak, and decline time ($t_{\rm decay}$) determined by how long it takes to decline from peak 
by 0.75\,mag (corresponding to half of maximum flux). Since \name\ shows no evidence of hydrogen 
(Section \ref{subsec:spec_properties}) and exhibits a fast rise (Figure~\ref{fig:lightcurve}), we compare 
the $t_{\rm rise}$, $t_{\rm decay}$, and peak absolute magnitude between 
\name\ and two other groups of transients: 
\begin{itemize}
	\item Fast-evolving hydrogen-deficient transients that are fainter than normal SNe Ia (i.e. 
	$<-19$\,mag), including
	SN2002bj \citep{Poznanski2010},
	SN2005ek \citep{Drout2013},  
	PTF09dav \citep{Sullivan2011},
	SN2010X \citep{Kasliwal2010},
	PTF10iuv \citep{Kasliwal2012},
	iPTF14gqr \citep{De2018},
	iPTF16hgs \citep{DeKC2018},
	SN2018kzr \citep{McBrien2019}, 
	and SN2019bkc \citep{Chen2020}.
	\item ``Fast evolving luminous transients'' (FELT, \citealt{Rest2018}) or ``fast blue optical 
	transients'' (FBOT, \citealt{Margutti2019}).
	We select well-studied representative objects of this population, including
	KSN2015K \citep{Rest2018},
	iPTF16asu \citep{Whitesides2017}, 
	AT2018cow \citep{Prentice2018, Perley2019},
	SN2018gep \citep{Ho2019},
	and ZTF18abvkwla (also known as the Koala; \citealt{Ho2020}).
\end{itemize}

In Figure \ref{fig:compare_mag}, peak magnitudes are given in (observer-frame) $r$-band, except for 
KSN2015K where we only have observations in the \textit{Kepler} ``white'' filter, and iPTF16asu where 
the rise was only caught in $g$ band. We only correct for Galactic extinction to compute $M_{r\rm , 
peak}$ (assuming no host extinction). Note that iPTF14gqr and iPTF16hgs are SNe exhibiting 
double peaked light curves. Since the rise to first peak was not captured, an upper limit of $t_{\rm 
rise}$ is calculated by taking the time difference between the first $r$-band detection and the latest 
pre-discovery non-detection\footnote{For the second peak, $t_{\rm rise}\sim5$\,d for iPTF14gqr and 
$8<t_{\rm rise}<20$\,d for iPTF16hgs.}, and absolute magnitude of the first $r$-band detection is 
considered to be a fainter limit of $M_{r\rm , peak}$ (plotted in the upper panel). In the lower panel, 
since observations of iPTF14gqr do not extend to 0.75\,mag below its second peak, we present a 
lower limit of its $t_{\rm decay}$.

It is clear from the upper panel of Figure \ref{fig:compare_mag} that \name\ rose faster than normal 
Ca-rich events such as PTF09dav and PTF10iuv. The $t_{\rm rise}$ of $\approx 2.0$\,d is similar to 
the population of FELTs/FBOTs, but \name\ is substantially fainter. In the subluminous regime, 
iPTF14gqr has $t_{\rm rise}$ comparable to \name, and its first peak has been postulated to be caused 
by the diffusion of shock-deposited energy out of an envelope around the progenitor star 
\citep{De2018}.
 
The bottom panel of Figure \ref{fig:compare_mag} shows that $t_{\rm decay}$ of \name\ is 
longer than that for the most rapid-fading SNe Ibc, such as SN2005ek, SN2018kzr, 
and SN2019bkc. Its decay timescale is more similar to SN2002bj, SN2010X, the population of Ca-rich 
transients (PTF09dav, PTF10iuv, iPTF16hgs), and likely iPTF14gqr. It has been suggested that the latter 
group of events have radioactivity powered main peak with low mass of nickel ($M_{\rm Ni} \lesssim 0.1 
\,M_\odot$).

\subsubsection{Bolometric Evolution} \label{subsubsec:bolometric_evolution}
\begin{figure}[!htbp] 
	\centering
	\includegraphics[width=\columnwidth]{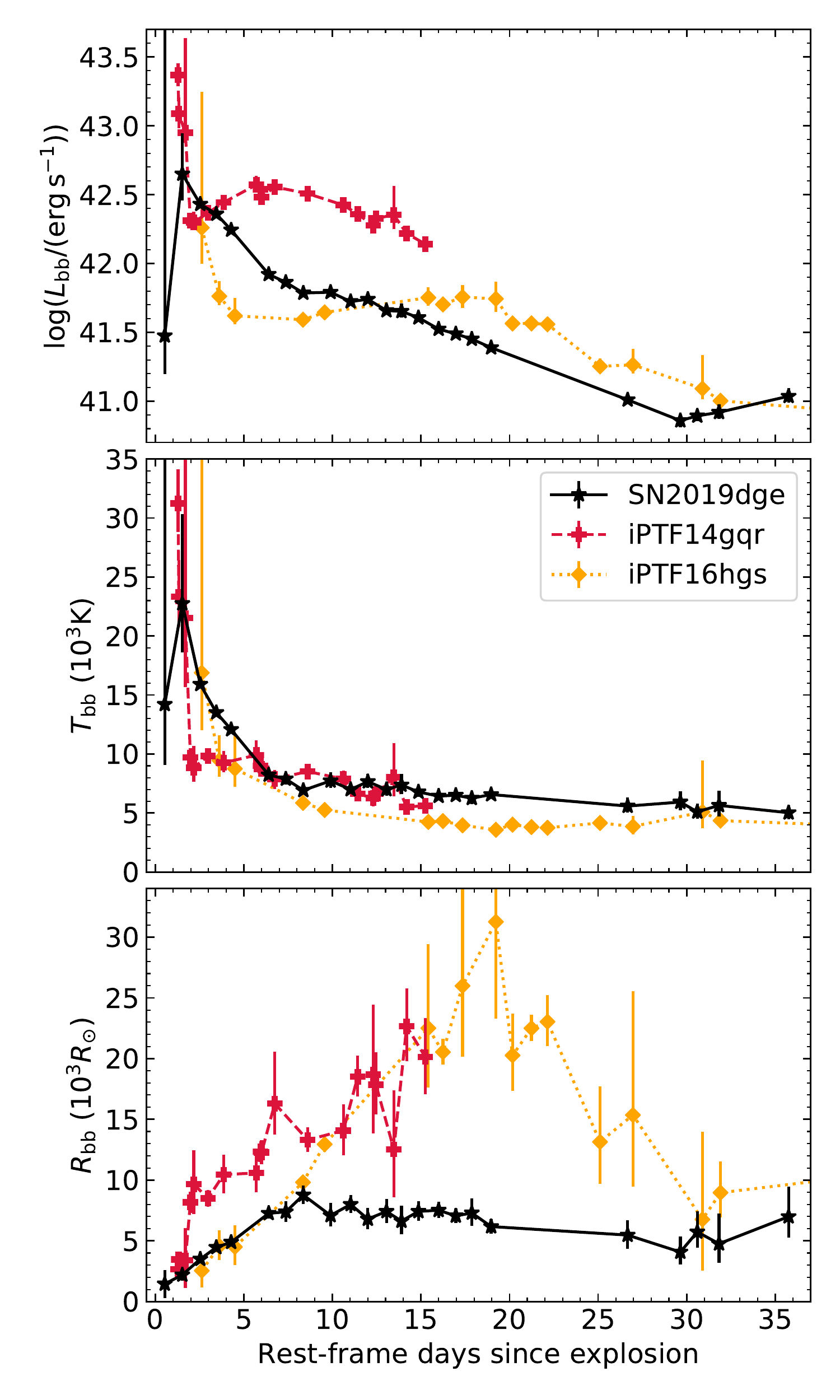}
	\caption{Evolution of blackbody properties (luminosity, temperature, radius) over time of 
		\name\ compared to iPTF14gqr and iPTF16hgs. We use the same method as applied in \name\ to 
		derive $L_{\rm bb}$, $T_{\rm bb}$, and $R_{\rm bb}$ for iPTF14gqr and iPTF16hgs. }
	\label{fig:Tbb_Rbb_Lbb}
\end{figure}
\begin{figure*}[htbp!]
	\centering
	\includegraphics[width=\textwidth]{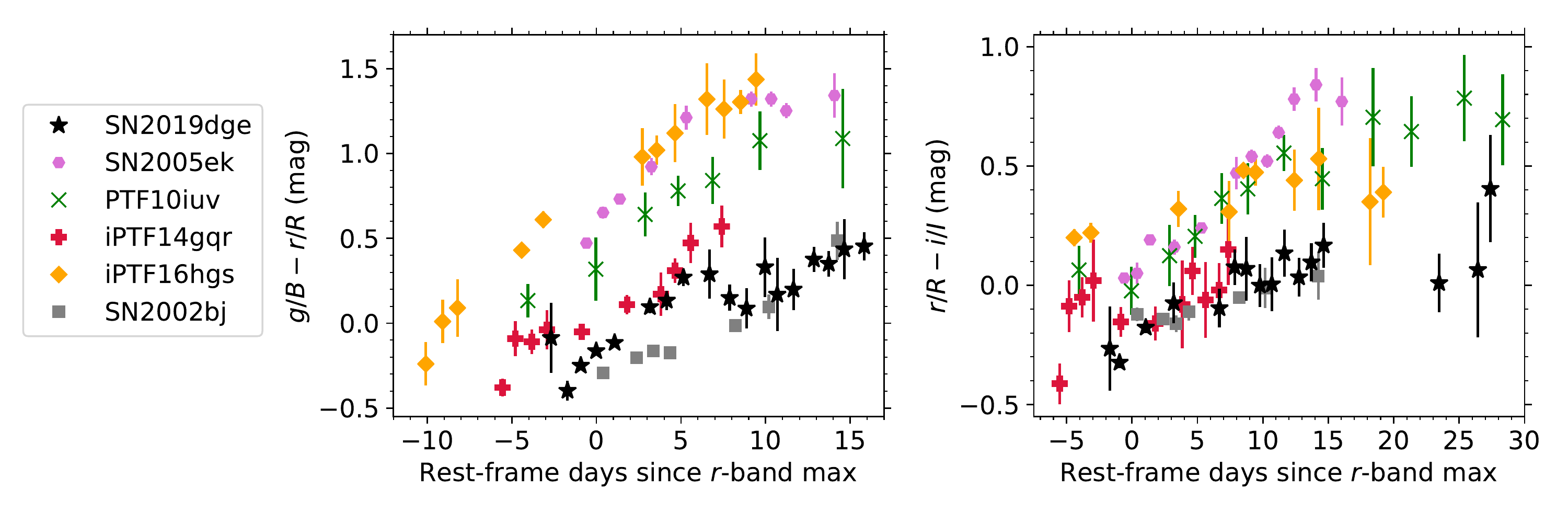}
	\caption{Comparison of the color evolution of \name\ with a subset of fast SNe shown in 
		Figure~\ref{fig:compare_mag}. All colors have been corrected for Galactic extinction. Due to 
		absence of photometry in identical filters, we compare colors in corresponding filter pairs of 
		$B$/$g$, $R$/$r$ and $I$/$i$. Since all SNe shown here are at relatively low redshifts ($z\leq 
		0.063$), the observed colors probe similar rest-frame bands. \label{fig:compare_color}}
\end{figure*}

We constructed the bolometric light curve at epochs where at least detections in two 
filters are available by fitting the spectral energy distribution (SED) with a blackbody function (see 
details of model fitting in Appendix \ref{subsec:bbfit}). We plot the physical evolution of \name\ 
with comparisons to iPTF14gqr and iPTF16hgs in Figure~\ref{fig:Tbb_Rbb_Lbb}, where we have 
adopted the explosion epoch of iPTF14gqr, iPTF16hgs, and \name\ estimated by \citet{De2018}, 
\citet{DeKC2018}, and Section \ref{subsec:fastrise} of this paper, respectively. The bolometric 
luminosity of \name\ reaches $\sim 5\times 10^{42}\,{\rm erg\, s^{-1}}$ 
at $\sim 1.5$\,d after the explosion epoch. The subsequent decline displays an initial 
fast drop of $0.36\,{\rm mag\,d^{-1}}$ at age 2--9\,d, and transitions to a slower drop of $0.11\,{\rm 
mag\,d^{-1}}$ at age 10--30\,d. 

The bolometric temperature of \name\ reaches as high as $\sim 2.3\times 10^4$\,K at age $1.5$\,d 
and rapidly falls afterwards. The maximum $T_{\rm bb}$ is much hotter than that observed in normal 
SNe Ibc (6000--10000\,K, \citealt{Taddia2018}). Its early light curve evolution is slower than iPTF14gqr, 
but similar to iPTF16hgs and several other stripped envelope SNe displaying double-peaked light curves 
(e.g., see Figure~2 of \citealt{Fremling2019}). Their first peaks have been modelled by cooling emission 
from an extended envelope around the progenitor after the core-collapse SN (CCSN) shock breaks out 
\citep{Modjaz2019}. After $\sim 8$\,d past explosion, $T_{\rm bb}$ flattens to $6000\pm1000$\,K, 
similar to the behavior of normal SNe Ibc at a much later phase ($\sim30$\,d after explosion, 
\citealt{Taddia2018}).

Assuming that the photospheric radius can be approximated by $R_{\rm bb}$ and linearly expands at 
early phase, we fit a linear function to the first few $R_{\rm bb}$ vs.~time measurements of \name\, 
which gives $\approx 8000\, {\rm km\,s^{-1}}$. The radius then remains flat at $\sim 6.7\times 
10^3\,R_\odot$ during age 8--30\,d, and even appears to slowly recede. The reason for this is that the 
temperature drops to a recombination temperature for helium and the opacity becomes small. As a 
result, the outer layers of SN ejecta becomes more transparent, and deeper regions of the ejecta are 
being probed \citep{Piro2014}.

\subsubsection{Color Evolution}
We compare the color curves of other fast transients to that of \name\ in 
Figure~\ref{fig:compare_color}, in corresponding pairs of $B$/$g-R$/$r$ and $R$/$r-I$/$i$ colors. For 
the double-peaked events iPTF14gqr and iPTF16hgs, ``maximum'' time corresponds to epoch of 
maximum light in the second peak.

The early-time blue color of \name\ arises from the high-temperature peak. Among other events, 
SN2002bj, iPTF14gqr and iPTF16hgs exhibit early colors as blue as \name. Subsequently, 
\name\ displays a color starting out blue and turning redder with time, consistent with a cooling 
process. 

One unusual feature of \name\ is that at $\sim6$--9\,d after maximum light, the $g-r$ color becomes 
bluer by $\approx 0.2$\,mag, while after that the color continues to redden. We notice that iPTF14gqr
exhibits a similar trend --- around 4\,d before the second peak, its $g-r$ color stays flat before getting 
redder afterwards, while around 2\,d before the second peak, its $r-i$ color also turns bluer by 
$\approx 0.2$\,mag.

\begin{figure*}[htbp!]
	\centering
	\includegraphics[width=\textwidth]{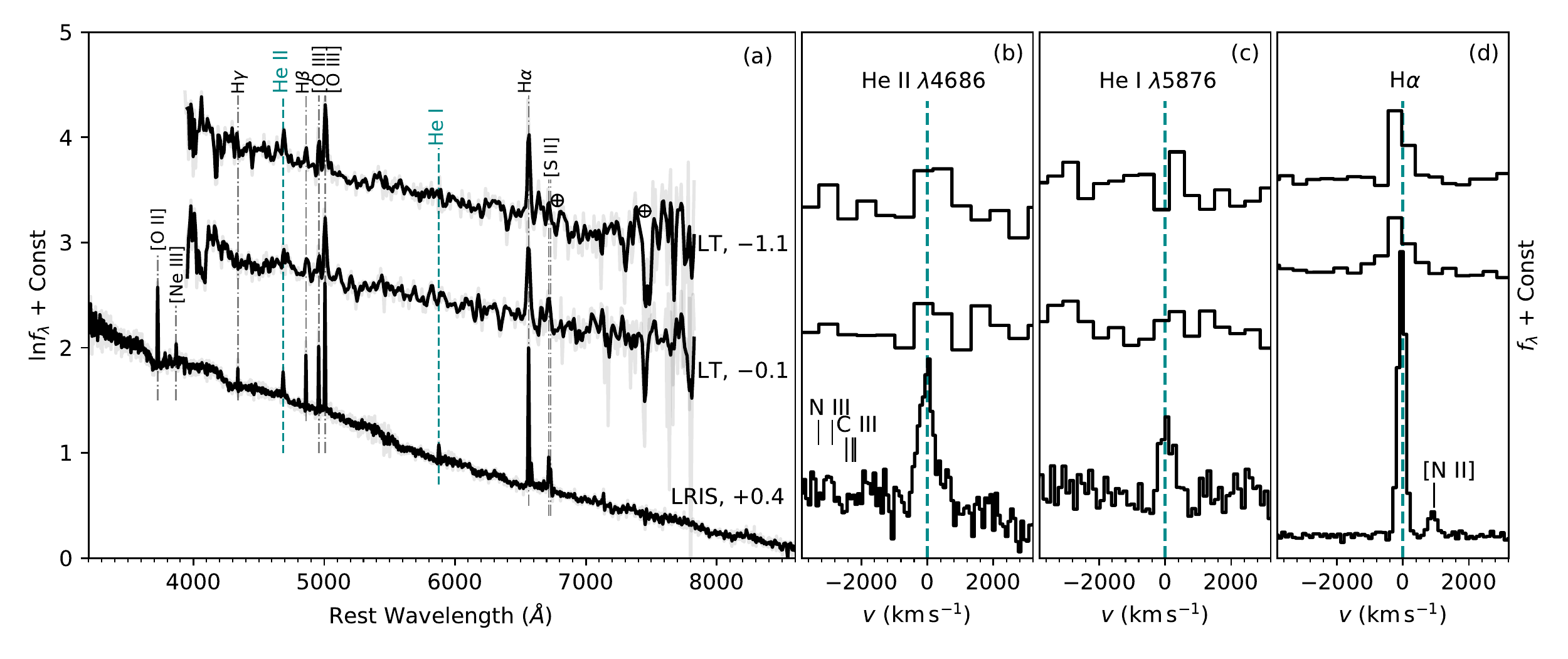}
	\caption{Early-time spectra of \name.  In panel (a), the original spectra are	
		shown in translucent colors, with the overlying black lines showing the same spectra convolved 
		with an ${\rm FWHM} = 800\, {\rm km\, s^{-1}}$ (for LT)  or ${\rm FWHM} = 200\, {\rm km\, 
			s^{-1}}$ (for LRIS) Gaussian kernel. Prominent galaxy lines are marked by the 
		dash-dotted lines. In panel (b) (c) and (d), we show the observed spectra (not convolved) in 
		velocity space around the \ion{He}{II} $\lambda 4686$, \ion{He}{I} $\lambda 5876$ and 
		H$\alpha$ emission lines.
		\label{fig:spectra_early}}
\end{figure*}

\subsection{Spectroscopic Properties}\label{subsec:spec_properties}
%The phases of the spectra indicated in this section are relative to $g$-band peak. 
\begin{deluxetable}{cccc}[htpb!]
	\tablecaption{Rest-frame FWHM ($\rm km\, s^{-1}$) of narrow emission lines in the $+0.4$\,d, 
	$+85.3$\,d and $+314.4$\,d LRIS spectra. 
	\label{tab:eml_fwhm}}
	\tablehead{
		\colhead{Transition}
		&\colhead{$+0.4$\,d}
		&\colhead{$+85.3$\,d}
		&\colhead{$+314.4$\,d}
	}
\startdata
\ion{He}{II} $\lambda4686$ & $552\pm36$&---& ---\\
\ion{He}{I} $\lambda5876$	& $582\pm86$ & $272\pm 19$ & $298\pm24	$ \\
\ion{He}{I} $\lambda 6678$ &--- & $282\pm28$ &$339\pm59$ \\
\ion{He}{I} $\lambda 7065$ &--- & $230\pm37$ &$218\pm26$ \\
H$\alpha$                          &$291\pm49$& $263\pm 13$ & $280\pm14$\\ \relax
[\ion{S}{II}] $\lambda 6716$ & $285\pm 52$&$254\pm14$ & $274\pm18$\\ \relax
[\ion{S}{II}] $\lambda 6731$ &$263\pm78$ &$251\pm14$ & $264\pm15$\\\relax
[\ion{O}{I}] $\lambda 6300$ &--- & $263\pm28$ & $231\pm24$\\
\enddata
\end{deluxetable}

\subsubsection{Early Spectral Evolution} \label{subsubsec:spec_early}
The very early spectra at $-1.1$, $-0.1$, and $+0.4$\,d show a blue continuum and strong galaxy 
emission lines from the underlying \ion{H}{II} region (see Figure~\ref{fig:spectra_early}). In 
addition, these spectra also show prominent \ion{He}{I} $\lambda5876$ and high-ionization \ion{He}{II} 
$\lambda4686$ narrow emission lines. We computed the equivalent width (EW) of \ion{He}{II} emission 
using the spectral line and continuum wavelength ranges given by \citet{Khazov2016}. The EW is found 
to be $-7.56\pm 1.07$\,\AA, $-2.66\pm 1.30$\,\AA, and $-3.77\pm 0.16$\,\AA\ in the $-1.1$\,d, 
$-0.1$\,d, and $+0.4$\,d spectra. 

In Table~\ref{tab:eml_fwhm}, we show the measured full width at 
half-maximum intensity (FWHM) velocities of some emission lines by fitting a 
Gaussian to the line profile. Since the [\ion{S}{II}] $\lambda\lambda 6716$, 6731 doublet is definitely 
from the host galaxy, their line widths serve as a practical measurement of instrumental 
line-broadening. As shown in column 2, FWHM velocities of the \ion{He}{II} and \ion{He}{I} emission 
lines are $\sim 550\,{\rm km\,s^{-1}}$ and $\sim 580\,{\rm km\,s^{-1}}$, much broader than the 
resolution of $\approx 270\,{\rm km\,s^{-1}}$, whereas H$\alpha$ is not well resolved. Thus, we infer 
that the hydrogen emission is from the host galaxy, while the helium lines are from photoionized 
material exterior to the SN.

\begin{figure*}[htbp!]
	\centering
	\includegraphics[width=\textwidth]{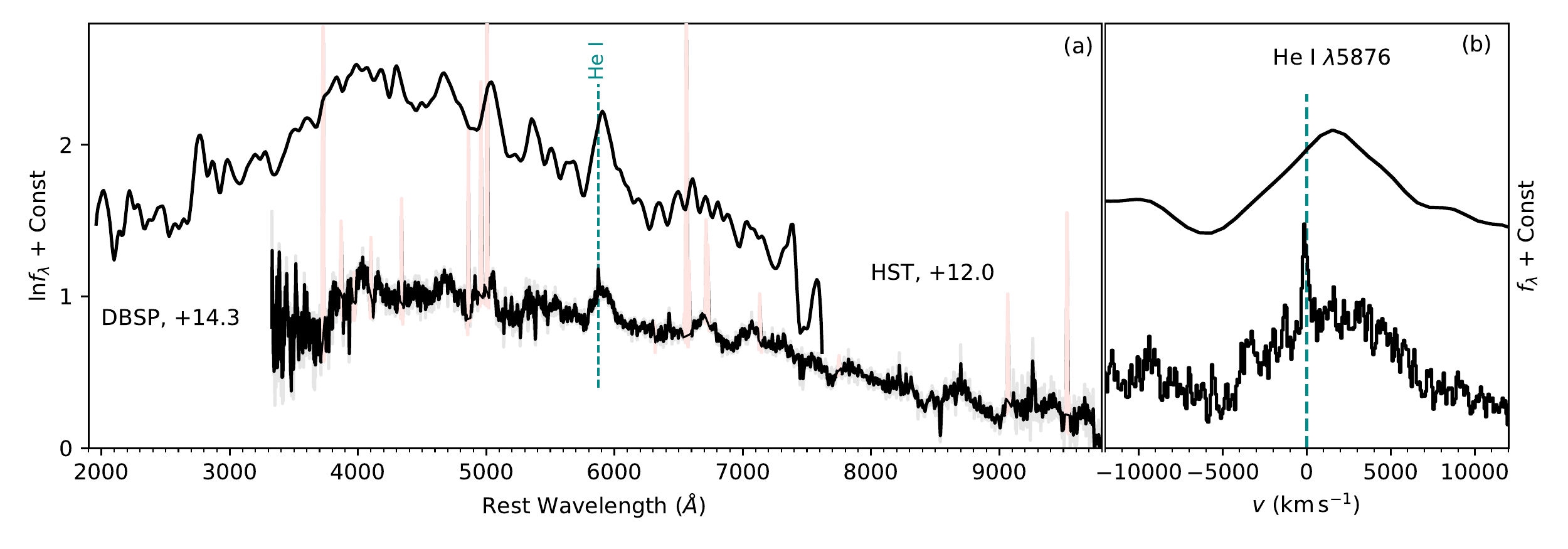}
	\caption{Photospheric phase spectra of \name. In panel (a), the original DBSP 
		spectrum is shown in translucent colors, with the overlying black lines showing the same 
		spectrum 
		convolved with an ${\rm FWHM} = 200\, {\rm km\, s^{-1}}$ Gaussian kernel. We mask prominent 
		galaxy lines in the DBSP spectrum in light red. In panel (b), we show the observed spectra (not 
		convolved with any kernels) in velocity space around \ion{He}{I} $\lambda 5876$ .
		\label{fig:spectra}}
\end{figure*}

\subsubsection{Photospheric Phase Spectral Evolution}  \label{subsubsec:spec_middle}

\begin{figure}[htbp!]
	\centering
	\includegraphics[width=\columnwidth]{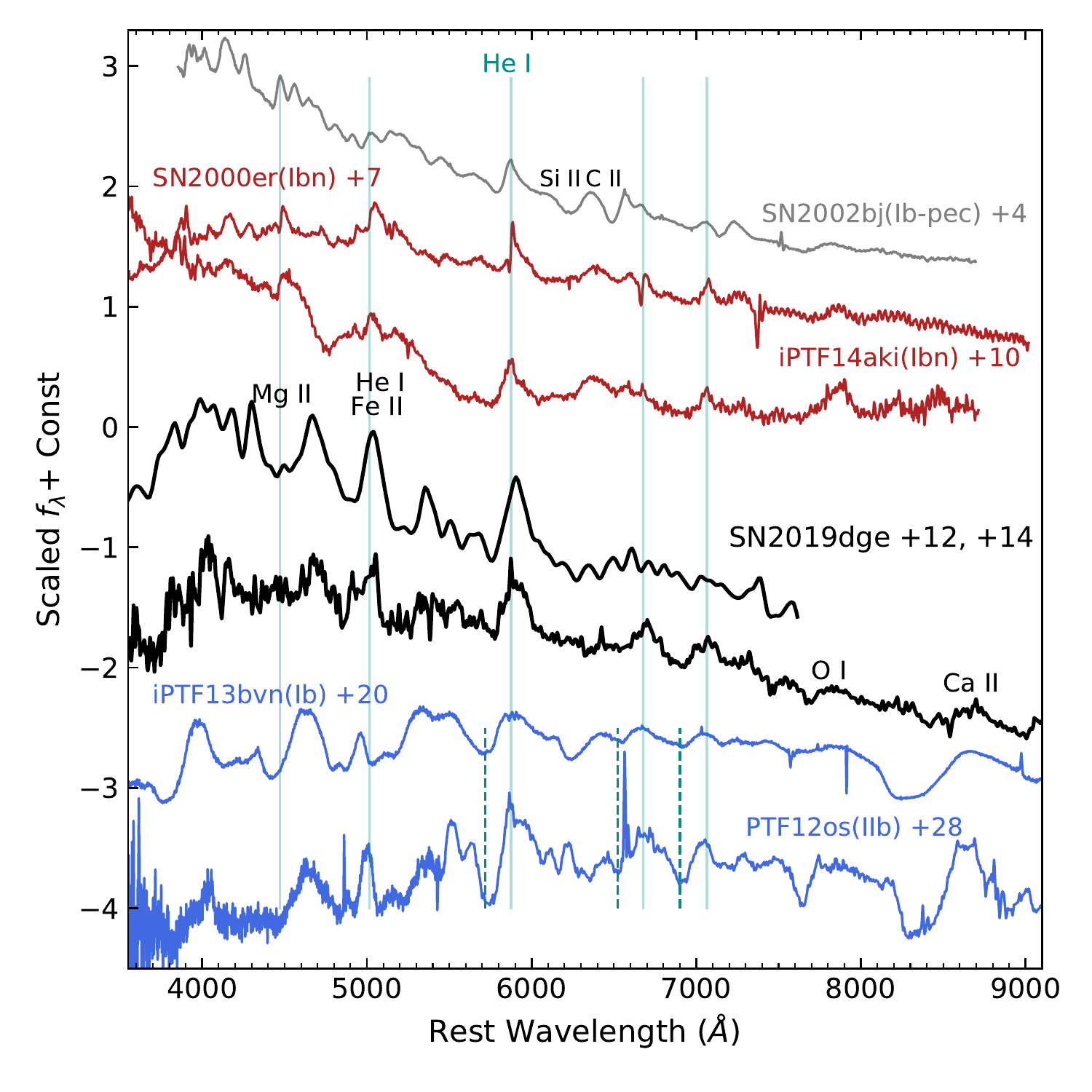}
	\caption{Photospheric phase spectra of \name\ compared with 
		other SNe, including SN2000er \citep{Pastorello2008}, SN2002bj \citep{Poznanski2010}, iPTF14aki 
		\citep{Hosseinzadeh2017}, PTF12os and iPTF13bvn \citep{Fremling2016}. \ion{He}{I} transitions at 
		rest wavelength are marked by the vertical cyan lines (though note that not all of these lines are 
		visible in all spectra shown here).
		\label{fig:hst_opt}}
\end{figure}

\begin{figure}[htbp!]
	\centering
	\includegraphics[width=\columnwidth]{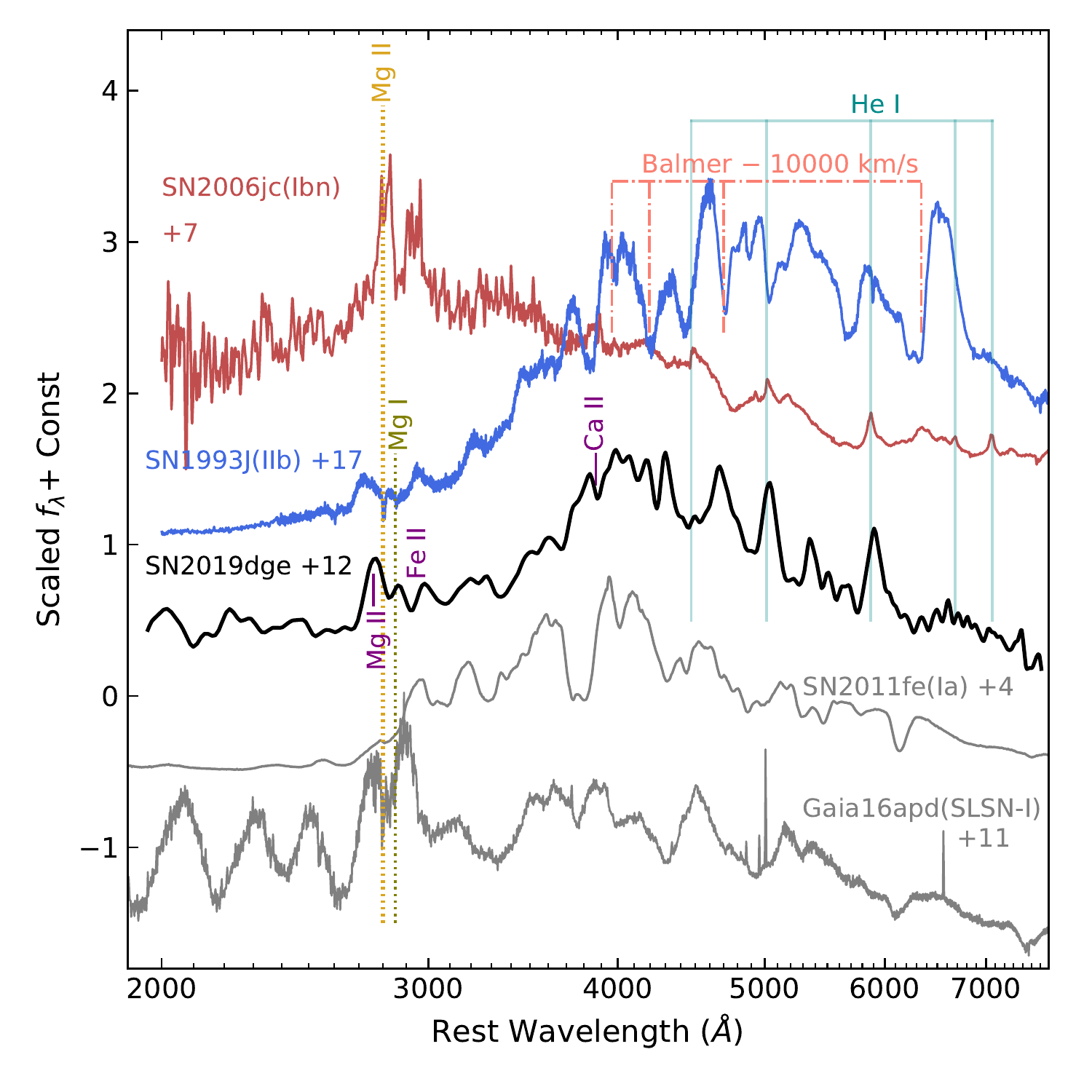}
	\caption{\textit{HST} spectrum of \name\ compared with other SNe, including SN2006jc 
		\citep{Bufano2009}, SN1993J \citep{Jeffery1994}, SN2011fe \citep{Mazzali2014}, and Gaia16apd 
		\citep{Yan2017}. The slightly blueshifted \ion{Mg}{II} $\lambda 2800$ emission feature is observed 
		in SN2006jc, SN1993J, SN2019dge, and Gaia16apd.
		\label{fig:hst}}
\end{figure}

Broad transient features are present in the $+12.0$ and $+14.3$\,d spectra (Figure~\ref{fig:spectra}). 
These spectra are taken at the photospheric phase where emission comes from a photosphere 
receding (in mass coordinates) back through freely expanding SN ejecta. 
The \textit{HST} spectrum contains little host-galaxy contamination 
due to its high angular resolution. Prominent galaxy emission lines in the DBSP 
spectrum are identified and plotted in light red to emphasize transient features. The existence of the
P-Cygni \ion{He}{I} $\lambda5876$ profile and non-existence of hydrogen nominally classify \name\ as 
a Type Ib SN. We measure the velocity of the \ion{He}{I} $\lambda5876$ line by fitting a parabola to the 
absorption minimum. The resulting fits give velocities of $\approx 6000\, {\rm km\,s^{-1}}$ and $ 5900\, 
{\rm km\,s^{-1}}$ for the $+12.0$\,d and $+14.3$\,d spectra, respectively. This is lower than velocities 
of normal SNe Ib measured from the \ion{He}{I} $\lambda5876$ absorption minimum ($\sim 10^4\, \rm 
km\, s^{-1}$, \citealt{Liu2016}), but higher than that in Type Ibn SNe ($\sim 3000\, \rm km\, s^{-1}$, 
\citealt{Hosseinzadeh2017}). 

In Figure~\ref{fig:hst_opt}, we compare the photospheric phase optical spectra of 
\name\ with other helium-rich events. Note that the DBSP spectrum has host emission lines 
masked. \name\ is different from normal helium-rich stripped envelope SNe Ib/IIb or SNe Ibn in the 
sense that its P-Cygni absorption minimum in the \ion{He}{I} $\lambda5876$ line is weaker. The 
feature at $\sim5000{\rm \AA}$ is often attributed to \ion{He}{I} $\lambda 5016$ and \ion{Fe}{II} triplet 
$\lambda\lambda\lambda4924$, 5018, and 5169 \citep{Liu2016}. The shape of this feature in 
\name\ is similar to those in normal SNe Ib/IIb at much later phases ($\sim 20$\,d post maximum), 
indicating that the spectral evolution of \name\ is faster. The complex absorption profile at 
$\sim4500\,{\rm \AA}$ has been identified as a blend of \ion{Fe}{II}, \ion{Mg}{II} $\lambda 4481$ 
and \ion{He}{I} $\lambda 4472$ \citep{Hamuy2002}. In the DBSP spectrum, we detected \ion{O}{I} 
$\lambda 7774$ and broad \ion{Ca}{II} at $\sim8500\,{\rm \AA}$ (due to the
$\lambda\lambda$ 8498, 8542, and 8662 triplet) with clear P-Cygni profiles; Both are major features 
of stripped envelope SNe 
\citep{GalYam2017}. 

In Figure~\ref{fig:hst}, we compare the \textit{HST} NUV spectrum with spectra of other types of SNe. 
The UV part of \name\ is much weaker than a blackbody extrapolation of the optical spectra would 
predict. This has also been seen in normal thermonuclear and CCSNe, and interpreted as strong 
metal-line blanketing caused by iron-peak elements, particularly \ion{Fe}{II} and \ion{Co}{II} 
\citep{Gal-Yam2008}. \name\ bears a close resemblance to SN1993J between 2000\,\AA and 4000\,\AA. 
In Figure~\ref{fig:hst}, we also marked the rest wavelength of \ion{Mg}{I} $\lambda2852$ and 
\ion{Mg}{II} $\lambda \lambda 2796$, 2803. The emission features at $\sim2760\,{\rm \AA}$ in \name\ 
and Gaia16apd are similar to the bump at $\sim2730\,{\rm \AA}$ in SN1993J, which was found to be a 
NLTE \ion{Mg}{II} emission line \citep{Jeffery1994}. This resonance line is blueshifted from its rest 
wavelength, and is suggested to come from a circumstellar region that is distinctly separated from the 
SN photosphere in velocity and excitation conditions \citep{Panagia1980, Fransson1984}.

\subsubsection{Late-time  Spectral Evolution}
\begin{figure*}[htbp!]
	\centering
	\includegraphics[width=\textwidth]{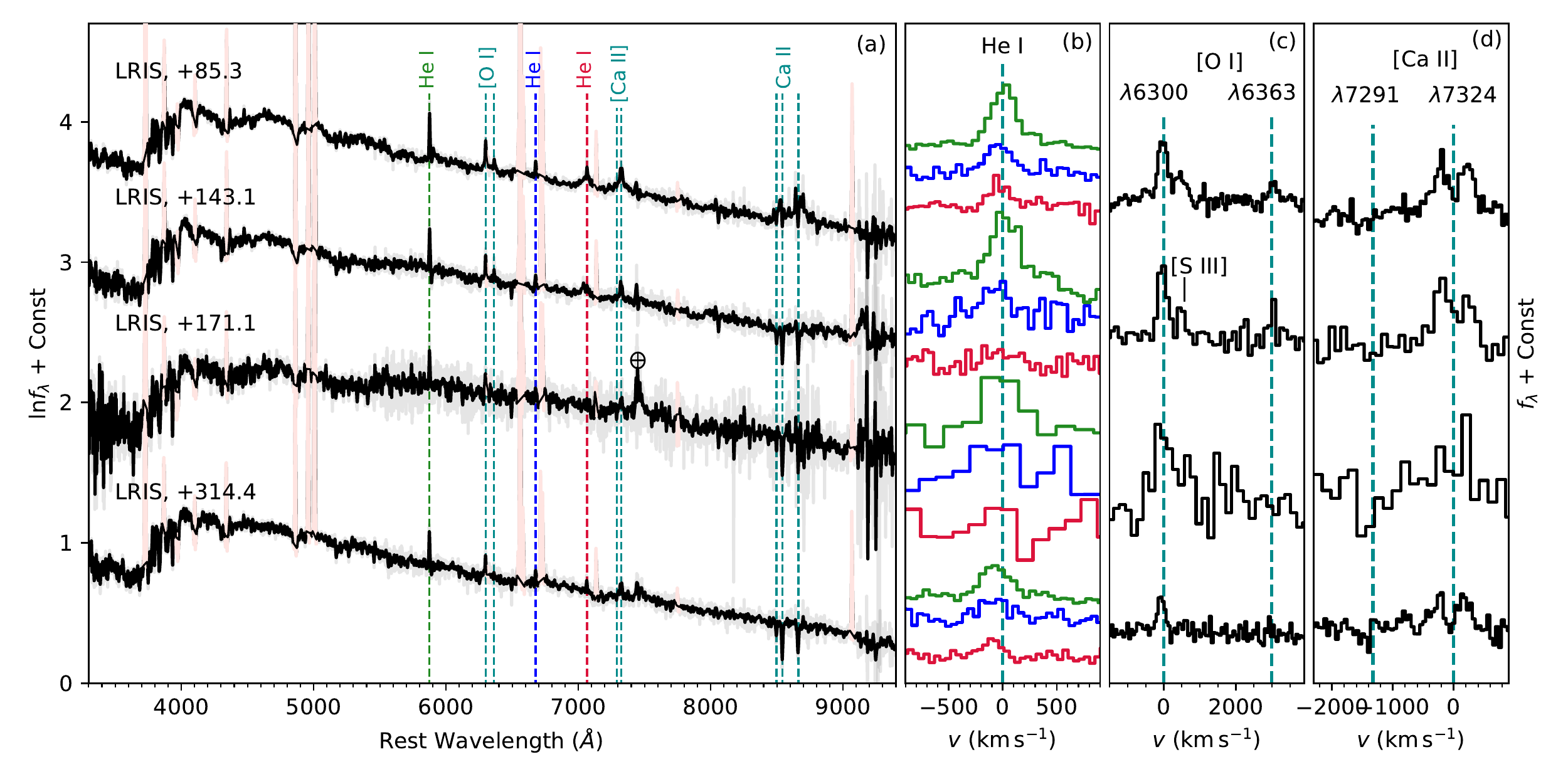}
	\caption{Late-time spectra of \name. In panel (a), the original spectra are	
		shown in translucent colors, with the overlying black 
		lines showing the same spectra convolved with ${\rm FWHM} = 200\, {\rm km\, 
			s^{-1}}$ Gaussian kernels. We mask prominent galaxy lines in light red. Possible SN features are 
		marked by the dashed lines. In panel (b) (c) and (d), 
		the spectra at phase $+85.3$\,d, $+143.1$\,d, $+171.1$\,d, and $+314.4$\,d, are 
		binned by 1, 2, 3, and 1 pixel(s), respectively (1.16\,\AA\ per pixel). The binning 
		factors are chosen based on the 
		different signal-to-noise ratio (SNR) in these spectra (see exposure times in 
		Table~\ref{tab:spec}). Note that in panel (b), we plot evolution of \ion{He}{I} $\lambda 5876$, 
		$\lambda 6678$, and $\lambda 7065$ emissions in green, blue, and crimson, respectively.
		\label{fig:spectra_late}}
\end{figure*}
Figure~\ref{fig:spectra_late} shows late time spectra of \name\ obtained at $+85.3$, 
$+143.1$, $+171.1$, and $+314.4$\,d.  The general shape of the spectra is determined by the host 
galaxy, while possible SN features are marked by the dashed lines. The right panels (b), (c), and (d) 
highlight emission lines at wavelengths of \ion{He}{I}, [\ion{O}{I}], and [\ion{Ca}{II}]. In panel (c) of 
Figure~\ref{fig:spectra_late}, the [\ion{O}{I}] $\lambda \lambda 6300, 6363$ feature 
consists of two narrow emission peaks. This doublet transitions share the same upper level ($\rm 
^{3}P_{1,2}$--$\rm ^{1}D_2$). The observed intensity ratio $R \equiv F(6300/6364) \sim 3.1$ agrees 
with the nebular condition, as one would expect in the optically thin regime \citep{Leibundgut1991, 
Li1992}. In panel (d), we mark the position of the [\ion{Ca}{II}] doublet in dashed lines, but only the 
$\lambda 7324$ line is clearly detected. It presents a double-peaked profile with a peak separation of 
$\sim 400\,{\rm km\,s^{-1}}$. 

\begin{figure*}
	\centering
	\includegraphics[width=\textwidth]{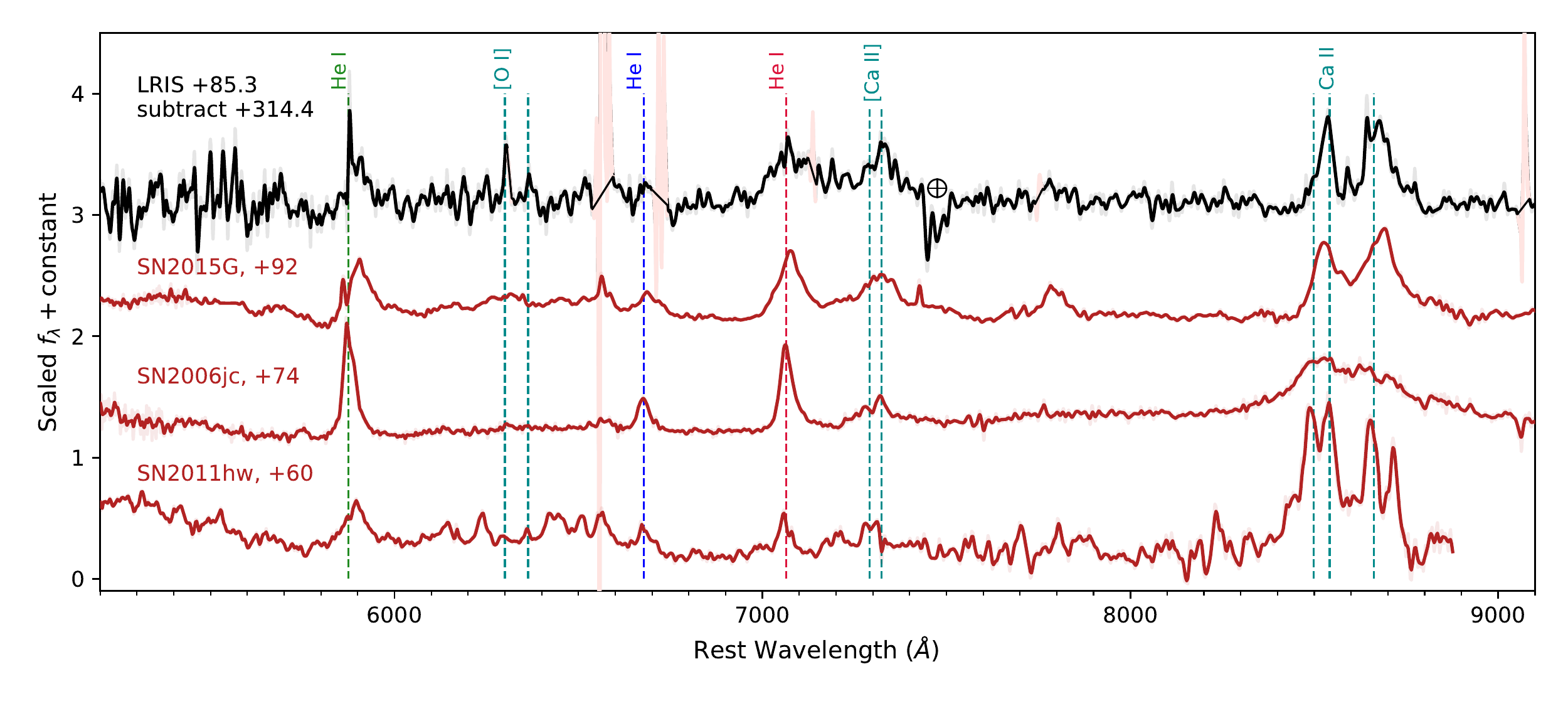}
	\caption{Subtracted late-time spectrum of \name\ compared with Type Ibn SNe SN2006jc 
		\citep{Shivvers2019}, 
		SN2011hw \citep{Pastorello2015}, and SN2015G \citep{Shivvers2017}.
		\label{fig:spec_subtract}}
\end{figure*}

From panel (a) of Figure~\ref{fig:spectra_late}, it is also clear that in the $+85.3$\,d spectrum, the 
\ion{He}{I} and [\ion{Ca}{II}] lines have broader emission components with Lorentzian profiles at the 
bases of the narrow emission lines. These Lorentz-shape components are not visible in the 
$+314.4$\,d spectrum. Therefore, to further investigate the broader features, we subtract the 
$+314.4$\,d spectrum from the $+85.3$\,d spectrum. The resulting subtraction 
(Figure~\ref{fig:spec_subtract}) reveals intermediate-width (FWHM $\sim 2000\,{\rm km\,s^{-1}}$) 
components of \ion{He}{I}, [\ion{Ca}{II}], and the \ion{Ca}{II} IR triplet. It shares a close resemblance to 
some Type Ibn SNe, such as SN2011hw \citep{Pastorello2015} and SN2015G \citep{Shivvers2017}. 
These intermediate-width features are too narrow to be explained by emission from
%radioactivity-heated optically thin 
SN ejecta. Instead, they are probably emitted by a cold dense CSM 
shell formed by radiative cooling from the post-shock material, as has been proposed to be the case in 
interacting Type IIn/Ibn SNe \citep{Chugai1994, Smith2017}. 

Table~\ref{tab:eml_fwhm} (column 3 and 4) gives the measured FWHM velocities of narrow emissions 
shown in panel (b), (c), and (d) of Figure~\ref{fig:spectra_late}. It can 
be seen that the measured FWHM of other emission lines are similar to the [\ion{S}{II}] line-width. 
Therefore, we conclude that the observed narrow emissions are not resolved.

Due to the low resolution of our LRIS spectra, we cannot directly rule out the possibility that the 
narrow lines are emanating from the host galaxy. However, there are evidence indicating that they are 
not merely a background contamination of an underlying \ion{H}{II} region: 
\begin{enumerate}[label=(\roman*)]
	\item In the $+85.3$\,d spectrum, the \ion{He}{I} and [\ion{Ca}{II}] narrow emissions are on top of 
	intermediate-width Lorentzian components characteristic of electron scattering \citep{Huang2018}, 
	which fades away in the $+314.4$\,d spectrum. However, the hydrogen Balmer lines do not have a 
	broader base in any of our spectra.
	\item The flux intensities of \ion{He}{I}, [\ion{O}{I}], and [\ion{Ca}{II}] lines decrease by a 
	factor of approximately two from $+85.3$\,d to $+314.4$\,d, consistent with the temporal 
	evolution from an emission mechanism connected to the aging supernova. As a comparison, the line 
	strengths  of the strongest emissions in normal ionized nebulae (H$\alpha$, [\ion{O}{III}], 
	[\ion{O}{II}], [\ion{S}{II}], etc) do not follow this behavior.
\item Although the \ion{He}{I} and [\ion{O}{I}] lines labelled in panel (a) of Figure~\ref{fig:spectra_late} 
have been observed in \ion{H}{II} regions \citep{Peimbert2000, Peimbert2017}, the doublet [\ion{Ca}{II}] 
$\lambda \lambda 7291$, 7324 has not been detected in gaseous nebulae \citep{Kingdon1995}.
\end{enumerate}

Taken together, we suggest that the narrow components ($\lesssim 270\,{\rm km\,s^{-1}}$) of 
\ion{He}{I}, [\ion{Ca}{II}], [\ion{O}{I}] and \ion{Ca}{II} are also associated with the transient. Their widths 
might be consistent with the typical velocities of pre-shock CSM. The detection of these lines at 
$>300$\,d after the SN explosion suggests that the circumstellar shell extends to $\gtrsim 2\times 
10^{16}\,{\rm cm}$ ($\sim 1000$\,AU) from the progenitor.\footnote{Adopting a conservative shock 
velocity estimation of 
	$v_s\approx10^4\,{\rm km\,s^{-1}}$, the forward shock travels $2.6\times 10^{16}\,{\rm cm}$ after 
	300\,d.}

\subsection{Host Galaxy Properties} \label{subsec:host}
We measure properties of the host galaxy using the spectrum obtained at phase $+314.4$\,d, 
assuming that the most prominent nebular line emissions of H$\alpha$ and [\ion{N}{II}] 
are from the host. The Galactic extinction corrected emission line fluxes of H$\alpha$ and [\ion{N}{II}] 
$\lambda6584$ are $(24.15 \pm 0.54) \times 10^{-16}~{\rm erg\,cm}^{-2}\,{\rm s}^{-1}$ and $(1.92 \pm 
0.10 ) \times 10^{-16}~{\rm erg\,cm}^{-2}\,{\rm s}^{-1}$, respectively. The fluxes were 
measured by fitting a Gaussian profile to the emission line profiles, 
measuring the integrated flux under the profile.

Using the \citet{Kennicutt1998} relation converted to a Chabrier initial mass function 
\citep{Chabrier2003, Madau2014}, we infer a star-formation rate of $\approx 0.012 M_\odot\, {\rm 
yr^{-1}}$ from the H$\alpha$ emission line. Note that this is a lower limit since the slit width in the 
LRIS spectrum is $1.0^{\prime\prime}$ ($\sim 0.44$\,kpc at the distance of the host) and the 
extraction aperture 
is $0.76^{\prime\prime}$, whereas the host diameter is about $4^{\prime\prime}$.

We also compute the oxygen abundance using the 
strong-line metallicity indicator N2\footnote{N2 $\equiv$ log\{[\ion{N}{II}] $\lambda 6583$/H$\alpha$\}} 
\citep{Pettini2004} with the updated calibration reported in 
\citet{Marino2013}. The oxygen abundance in the N2 scale is 8.23 $\pm$ 0.01 (stat) $\pm$ 0.05 (sys). 
We choose not to use the O3N2 index\footnote{O3N2 $\equiv$ log\{([\ion{O}{III}] $\lambda 
5007$/H$\beta$)/([\ion{N}{II}] $\lambda6583$/H$\alpha$)\}} since it requires line flux 
measurement of H$\beta$. As can be 
seen in panel (a) Figure~\ref{fig:spectra_late}, there is substantial stellar absorption around H$\beta$ 
(4861\,\AA). Compared to $12+{\rm log(O/H)_{\rm solar}} = 8.69$ 
\citep{Asplund2009}, the derived N2 index suggests a significantly subsolar metallicity of $\approx 
0.35 Z_\odot$ ($Z\approx 0.005$). This estimate places \name's host galaxy in the lowest 
10\% of the distribution of SNe Ibc host galaxy metallicities \citep{Sanders2012}, and it is on the lowest 
15\% in the range of Type Ic-BL SNe host galaxy metallicities \citep{Modjaz2020}

\begin{figure}[htbp!]
	\centering
	\includegraphics[width=\columnwidth]{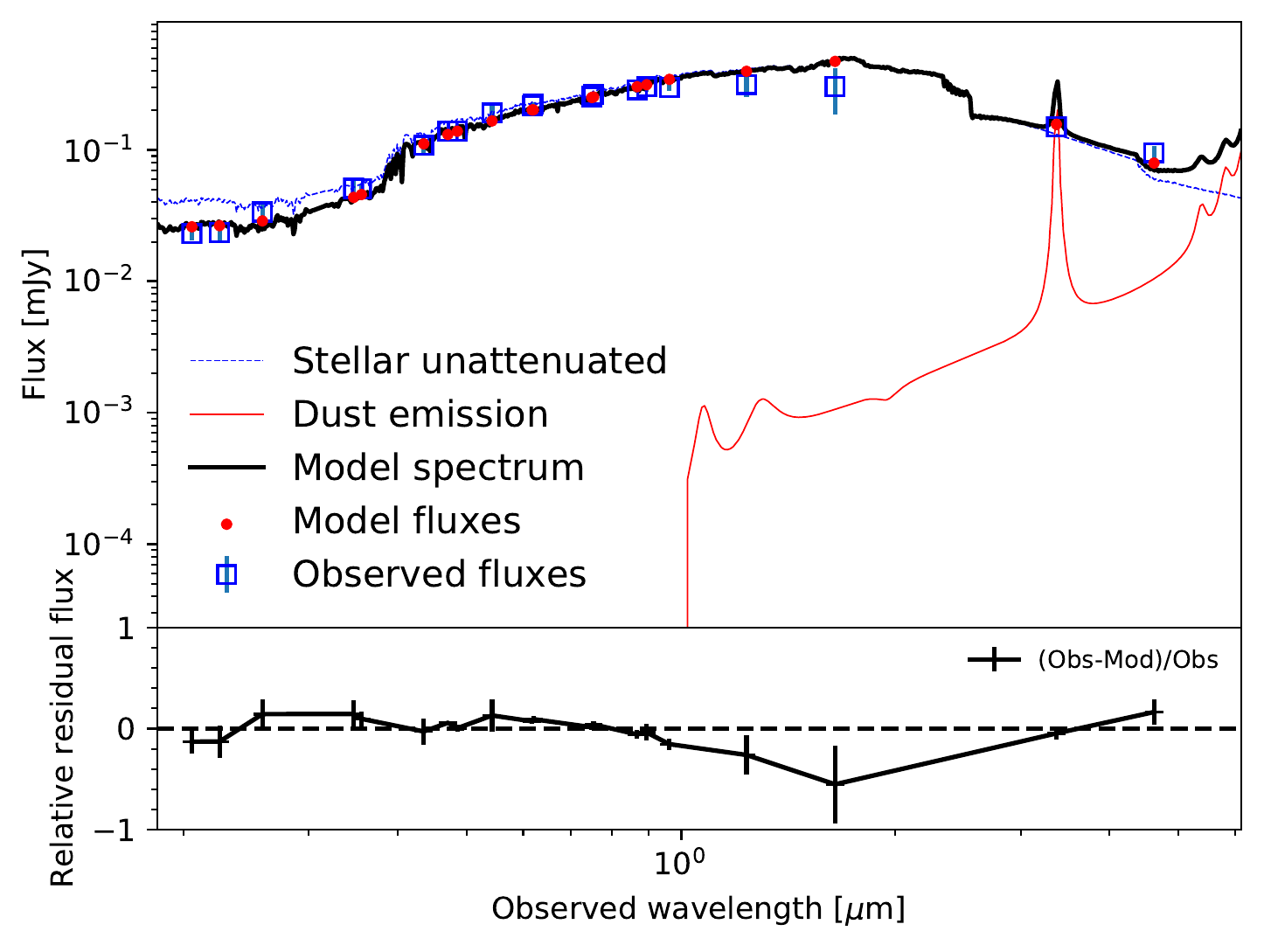}
	\caption{Spectral energy distribution of the host galaxy of \name. The observed photometric 
		data (with 1$\sigma$ error bars) are shown in blue open squares, and the 
		model is shown as a black curve (reduced $\chi^2 = 0.64$). The relative residual flux is shown in 
		the 
		bottom panel.
		\label{fig:SEDfit}}
\end{figure}

We determine the stellar mass ($M_{\star}$) of the host galaxy by SED modeling using 
\texttt{CIGALE} \citep{CIGALE19}. We adopt the stellar population synthesis models from \citet{BC03} 
with the Kroupa IMF \citep{Kroupa01}, and assume a double declining  exponential star formation 
history (SFH). In addition, a dust component is added using the \citet{DL07} model to account for dust 
emission. Finally, the total SED model is attenuated by a modified Calzetti extinction law 
\citep{Calzetti2000}. It assumes that the young stellar population is extincted by the normal Calzetti 
law, and the old stellar population is extincted less heavily than that by a certain factor ($<$ 1, 
\citealt{Charlot2000}). 

The fitted SED is shown in Figure 
\ref{fig:SEDfit}. The derived total stellar mass is ${\rm log}(M_{\star, \rm tot}/M_{\odot}) = 8.5 \pm 0.1$, 
the mass of the stars alive is ${\rm log}(M_{\star, \rm alive}/M_{\odot}) = 8.4 \pm 0.1$, and the 
inferred SFR is $0.030 \pm 0.005\, M_\odot\, {\rm yr^{-1}}$, about 2.5 times the SFR inferred 
from the H$\alpha$ flux measurement. The host extinction, $E(B - V)$, is $0.07 \pm 
0.02$\,mag and $0.03 \pm 0.01$\,mag for the young and old stellar population, respectively, 
both of which are insignificant. The stellar mass and the SFR of the this galaxy are in the lower half of 
the hosts of Type Ibc SN in the PTF sample (Schulze et al. in prep.). 

% The inferred specific SFR (sSFR) is calculated to be $6\times 10^{-11}\,{\rm yr^{-1}}$ (not checked). 

\section{Modeling} \label{sec:modelling}
\subsection{Shock Cooling Powered Fast Rise} \label{subsec:fastrise}
Supernovae light curves are mainly powered by shock energy or radiative diffusion from a heating 
source. We first examine if the peak of \name\ is likely to be powered by the radioactive decay of 
$^{56}$Ni$\rightarrow ^{56}$Co$\rightarrow ^{56}$Fe. With a peak luminosity of $L_{\rm 
peak}\approx 5\times 10^{42}\,{\rm erg \, s^{-1}}$ and a rise time of  $t_{\rm peak}\approx 2$--$4\,{\rm 
d}$, \name\ falls into the unshaded region of \citet[][their Figure~1]{Kasen2017}, where an unphysical 
condition of $M_{\rm Ni} > M_{\rm ej}$ is required. Therefore, we rule out radioactivity as the 
power source for the fast rise of the light curve.

There have been clues for the early emission mechanism of \name,:
\begin{enumerate}[label=(\roman*)]
	\item The fast $t_{\rm rise}$ (Figure~\ref{fig:compare_mag}), high initial high temperature (middle 
	panel of Figure~\ref{fig:Tbb_Rbb_Lbb}), blue color (Figure~\ref{fig:compare_color}), and relatively 
	fast color evolution of \name\ are reminiscent of shock cooling emission \citep{Nakar2014, Piro2015}.
	\item The color jump in $g-r$ is observed 6--9\,d after maximum (left panel of 
	Figure~\ref{fig:compare_color}). It is roughly at this phase that the change in bolometric luminosity 
	decline rate transitions from 0.36\,$\rm mag\, d^{-1}$ to 0.11 $\rm mag\, d^{-1}$
	(upper panel of Figure~\ref{fig:Tbb_Rbb_Lbb}). This supports the idea that the dominant power 
	mechanisms before and after this transition are different.
\end{enumerate}

Therefore, we model the early light curve as cooling emission from shock-heated extended material, 
which is located at the outer layers of the progenitor or outside of the progenitor. We use models 
presented by \citet[][hereafter P15]{Piro2015} to constrain the mass and radius of the extended 
material ($M_{\rm ext}$ and $R_{\rm ext}$, respectively), where $M_{\rm ext}$ includes only mass 
concentrated around $R_{\rm ext}$. This model is built on analytical results of \citet{Nakar2014}. 
Details of the model fitting to multi-band observations are illustrated in Appendix \ref{subsec:p15fit}. 
\begin{figure}
	\centering
	\includegraphics[width=\columnwidth]{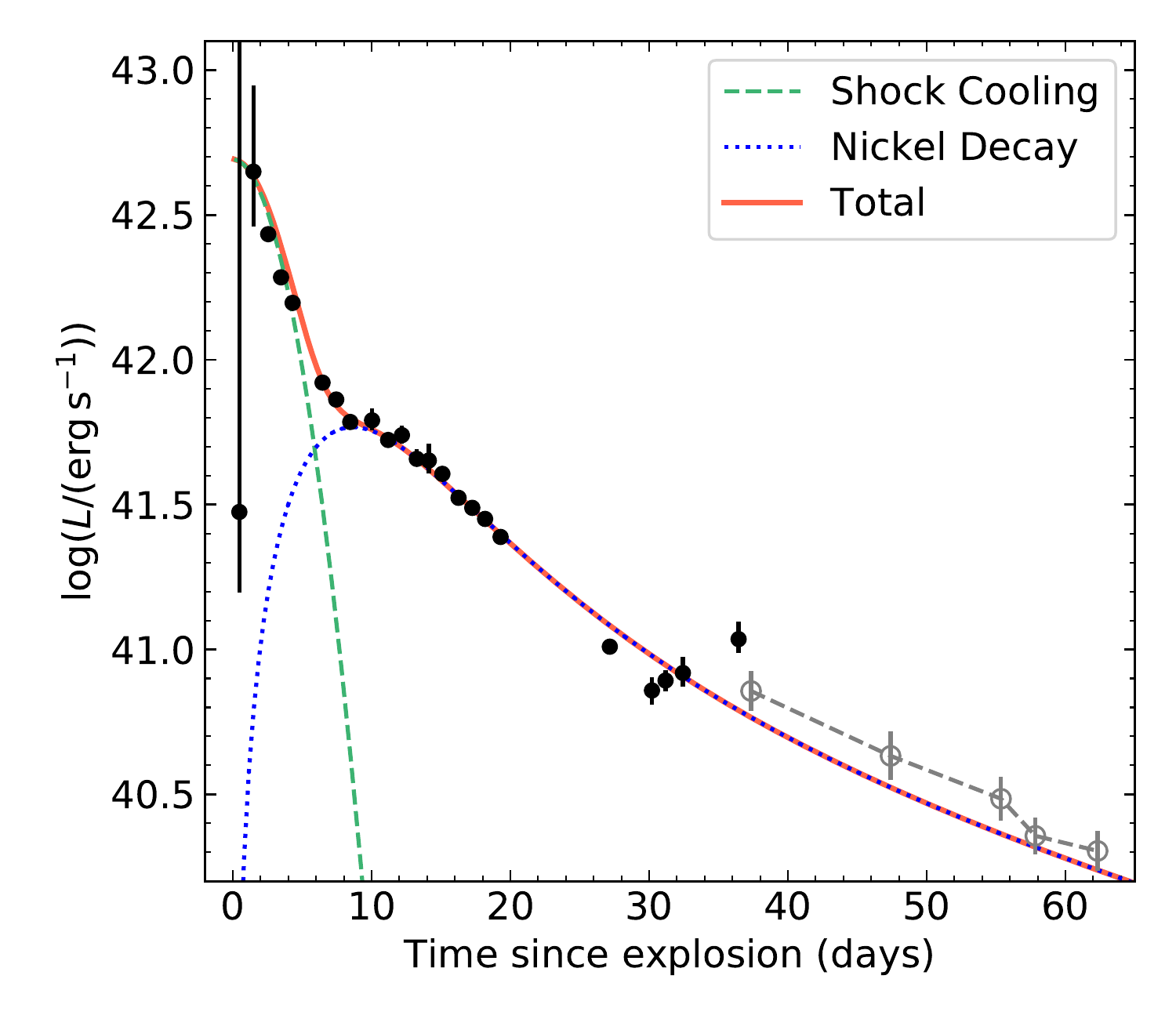}
	\caption{Bolometric light curve for \name. Late-time quasi-bolometric light curve 
	estimated by computing $\nu L_\nu$ in $r$-band is shown as empty grey circles. The dashed green 
	and dotted blue lines show the best fits of shock cooling and nickel decay models. The solid red line 
	shows the combination of the two components.}
	\label{fig:Lbb}
\end{figure}

In Figure~\ref{fig:Lbb}, the bolometric light curve measured in Section \ref{subsec:lc_properties} is 
shown in black. We also show late-time $r$-band $\nu L_{\nu}$ measurements in grey empty circles as 
a proxy of bolometric light curve evolution. The dashed green line shows the best-fit model of 
$M_{\rm ext} = 9.34 \pm 0.36 \times 10^{-2} M_\odot$,
 $R_{\rm ext} =2.71_{-0.17}^{+0.19}\times10^{12}$\,cm (i.e., $39.0_{-2.5}^{+2.7} R_\odot$), 
 and explosion epoch at phase $t_{\rm exp}= -3.21 \pm 0.04$\,d (i.e., the explosion occurred 0.45\,d 
 before the first detection in $g$-band). The amount of energy in the extended material is 
 well constrained to be $E_{\rm ext} = (1.15\pm 0.07) \times 10^{50}\,{\rm erg}$.

Given the simple assumptions of the model, we expect the constraints on $M_{\rm ext}$ and $R_{\rm 
ext}$ to be only approximately accurate. We thus conclude that the early shock cooling emission was 
produced by an extended envelope with a mass of $\sim 0.1 \,M_\odot$ locating at a radius of $\sim 
3\times 10^{12}\,{\rm cm}$ ($40\,R_\odot$). There are now numerous cases of early cooling envelope 
emission observed in CCSNe, where the extended material is estimated to have lower mass ($\sim 
0.001$--$0.01 \, M_\odot$) and larger radius ($\sim 10^{13}\, {\rm cm}$) compared to \name\ 
\citep{Modjaz2019}. 

\subsection{Mass-Loss Estimate from \ion{He}{II}} \label{subsec:flash}
Early-time low-velocity \ion{He}{II} $\lambda4686$ emission (Section \ref{subsubsec:spec_early}) has 
been detected in nearly twenty hydrogen-rich CCSNe and one hydrogen-poor SN iPTF14gqr. This 
feature often fades away within a few hours to a few days after the explosion \citep{Yaron2017}. The 
high ionization potential of this line requires high temperature or an ionizing flux, which might come 
from either shock breakout or CSM interaction \citep{GalYam2014, Smith2015}. Due to the rapid 
decrease in $T_{\rm bb}$ at the three epochs of our early-time spectra and the similarity between 
\name\ and iPTF14gqr, we favor shock cooling emission as the origin of recombination helium 
lines. Therefore, we can use the luminosity of the \ion{He}{II} $\lambda4686$ line to make an 
order-of-magnitude estimate on properties of the emission material, following the procedure given by 
\citet{Ofek2013} and \citet{De2018}.

Assuming that the immediate CSM around the progenitor has a spherical wind-density profile of the 
form $\rho = K r^{-2}$, where $r$ is distance from the progenitor, $K\equiv \dot M / (4\pi v_{\rm w})$ is 
the wind density parameter, $v_{\rm w}$ is the wind velocity, and $\dot M$ is the mass-loss rate. The 
integrated mass of the emitting material from $r$ to $r_1$ is 
\begin{align}
M_{\rm He} = \int_{r}^{r_1}4\pi r^2 \rho(r) {\rm d}r=4\pi K \beta r
\end{align}
where $\beta \equiv (r_1 - r) /r $ is assumed to be of order unity.

We can relate the mass of the \ion{He}{II} region to the \ion{He}{II} 
$\lambda4686$ line luminosity using 
\begin{align}
L_{\lambda 4686} \approx \frac{A n_e M_{\rm He}}{m_{\rm He}} \label{eq:L4686}.
\end{align}
Here
\begin{align}
A = \frac{4\pi j_{\lambda 4686}}{n_e n_{\rm He^{++}}},
\end{align}
$ j_{\lambda4868}$ (in ${\rm erg \, cm^{-3}\, s^{-1}\, sr^{-1}}$) is the emission coefficient for the 
$\lambda4686$ transition. $m_{\rm He}$ is mass of a helium nucleus, $n_{\rm He^{++}}$ is the number 
density of doubly ionized helium and $n_e$ is the number density of electrons.

Assuming a temperature of $10^4$\,K, electron density of $10^{10}\,{\rm cm^{-3}}$, and Case B 
recombination, we get $A=1.32\times 10^{-24}\,{\rm erg\, cm^{3}\, s^{-1}}$ \citep{Storey1995}. 
Using $n_e = 2 n_{\rm He^{++}}$ and the density profile, Eq.~(\ref{eq:L4686}) can be written as
\begin{align}
L_{\lambda 4686} \approx \frac{8\pi A \beta}{m_{\rm He}^2} \frac{K^2}{r}.
\end{align}

The location of the emitting region can be constrained by requiring that the Thompson optical depth 
($\tau$) in the region must be small for the lines to escape. We require
\begin{align}
\tau = n_e \sigma_{\rm T} \int_{r}^{r_1} {\rm d}r = \frac{2\sigma_{\rm T}K\beta}{m_{\rm He}r} \lesssim 1
\end{align}
Thus
\begin{subequations}
\begin{align}
r^2 &\gtrsim  \left (\frac{2\sigma_{\rm T}\beta}{m_{\rm He}} \right)^2 \frac{L_{\lambda 4686} m_{\rm 
		He}^2 r }{8\pi A \beta }  \\
r & \gtrsim L_{\lambda 4686} \frac{\sigma_{\rm T}^2  \beta }{2\pi A }
\end{align}
\end{subequations}

The $+0.4$\,d emission line flux is measured to be $F = (8.99\pm 0.71)\times10^{-16}\,{\rm erg\, 
	cm^{-2}\,s^{-1}}$, corresponding to $L_{\lambda 4686} = 9.0\times 10^{38}\,{\rm erg\, s^{-1}}$. 
	Hence, 
we get $r \gtrsim 4.8 \times 10^{13} \beta \,{\rm cm}$, $K \gtrsim 1.2\times 10^{14} \, {\rm 
	g\,cm^{-1}}$, and $M_{\rm He} \gtrsim 3.7\times 10^{-5} \beta^2\, M_{\odot}$. Adopting a wind 
	velocity of $v_{\rm w} \approx 550 \, {\rm km\, s^{-1}}$ as measured from the \ion{He}{II} FWHM, 
	the mass-loss rate can be constrained to be $\dot M \gtrsim 1.1\times 10^{-4}\,{M_\odot\, \rm 
	yr^{-1}}$. Note that these 
	estimates can be affected if the CSM cannot be well characterized by a spherically 
	symmetric $\rho(r) \propto r^{-2}$ density profile, or if the emitting region was confined to a thin 
	shell ($\beta \ll 1$).
	
\subsection{Constraints from Radio Upper Limits}
\begin{figure}[htbp!]
	\centering
	\includegraphics[width=\columnwidth]{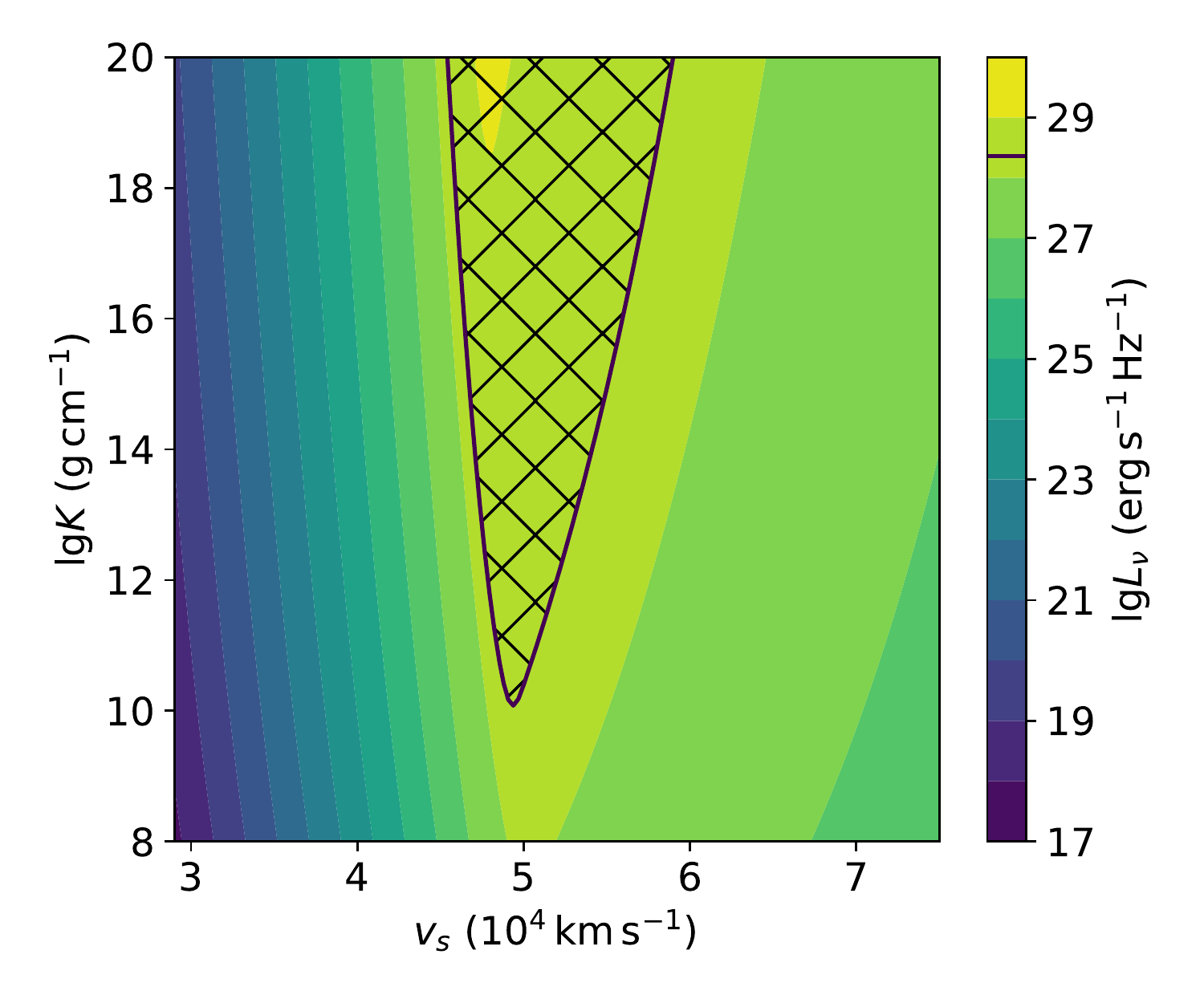}
	\includegraphics[width=\columnwidth]{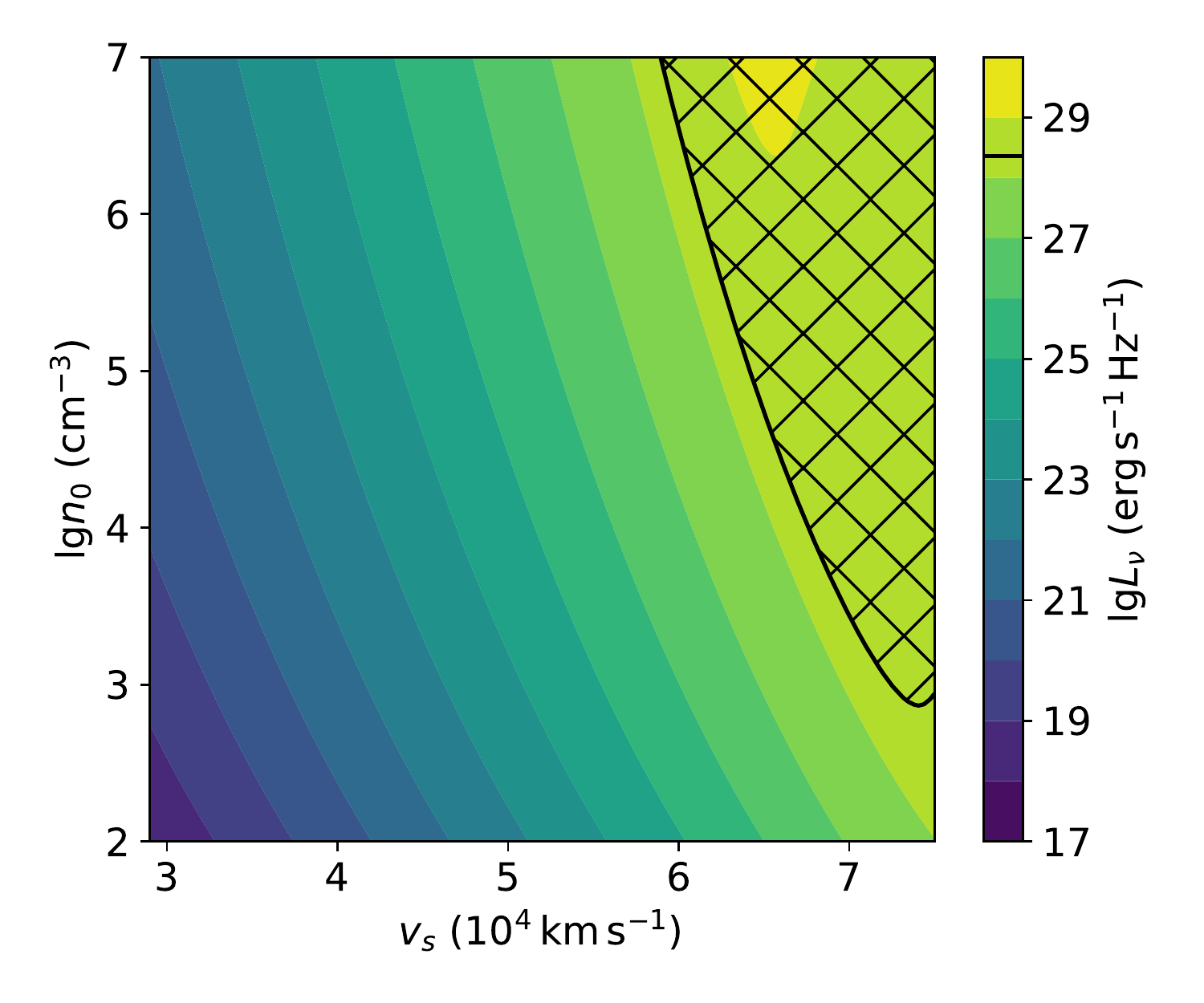}
	\caption{Maps of expected radio luminosity at 230\,GHz. The $x$-axis is the shock velocity $v_s$. 
	The $y$-axis is wind mass-loss parameter $K$ in the case of $\rho \propto r^{-2}$ 
	CSM environment in the upper panel, while in the bottom panel it is the number density $n_0$ in the 
	constant-density case. The black contour in each panel shows the location of 3$\sigma$ upper limit 
	at 230\,GHz on \name. The phase space with a luminosity higher the black line in each panel is ruled 
	out by the 
	observation.
		\label{fig:radio}}
\end{figure}
Radio emission in SNe is produced by shock accelerated electrons in the circumstellar material as they 
gyrate in the post-shock magnetic field when the shock freely expands. Should the circumstellar 
medium be formed by a pre-SN stellar wind, the radio synchrotron radiation can be used to probe the 
pre-explosion mass-loss \citep{Chevalier1982}. High frequency ($\nu>90$\,GHz) bright 
($\nu L_\nu \gtrsim 10^{40} \, \rm erg\, 
s^{-1}$) radio sources are often found to be associated with gamma-ray bursts (GRBs), TDEs, and 
relativistic transients (see Figure~6 of \citealt{HoPhinney2019}). Among normal SNe Ibc, moderate 
submillimeter luminosity at $\sim 5\times 10^{37}\, \rm erg\, s^{-1}$ has been observed in SN1993J 
\citep{Weiler2007} and SN2011dh \citep{Horesh2013}.

Our SMA observations constrain the submillimeter luminosity of \name\ to $\nu L_{\nu \rm , 
230GHz} < 5.3\times 10^{39}\, \rm erg\, s^{-1}$ and $\nu L_{\nu \rm , 345GHz} < 3.0\times 10^{40}\, \rm 
erg\, s^{-1}$. We place these upper limits in physical context using the synchrotron self-absorption 
model given by \citet{Chevalier1998}. The expected radio luminosities are computed at 230 and 
345\,GHz for two types of circumstellar environments --- one with a wind-density with the same 
parameterization as that adopted in Section \ref{subsec:flash} and the other with a constant-density 
environment ($\rho = \,{\rm constant}$). 

Adopting the explosion epoch found in Section \ref{subsec:fastrise}, our SMA observations were 
obtained at 2.75\,d after explosion. Given the early time of these observations, we consider constant 
shock velocities at 0.1--0.25$c$, as found to be typical in SNe Ibc \citep{Wellons2012}. We assume an 
electron energy power-law index of $p = 3$, a volume filling factor $f=0.5$, and that the electrons 
and magnetic field in the post-shock region share constant fractions of the post-shock energy 
density, i.e., $\epsilon_e = \epsilon_B = 0.1$.

The expected radio luminosity predicted by the Chevalier model in the two environments 
at 230\,GHz are shown in Figure~\ref{fig:radio} by the color maps, and the black contours indicate our 
$3\sigma$ limits. As can be seen, only small regions have expected luminosity higher than the 
$3\sigma$ limits (indicated by the hatched regions), and thus our observations are not deep enough to 
provide stringent constrains on the circumstellar properties. Compared with 230\,GHz, the parameter 
space is even more poorly constrained at 350\,GHz and are thus not shown.

\subsection{Radioactivity Powered Main Peak} \label{subsec:radioactivity}
After subtracting the shock cooling emission from the bolometric light curve, the remaining light curve 
has a peak luminosity of $L_{\rm peak}\approx 6\times 10^{41}\,{\rm erg \, s^{-1}}$ and a rise time of  
$t_{\rm peak}\approx 9\,{\rm d}$. In the shaded region of \citet[][Fig.~1]{Kasen2017}, this falls between 
the $M_{\rm Ni} = 0.1 \, M_{\rm ej}$ and $M_{\rm Ni} = 0.01 \, M_{\rm ej}$ lines, indicating that the 
remaining component can be powered by $^{56}$Ni decay. Apart from this, the moderate $t_{\rm 
decay}$ of \name\ (bottom panel in Figure~\ref{fig:compare_mag}) is similar to that found in a few 
Ca-rich transients, and consistent with coming from radioactivity. Here we use two methods to 
estimate $M_{\rm ej}$ and $M_{\rm Ni}$.

First of all, we use analytical models \citep{Arnett1982, Valenti2008, Wheeler2015} to constrain the 
nickel mass ($M_{\rm Ni})$, a characteristic photon diffusion timescale ($\tau_{\rm m}$), and a 
characteristic $\gamma$-ray escape timescale ($t_0$). Details of the model fitting are given in 
Appendix \ref{subsec:arnettfit}. The dotted blue line in Figure~\ref{fig:Lbb} shows the best-fit model of 
$M_{\rm Ni} = 1.61_{-0.03}^{+0.04}\times 10^{-2} M_\odot$, $\tau_{\rm m} = 6.35\pm 0.18$\,d, and $t_0 
= 24.04_{-0.73}^{+0.76}$\,d. Thus, using Equation~(\ref{eq:taum}), the ejecta mass can 
be estimated to be
\begin{align}
M_{\rm ej} = 0.36\pm 0.02 \left( \frac{v_{\rm ej}}{8000\,{\rm km\,s^{-1}}} \right) \left( 
\frac{0.07\,{\rm 
		cm^2\,g^{-1}}}{\kappa_{\rm 	opt}}\right) \, M_\odot 
\end{align}
Here we adopt the the mean opacity of SNe Ibc found by \citet{Taddia2018}. In Section 
\ref{subsubsec:spec_middle}, the the photospheric velocity of $\approx 6000\,{\rm km\,s^{-1}}$ is 
measured at phase $\Delta t\sim12$\,d (i.e., $\sim15$--16\,d post explosion). At that time, 
Figure~\ref{fig:Tbb_Rbb_Lbb} shows that $R_{\rm bb}$ stays roughly flat, indicating that a certain 
amount of ejecta in the outer layers should have a velocity greater than 
that measured from the \ion{He}{I} absorption minimum. Therefore, we adopt the $\approx 8000\,{\rm 
km\,s^{-1}}$ measured from early $R_{\rm bb}$ evolution (see Section 
\ref{subsubsec:bolometric_evolution}) to be a more appropriate estimate for $v_{\rm ej}$. 
The kinetic energy is then calculated to be 
\begin{align}
E_{\rm kin} = \frac{3}{10}M_{\rm ej} v_{\rm ej}^2 = (1.36 \pm 0.08)\times 10^{50}\,{\rm erg}
\end{align}

\citet[][hereafter KK19]{Khatami2019} presented improved analytic relations (compared with 
the original \citealt{Arnett1982} model) between $t_{\rm peak}$ and $L_{\rm peak}$. When $t<10$\,d, 
$\varepsilon_{\rm Ni}(t) \gg \varepsilon_{\rm Co}(t)$ (see Equations~\ref{eq:heatNi}, \ref{eq:heatCo}), 
and hence we have an exponential heating function 
\begin{equation}
L_{\rm heat}(t) = L_0 e^{-t/\tau_{\rm Ni}}
\end{equation}
where $L_0 = M_{\rm Ni}\times \epsilon_{\rm Ni}$. In this case, KK19 (Eq.~21) shows that 
the relation between peak time and luminosity is:
\begin{equation}
L_{\rm peak} = \frac{2L_0 \tau_{\rm Ni}^2}{\beta^2 t_{\rm peak}^2} \left[ 1 - (1 + \beta t_{\rm 
peak}/\tau_{\rm Ni} ) e^{-\beta t_{\rm peak}/ \tau_{\rm Ni}} \right]
\end{equation}
where $\beta \sim 4/3$ gives a reasonable match to numerical simulations. With $L_{\rm 	peak}\approx 
6\times 10^{41}\,{\rm erg \, s^{-1}}$ and $t_{\rm peak} \approx 9$\,d, we get an estimate of $M_{\rm 
Ni}\sim 0.017 \, M_{\odot}$.

$M_{\rm ej}$ can be estimated using Eq.~23 of KK19:
\begin{align}
\frac{t_{\rm peak}}{t_{\rm d}} = 0.11\,{\rm ln} \left( 1 + \frac{9\tau_{\rm Ni}}{t_{\rm d}} \right)+ 0.36,
\label{eq:kk19_23}
\end{align}
where $t_{\rm d}$ is the characteristic timescale without any numerical factors
\begin{align}
t_{\rm d} = \left(\frac{\kappa_{\rm opt} M_{\rm ej}}{v_{\rm ej}c}\right)^{1/2}. \label{eq:kk19_12}
\end{align}
We derive $t_{\rm d} \approx 15.4$\,d, which implies 
\begin{align}
M_{\rm ej} \approx 0.30 \left( \frac{v_{\rm ej}}{8000\,{\rm km\,s^{-1}}} \right) \left( \frac{0.07\,{\rm 
cm^2\,g^{-1}}}{\kappa_{\rm opt}} \right) \, M_\odot
\end{align}
The kinetic energy of the ejecta is then $E_{\rm kin} \approx 1.2\times 10^{50}\,{\rm erg}$.

In conclusion, the estimates derived from simplified model fitting and new analytic relations from 
KK19 are roughly the same. The ejecta mass ($M_{\rm ej}\sim 0.3M_\odot$), the nickel mass ($M_{\rm 
Ni} \sim 0.017M_\odot$), and the total kinetic energy ($E_{\rm kin}\sim 1.2\times 10^{50}\,{\rm erg}$) of 
\name\ are very small.

\section{Interpretation} \label{sec:interpretation}
\subsection{A Core-Collapse Supernova}
At early times, the cooling emission from shock-heated surrounding material of $M_{\rm 
ext}\sim0.1\,M_\odot$ and $R_{\rm ext}\sim 3\times 10^{12}\,{\rm cm}$ ($40\,{ R_\odot}$) corroborates 
that the progenitor of \name\ is a star with an extended envelope. Indeed stellar evolution models 
predict envelope radii of 10--100\,$R_\odot$ for helium stars with zero-age helium core masses within 
2.5--3.2\,$M_\odot$ that have stripped all of the hydrogen-rich envelope \citep{Woosley2019, 
Laplace2020}. Therefore, the early-time shock cooling 
light curve serves as strong evidence that \name\ is the explosion of a star with inflated radius (not a 
compact object).

The $^{56}$Ni mass of $\sim 0.017\,M_\odot$ inferred from the radioactivity-powered decay 
is much greater than that produced in electron-capture SNe ($\sim10^{-3}\,M_\odot$, 
\citealt{Moriya2014}), whereas the ejecta velocity of $\approx8000\,{\rm km\, s^{-1}}$ is 
larger than that expected in fallback SNe ($\sim3000\,{\rm km\,s^{-1}}$; \citealt{Moriya2010}). 
Therefore, we conclude that \name\ is associated with the class of Fe CCSNe.

\subsection{An Ultra-Stripped Progenitor}
As noted in the introduction, the majority of SNe Ibc, with $M_{\rm ej}$ in the range of 
1--5\,$M_\odot$, are believed to come from binary evolution. The small amount of ejecta mass seen in 
\name\ ($M_{\rm ej}\sim 0.3\, M_\odot$) requires extreme stripping prior to the explosion in a 
binary system, which suggests an ultra-stripped progenitor \citep{Tauris2013}. 

Compared with iPTF14gqr, where the second peak of the light curve suggests $M_{\rm ej}\sim 
0.2\,M_\odot$, \name\ has a higher ejecta mass. In particular, the helium-rich photospheric 
spectra indicate that \name\ has a greater amount of helium in the ejecta. \ion{He}{I} emission lines are 
non-thermally excited by collisions with fast electrons, which result from Compton processes with 
$\gamma$-rays from $^{56}$Ni decay \citep{Dessart2012, Hachinger2012}. On the other hand, the 
weak absorption strength in the \ion{He}{I} P-Cygni profile (Figure~\ref{fig:spectra} and 
Figure~\ref{fig:hst_opt}) suggests that the helium envelope mass of \name\ is substantially lower than 
that in a canonical Type Ib SN \citep{Fremling2018}. While the stripping in \name\ is less extreme than 
for iPTF14gqr, the striking similarities between these two events indicate that they probably originate 
from similar channels. 

The \ion{He}{II} $\lambda4686$ flash ionized emission comes from optically thin material located at 
$\sim 5\times 10^{13}\,{\rm cm}$ ($700\,R_\odot$). This is even larger than the expected orbital 
separation required for extreme stripping. Therefore, material at such a large radius might be ejected 
prior to the explosion, with a mass-loss timescale $t \sim 5 \times 10^{13} \, {\rm cm}/(500 \, {\rm 
km}\,{\rm s}^{-1}) \sim 10 \, {\rm d}$. The inferred mass-loss rate of $\dot M \gtrsim 10^{-4}\,{ 
	M_\odot}\, \rm yr^{-1}$ is much higher than that observed in Galactic Wolf-Rayet stars 
\citep{Smith2014}. Additionally, the photospheric and late-time spectra of \name\ signify interaction 
with a helium-rich extended dense shell, which may also consist of gas originally ejected by 
the progenitor as a stellar wind or deposited by binary interaction. The high mass-loss rate and 
short ejection timescale can be achieved in the final stages of stellar evolution by several mechanisms: 
1. a powerful outflow driven by super-Eddington wave energy deposition
%with mass-loss rates in the range of $10^{-3}$--$10\,M_\odot\,{\rm yr^{-1}}$ 
during the last few years before explosion \citep{Quataert2012}; 2. explosive mass ejection due to 
violent silicon flashes within a few weeks before the explosion of low-mass helium stars 
\citep{Woosley2019}; 3. nonconservative case BB
mass transfer in binary evolution of ultra-stripped stars \citep{Tauris2015}. 

\begin{deluxetable*}{cc|ccc|cc|cc}[htpb!]
	\tablecaption{ Model parameters for hydrogen-poor subluminous fast-evolving SNe where the 
	bolometric light curve can be fitted with a shock cooling powered component and a radioactivity 
	powered component.\label{tab:model_compare}}
	\tablehead{
		\colhead{Name}
		& \colhead{$v_{\rm ej}$}
		&\colhead{$\tau_{\rm m, A82}$}
		&\colhead{$M_{\rm ej, A82}$}
		& \colhead{$M_{\rm Ni, A82}$}
		%&\colhead{$E_{\rm kin}$ ($10^{49}\,{\rm erg\,s^{-1}}$)}
		&\colhead{$M_{\rm ej, KK19}$}
		& \colhead{$M_{\rm Ni, KK19}$}
		&\colhead{$R_{\rm ext}$}
		&\colhead{$M_{\rm ext}$}\\
		\colhead{}
		& \colhead{($\rm km\,s^{-1}$)}
		& \colhead{(d)}
		& \colhead{($M_\odot$)}
		& \colhead{($10^{-2} M_\odot$)}
		& \colhead{($M_\odot$)}
		&\colhead{($10^{-2} M_\odot$)}
		& \colhead{($10^{12}\,{\rm 	cm}$)}
		& \colhead{($10^{-2} M_\odot$)}
	}
	\startdata
	iPTF14gqr  & 10000 
	& $4.38^{+0.14}_{-0.15}$ & $0.21_{-0.01}^{+0.01}$ & $8.01^{+0.14}_{-0.15}$
	&0.22 & 8.35
	&$55.33^{+9.73}_{-9.68}$ &$1.46^{+0.37}_{-0.27}$  \\
	iPTF16hgs & 10000 
	& $11.09^{+1.08}_{-1.09}$  &$1.36^{+0.28}_{-0.25}$ & $2.33^{+0.23}_{-0.23}$ 
	& 0.78 &2.29
	&$1.67^{+2.24}_{-0.95}$ & $8.94^{+3.77}_{-2.61}$\\
	SN2018lqo & 8250
	& $9.71_{-0.37}^{+0.52}$ & $0.86_{-0.06}^{+0.09}$& $2.75^{+0.09}_{-0.07}$
	& 0.63 & 3.08
	&$26.80^{+108.58}_{-23.55}$  & $4.82^{+3.23}_{-1.15}$ \\
	SN2019dge  & 8000 
	&$6.35_{-0.18}^{+0.18}$ &$0.36_{-0.02}^{+0.02}$ &$1.61^{+0.04}_{-0.03}$ 
	&0.30 &1.70
	&  $2.71_{-0.17}^{+0.19}$ & $9.34^{+0.36}_{-0.36}$  \\
	\enddata
\end{deluxetable*}

\subsection{Stellar Evolution Pathways} \label{subsec:stellar_pathways}
Here we discuss possible evolution paths of \name's progenitor. 

We first consider the scenario where \name\ comes from a progenitor more massive than $\sim \! 15 \, 
M_\odot$ in a binary system that loses its mass in case B mass transfer. \citet{Yoon2010} showed that 
subsequent wind mass loss is weak at subsolar metallicity of $Z\approx 0.004$ (similar to the 
$Z\approx 0.005$ calculated in Section \ref{subsec:host}), such that the final mass of the primary at 
the time of core-collapse will be higher than 3.8\,$M_\odot$.
This will lead to $M_{\rm ej}\gtrsim 2.3\,M_\odot$, assuming that the explosion forms a neutron star of 
1.5\,$M_\odot$. This inferred ejecta mass is much higher than that observed ($M_{\rm ej}\sim 0.3 \, 
M_\odot$), so this scenario is not favored. 

We next consider the possibility that the primary has an initial mass $M_1 \lesssim 15 \, M_\odot$. In 
many binary scenarios involving a companion that is a  main sequence star, the primary will 
experience stable case B mass transfer that strips the hydrogen envelope, followed by case BB mass 
transfer that strips most of its helium envelope, resulting in an ultrastripped exploding star. 
\citet{Zapartas2017} performed population synthesis simulations, showing that for the pre-SN helium 
star to reach $\lesssim 2\, M_\odot$, a relatively wide range of companion mass is possible (initial mass 
4--10\,$M_\odot$). Therefore, this scenario is consistent with observations of \name. A compact object 
companion is less likely because its lower mass would likely lead to unstable case B mass transfer, with 
a post-common envelope orbital separation much smaller than the inferred radius of the progenitor of 
\name \, \citep{Laplace2020}.  

The final mass of the helium envelope depends on the initial mass of the helium star and the orbital 
period of the compact binary. To reconcile with the ejecta mass observed in \name, we expect a small 
final envelope mass ($\lesssim0.3\,M_\odot$) but large enough for optical helium 
features to be observed in the SN explosion ($\gtrsim 0.06\, M_\odot$, \citealt{Hachinger2012}). This 
can be achieved in a system where the progenitor is a helium star in a compact binary with $
P_{\rm orb} \gtrsim 0.2$\,d at the start of the helium 
burning phase \citep{Tauris2015}. However, the large inferred radius of the progenitor requires an 
orbital separation $a \gtrsim 40 \, R_\odot$, implying an orbital period of tens of days at the time of 
explosion, depending on the companion mass. 

%Taken together, we conclude that companion of \name\ can be best explained by a low-mass main 
%sequence star or a degenerate compact object.

%The outcome of \name\ --- an iron core-collapse ultra-stripped SN --- is a neutron star with mass 
%in the range 1.1--1.8\,$M_\odot$ \citep{Tauris2015}. The small ejecta mass and the small binding 
%energy of the stripped envelope imply a small kick velocity ($\sim 50\,{\rm km\, s^{-1}}$) imparted 
%onto 
%the newborn NS \citep{Tauris2015, Suwa2015, Bray2016, Muller2018}, which can prevent the binary 
%system from being disrupted or broken up, leaving a compact NS binary \citep{Tauris2017}. 
%Ultra-stripped iron CCSNe therefore serve as a natural formation channel for compact NS binaries 
%with 
%small eccentricities. 

\subsection{Comparison with Other Ultra-Stripped SN Candidates}

In addition to \name\ and iPTF14gqr, we search the literature for other subluminous fast-evolving 
hydrogen-poor SNe whose light curves can potentially be well fitted by an early-time shock-cooling 
component from an extended envelope and a radioactivity-powered second peak with small $M_{\rm 
ej}$. We recover iPTFF16hgs \citep{DeKC2018} and SN2018lqo \citep{De2020b} as ultra-stripped SN
candidates. Here we apply our modeling approach described in Section \ref{subsec:fastrise} and 
\ref{subsec:radioactivity} to iPTF16hgs and SN2018lqo to distill the physical parameters of these two 
events. We show the results in Table \ref{tab:model_compare}. The ejecta masses of 
iPTF16hgs and SN2018lqo are greater than that in \name\ and iPTF14gqr by a factor of $\sim3$, 
and falls inside the range of $M_{\rm ej}$ expected in explosions of a helium star orbiting a compact 
object, but is at the upper side of the boundaries \citep{Tauris2015}. 

A full discussion of the progenitors of iPTF16hgs and SN2018lqo is beyond the scope of this paper. 
Here we refer to a recent study conducted by \citet{De2020b}, which classify these two objects into 
the ``green Ca-Ib'' subclass in the Ca-rich SNe category. This class of objects is spectroscopically 
similar to SNe Ib at maximum light, and do not exhibit line-blanketed continua at 
$\sim3500$--5500\,\AA.  \citet{De2020b} proposed that pure helium-shell detonations or deflagrations 
can explain their photometric and spectroscopic properties. Although it has been suggested that the 
existence of early first peak can distinguish ultra-stripped SNe from other Ca-rich transients arising 
from helium dotonation on the surface of white dwarfs \citep{Nakaoka2020}, the early-time peak might 
also be caused by radioactive decay from short-lived isotopes in the outermost ejecta \citep{De2020b}. 
For example, the double-peaked Ca-rich transient SN2018lqo occurs in an elliptical galaxy, which is not 
expected to be the host for ultra-stripped SNe.

Although we only include hydrogen-poor events in this comparison, we note that SN2019ehk, a 
Ca-rich transient that exhibits flash ionized hydrogen in its early-time spectra 
\citep{Jacobson-Galan2020} as well as possible hydrogen photospheric features \citep{De2020b}, has 
also been suggested to be an ultra-stripped CCSN \citep{Nakaoka2020}. If both iPTF16hgs and 
SN2019ehk are bona fide ultra-stripped SNe, there might exist a continuum of ejecta mass from 
normal stripped envelope SNe ($1\,M_\odot \lesssim M_{\rm ej} \lesssim 5\,M_\odot$), to a higher 
degree of stripping in the progenitors of SN2019ehk and iPTF16hgs ($0.5\,M_\odot \lesssim M_{\rm 
ej}\lesssim 1\,M_\odot$), to a more extreme degree of stripping seen in SN2019dge ($M_{\rm ej}\sim 
0.3\,M_\odot$), to the most extreme stripping seen in iPTF14gqr ($M_{\rm ej}\sim 0.2\,M_\odot$). The 
remaining amount of helium, the companion mass, and final orbital separation might become smaller 
along this sequence.

\section{Rates} \label{sec:rates}
As progenitors of compact neutron star binaries, the volumetric rates 
of ultra-stripped SNe have implications for our understanding of the evolutionary pathways leading to 
these systems and the gravitational waves detected by existing and upcoming facilities such as 
LIGO/VIRGO \citep{GW170817}. 

Based on population synthesis calculation, \citet{Tauris2015} estimate 
that the volumetric rates of ultra-stripped SNe should be $\sim 0.1$--1\% of the rate of Core-collapse 
SNe. Using the properties of the promising ultra-stripped SN iPTF14gqr \citep{De2018}, 
\citet{Hijikawa2019} estimate the volumetric rates of iPTF14gqr-like ultra-stripped SNe to be $\sim 2 
\times 10^{-7}\,{\rm Mpc^{-3}\, yr^{-1}}$, or $\sim 0.2$\% of the local CCSNe rate 
\citep{Li2011a}. However, since existing ultra-stripped SN candidates were found outside of systematic 
SN classification efforts, observationally constraining the rates of ultra-stripped SNe has not been 
possible thus far. 

\subsection{Simple Estimation}
\subsubsection{Using the BTS Sample} \label{subsubsec:BTS}
\begin{figure}[htbp!]
	\centering
	\includegraphics[width=\columnwidth]{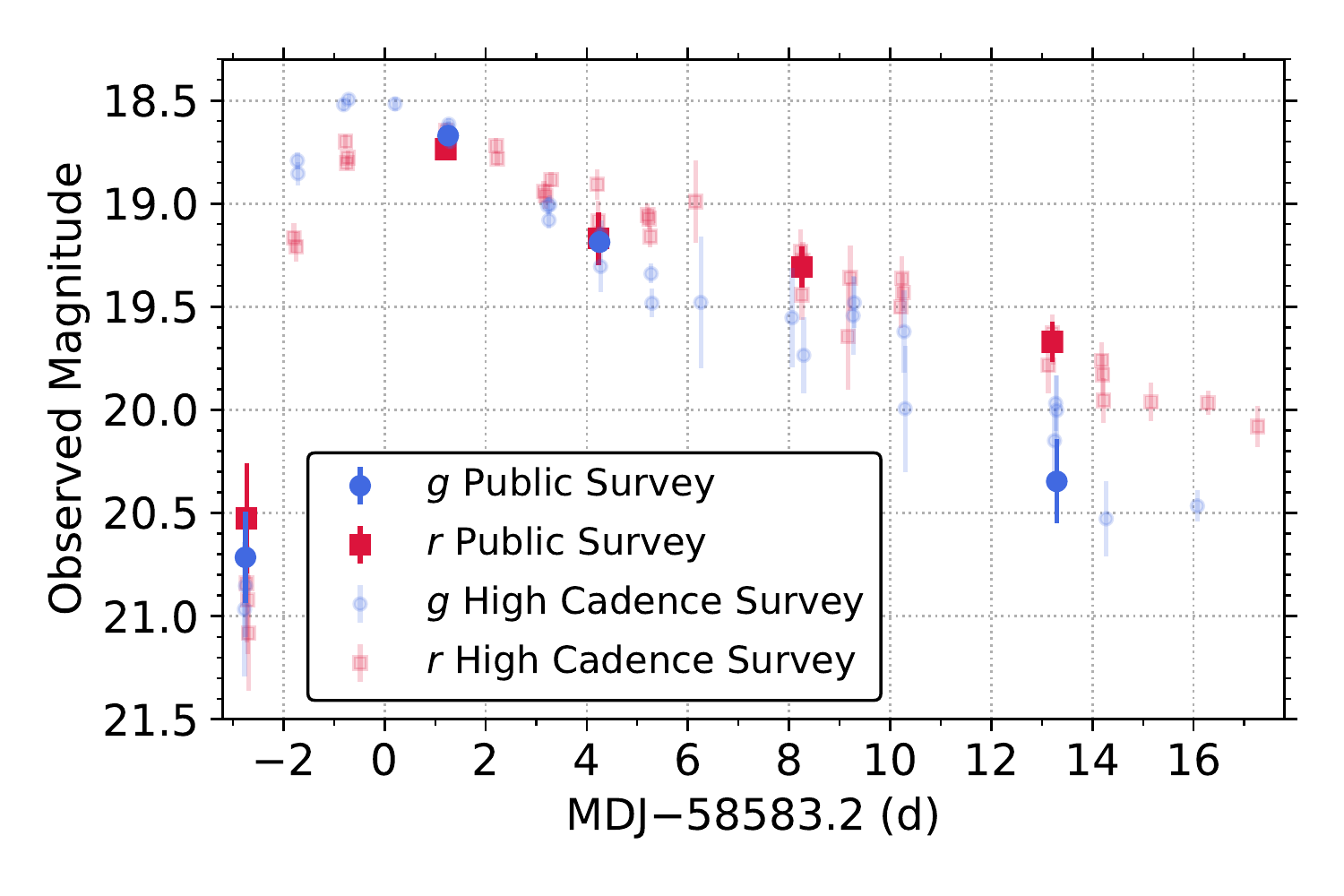}
	\caption{Un-binned P48 light curve of \name. We highlight observations obtained in the public 
	Northern Sky Survey in a greater marker size and high-opacity colors, while observations obtained in 
	the high-cadence survey are shown in semi-transparent.
		\label{fig:lc_pid}}
\end{figure}

\name\ was followed up as a part of the ZTF Bright Transient Survey 
\citep[BTS,][]{FremlingBTS2019} that aims to spectroscopically classify all extragalactic transients in 
ZTF brighter than 18.5\,mag at peak. Since BTS only reads from the ZTF public alert stream 
(highlighted with a greater marker size in Figure~\ref{fig:lc_pid}), \name\ peaks between 18.5 and 
19.0\,mag in the BTS sample. Thanks to the relatively high spectroscopic completeness ($\approx 
89$\%) at the brightness limit of 19.0\,mag, we can directly place constraints on the rates of 19dge-like 
ultra-stripped SNe using the BTS sample. 

\name\ peaked at an absolute magnitude of $-16.44$\,mag in $g$-band.
At the BTS peak brightness limit of 19.0\,mag, objects similar to \name\ would be detectable 
out to 123\,Mpc. Thus, taking only the local 123\,Mpc volume within redshift of $z = 0.028$, we 
compare the number of CCSNe brighter than 19.0\,mag at peak that were found in the BTS experiment 
in its first 12 months of operations (between 2018-06-01 and 2019-06-01). In this time period, BTS 
classified a total of 116 CCSNe in this volume. As such, the detection of one object in this sample 
constrains the rate of ultra-stripped SNe to be $\sim$0.86\% of the CCSNe rate brighter than $M = 
-16.44$\,mag in this volume. 

Taking the observed luminosity function of CCSNe in the local universe \citep{Li2011b}, we find that 
$\approx 50$\% of CCSNe are fainter than $M = -16.44$\,mag. The 
luminosity function corrected rate of 19dge-like events is then $\sim$0.43\% of the local CCSNe rate. 
The inferred rate is consistent with that estimated in 
\citet{Tauris2015}, but higher than that inferred for iPTF14gqr-like events \citep{Hijikawa2019}. 
Adopting the CCSNe volumetric rate of $0.7 \times 10^{-4}\,{\rm Mpc^{-3}\, yr^{-1}}$
\citep{Li2011a}, the volumetric rate of 19dge-like ultra-stripped SNe rate is $\sim$300\,${\rm Gpc^{-3}\, 
yr^{-1}}$. This rate estimation is only a lower limit, since the fast photometric evolution of objects 
similar to \name\ can be easily missed due to the slower 3-day cadence of the ZTF public survey.

\subsubsection{Using the CLU sample}
The ZTF team also conducts a campaign to spectroscopically 
classify all SNe within 200\,Mpc  by filtering transients occurring in galaxies with 
previously known redshifts within $z\leq0.05$ in the Census of the Local Universe (CLU) catalog 
\citep{De2020b}. Hereafter we refer this experiment as CLU. The spectroscopic completeness of 
transients in the CLU sample that had at least one detection brighter 
than 20\,mag is 89\%. Since CLU reads from the whole ZTF alert stream (e.g., all data points shown in 
Figure~\ref{fig:lc_pid}), the higher-cadence sub-surveys allow it to better characterize fast-evolving 
SNe. However, the uncertainty in this experiment is the incompleteness of the input galaxy catalog. 
The redshift completeness fraction (RCF) is $\approx80$\% at the lowest redshifts and decreases to 
$\approx50$\% at $z=0.05$, as measured by the BTS experiment \citep{FremlingBTS2019}.

At the CLU peak brightness limit of 20.0\,mag, objects similar to \name\ would be detectable out to 
195\,Mpc. Between 2018-06-01 and 2019-06-01, CLU classified a total of 273 CCSNe in this volume, 
whereas no good ultra-stripped SN candidates have been identified. We place an 
upper limit of ultra-stripped SNe rate to be $\sim$5100\,${\rm Gpc^{-3}\, yr^{-1}}$
% 1/273*7e-4*1e+9*2
following the simple calculation described in 
Section~\ref{subsubsec:BTS}. However, it is also susceptible to the fast 
evolution of 19dge-like SNe 
being missed by the observation gaps. In Section \ref{subsec:cadence} we attempt to place robust 
estimates of 19dge-like ultra-stripped SNe rate by running simulated surveys with the ZTF cadence.

\subsection{Estimation Based on Survey Simulations}\label{subsec:cadence}
We utilize \texttt{simsurvey} \citep{Feindt2019}, a \texttt{python} package designed for 
assessing the rates of transient discovery in surveys like ZTF. To simulate the expected yield of a 
specific type of transient given a volumetric rate, \texttt{simsurvey} requires three inputs: 1) A survey 
schedule. We use the actual ZTF observing history in $g$- and $r$-band between 2018-06-01 and 
2019-06-01 in any of the public or collaboration surveys as the input survey plan. 2) A transient model. 
We construct a light curve template of \name\ (see details in Appendix \ref{subsec:gaussian}). Using 
the template, we generate a \texttt{TimeSeriesSouce} model in the \texttt{sncosmo} package 
\citep{Barbary2016}. 3) A function to sample the transient model parameters. Transients are injected 
out to a redshift of $z=0.044$, since objects further out are not expected to peak brighter than 
20.0\,mag. 

We examine the expected number of detected 19dge-like SNe for a range of input rates. For each input 
rate, we performed 300 simulations of the ZTF observing plan. In order to select transient candidates 
that would have passed the selection criteria of the BTS or CLU experiment and been flagged as an 
object with photometric properties consistent with being a 19dge-like ultra-stripped SN, we apply cuts 
on the simulated light curves as described below.

For the BTS filter, we only use public survey pointings, and reject SNe at low Galactic latitudes ($|b|\leq 
7 \degree$) to be consistent with the BTS experiment \citep{FremlingBTS2019}. In either the $g$- or 
$r$-band light curve, we identify peak light as the brightest detection in the simulated light curve, and 
require:
\begin{enumerate}[label=(\roman*)]
	\item peak magnitude $<19.0$\,mag 
	\item within 4.1\,d before peak, there must be at least one detection or one upper limit 
	deeper than 1.5\,mag below peak
	\item within 15\,d after peak, there must be at least three detections, and the measured decline rate 
	must be greater than $0.07\,{\rm mag\,d^{-1}}$.
\end{enumerate}
Criterion (ii) is set to require that the fast rise of the light curve can be recognized from the 
observation. This is essential since if we only discover \name\ at the radioactive tail, we will 
probably classify it as a low-velocity SN Ib. Criterion (iii) is 
made to ensure that the rapid decline of the light curve can be captured, such that the small ejecta 
mass can be inferred.

For the CLU filter, we use all ZTF pointings, and require that in either the $g$- or $r$-band light curve:
\begin{enumerate}[label=(\roman*)]
	\item peak magnitude $<20.0$\,mag 
	\item the light curve must satisfy at least one of the following criteria: 1) within 4.1\,d before peak, 
	there must be at least one detection or one upper limit deeper than 1.5\,mag below peak, 
	2) within 2.5\,d before peak, there must be at least one detection deeper than 0.75\,mag below 
	peak
	\item same as criterion (iii) applied in the BTS filter.
\end{enumerate}
We apply the above criteria to the actual observations of CCSNe in the BTS and CLU sample. We 
identify one other SN --- ZTF18abwkrbl (SN2018gjx) --- that pass our criteria. However, ZTF18abwkrbl 
is a SN IIb that clearly shows hydrogen in the spectra, and can therefore be excluded as an 
ultra-stripped SN candidate \citep{Tauris2015}. 

\begin{figure}[htbp!]
	\centering
	\includegraphics[width=\columnwidth]{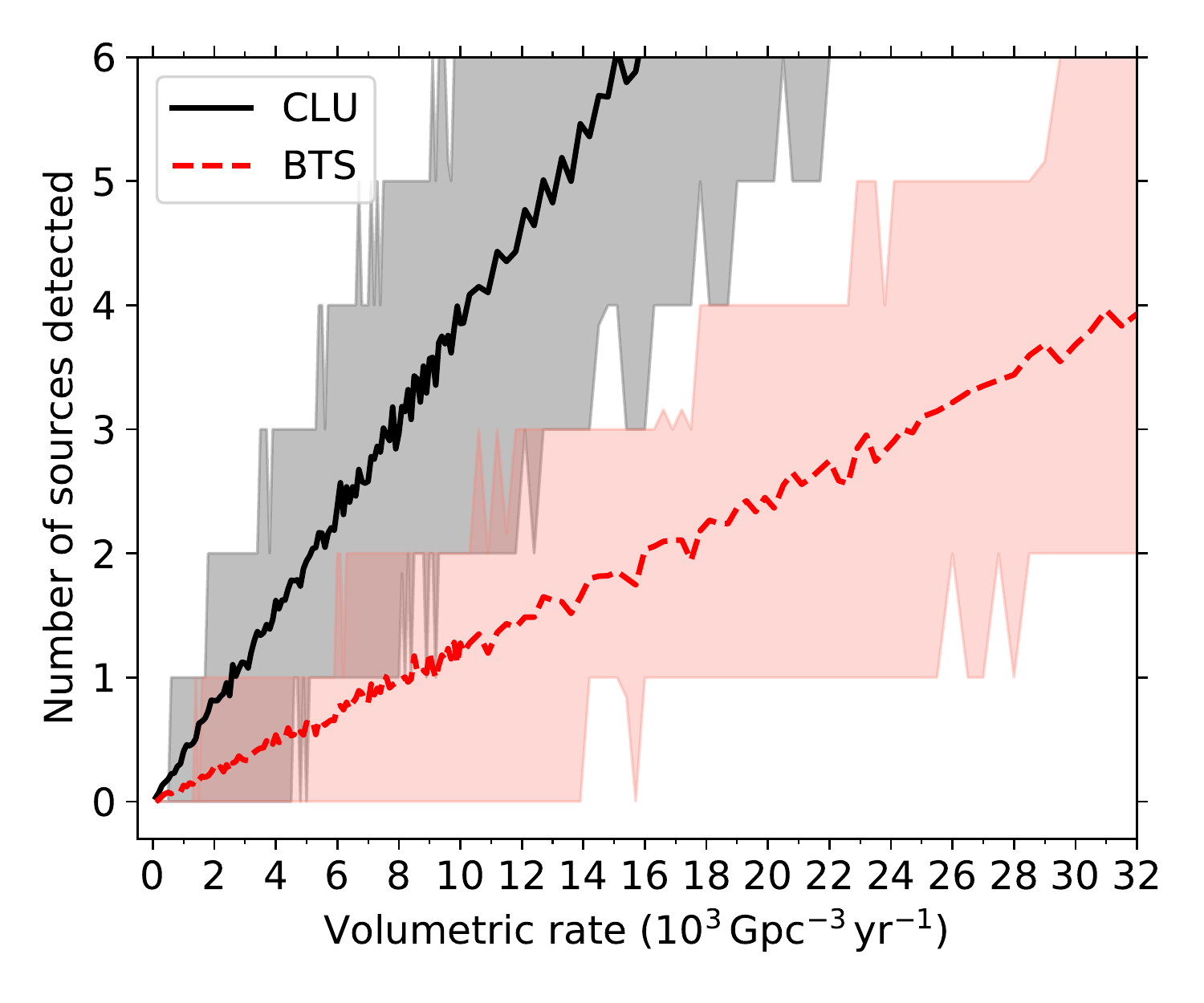}
	\caption{The number of sources passing criteria (described in text) as a 
		function of the input volumetric rate, in both the BTS and CLU experiments. The lines show the 
		mean of the 300 simulations, and the shaded boundaries indicate the 16th and 84th percentiles. 
		\label{fig:rate}}
\end{figure}

In Figure~\ref{fig:rate}, we show the number of transients that pass our selection criteria as a function 
of the input volumetric rate. The solid line and shaded region indicate the mean and 68\% credible 
region of the 300 simulations. In the actual BTS experiment, there was only \textit{one} detected 
ultra-stripped SN. Therefore, we consider the range of volumetric rate where \textit{one} falls within 
the shaded red region as a constraint on the rate of 19dge-like ultra-stripped SNe. 
This gives $R_{\rm 19dge}$ in the range of 1400--28000\,${\rm Gpc^{-3}\,yr^{-1}}$. 

Using the fact that there were \textit{zero} ultra-stripped SN detected in the actual CLU experiment, 
the 
grey shaded region in Figure~\ref{fig:rate} might suggest $R_{\rm 19dge}\lesssim 4500\,{\rm 
Gpc^{-3}\,yr^{-1}}$. However, this upper limit needs to be corrected for offset 
distribution and galaxy catalog incompleteness. First of all, as discussed in \citet{De2020b}, the CLU 
experiment is restricted to transients coincident within 100$^{\prime\prime}$ of the host galaxy nuclei. 
\name\ and iPTF14gqr are 0.5$^{\prime\prime}$ and 24$^{\prime\prime}$ from their host galaxies (all 
within 100$^{\prime\prime}$). Although a large sample of ultra-stripped SNe is needed to examine the 
host offset distribution of this class of objects, the fact that they arise from massive binary evolution 
suggest that the correction due to this factor should be small. Secondly, the incompleteness of the 
input galaxy catalog possibly leads to an underestimation of ultra-stripped SNe rate by a factor of 
55--80\%, as indicated by the RCF. We adjust for such an incompleteness by increase the upper limit 
from 4500 to 8200\,${\rm Gpc^{-3}\,yr^{-1}}$.

Combining results from the BTS and CLU experiments, we derive a 19dge-like ultra-stripped SNe rate of 
1400--8200\,${\rm Gpc^{-3}\,yr^{-1}}$, corresponding to 2--12\% of CCSNe rate.

\subsection{Effects of Different Envelope Masses and Radii} 
\label{subsubsec:physics}
\begin{figure}[htbp!]
	\centering
	\includegraphics[width=\columnwidth]{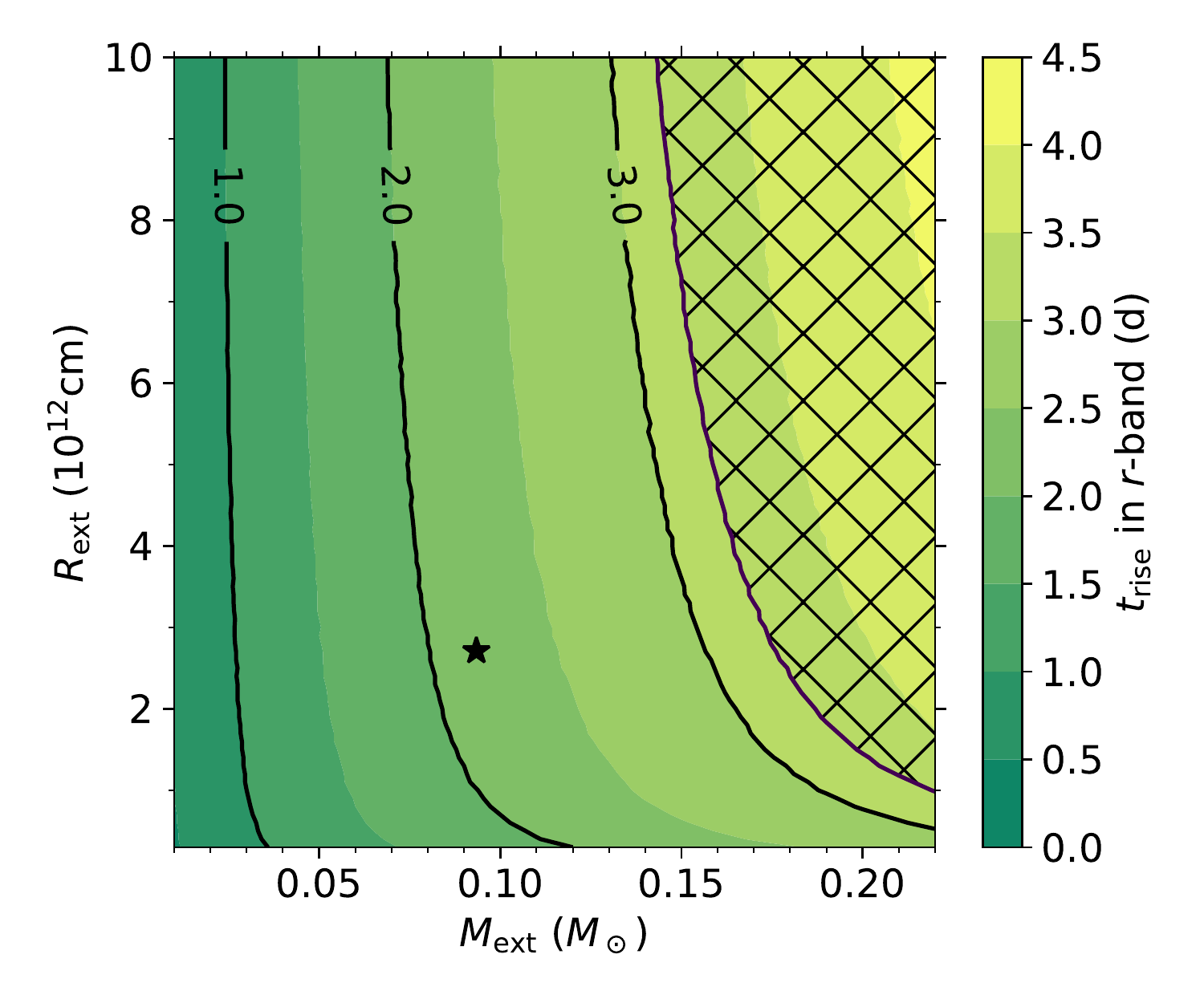}
	\includegraphics[width=\columnwidth]{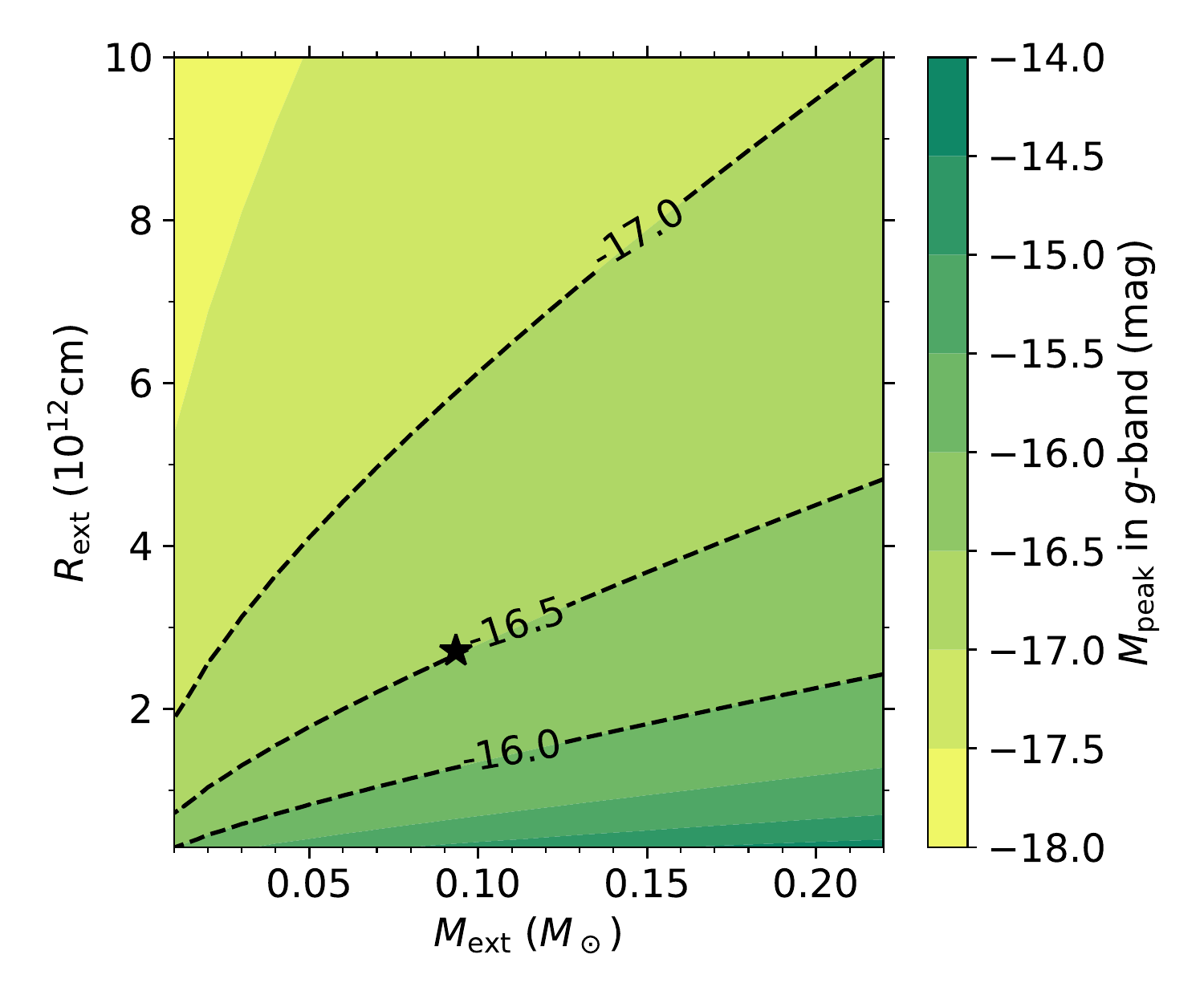}
	\caption{Expected $r$-band rise time (upper panel) and $g$-band peak absolute luminosity (bottom 
	panel) as a function of shock cooling model parameters $R_{\rm ext}$ and $M_{\rm ext}$. The 
	position of 
	\name\	is 	indicated by the black asterisks. In the upper panel, parameter space that could not pass 
	our criteria of ``rise from 1.5\,mag below peak to peak in less than 4.1\,d'' (Section 
	\ref{subsec:cadence}) is indicated by the hatched region. \label{fig:cooling}}
\end{figure}

Given the low mass of ultra-stripped progenitors, we expect to see shock cooling emission from the 
inflated pre-explosion star, as has been clearly seen in the case of iPTF14gqr and SN2019dge in the 
fast early-time evolution and blue colors of the optical light curve. As is shown by 
\citet[][Fig. 2]{Nakar2014}, rise time of the shock cooling light curve is determined by mass of the 
extended material $M_{\rm ext}$, while the peak luminosity is mainly modulated by $R_{\rm ext}$. We 
demonstrate this dependence in Figure~\ref{fig:cooling}. We simulate shock cooling light curves by 
varying $M_{\rm ext}$ and $R_{\rm ext}$, and at the same time setting $E_{\rm ext}=1.15\times 
10^{50}\,{\rm erg}$ (the value found in \name). 

In the upper panel, $t_{\rm rise}$ is defined in the same 
way as in Section \ref{subsubsec:compare_mag} (rise time from half-max to max). The rising part of 
the cooling light curve can be captured by a three-day, two-day, and one-day cadence optical survey 
at $M_{\rm ext}\gtrsim 0.14\,M_\odot$, $\gtrsim 0.07\,M_\odot$, and $\gtrsim 0.03\,M_\odot$, 
respectively. Transients with $M_{\rm ext} \gtrsim 0.15\,M_\odot$ will not pass our selection criteria in 
Section \ref{subsec:cadence}. In the bottom panel, we show the expected absolute luminosity at 
peak of the $g$-band cooling light curve. As is readily shown, for ultra-stripped progenitors with an 
extended radius $\lesssim 2\times 10^{12}\,{\rm cm}$, a survey like ZTF is only sensitive to objects
in the local universe ($\lesssim 100$--150\,Mpc) for the subsequent evolution of the light curve to be 
well-characterized. Taken together, we conclude that our estimation of the ultra-stripped SNe rate 
does not include ultra-stripped progenitors with $M_{\rm ext}\gtrsim 0.15\,M_\odot$ or $R_{\rm 
ext}\lesssim 2\times 10^{12}\,{\rm cm}$.  

\subsection{Discussion} 
\label{subsec:companion}
If the companion of the pre-explosion helium star is a low mass main sequence star, a white dwarf, or a 
black hole, \name\ will not be the progenitor of a double neutron star system, and thus the inferred 
$R_{\rm 19dge}$ is not connected with $R_{\rm DNS}$. Even in the case that the companion 
is a neutron star, if the forming DNS binary has orbital periods more than $\sim1$\,d, it will not merge
within the age of the Universe \citep{Tauris2015}. Therefore, the above estimation of the ultra-stripped 
SNe rate should provide an upper limit to the local coalescence rate density ($R_{\rm DNS}$) of double 
neutron stars not formed via dynamical capture in a globular cluster \citep{East2012, Andrews2019}. 

%Future theoretical work is needed to establish the relationship between orbital periods of the 
%pre-explosion binary systems and SNe observational properties. 

\section{Conclusion} \label{sec:conclusion}
In this paper we have presented the discovery, observation and modeling of the transient \name. We 
summarize the main characteristics of this object below:
\begin{enumerate}[label=(\alph*)]
	\item Peak absolute magnitudes are $M_{g\rm , peak}\approx -16.5$\,mag and $M_{r \rm ,peak } 
	\approx -16.3$\,mag. In $r$-band, rise time (half-max to max) is 2.0\,d and decay time (max to 
	half-max) is 8.6\,d. \name\ is one of the most rapidly rising subluminous SNe I discovered to date.
	\item Early-time spectra show a blue continuum and flash \ion{He}{II} features that indicate a 
	high mass-loss rate of $\gtrsim 10^{-4}\, M_\odot \, \rm yr^{-1}$.
	\item Photospheric spectra indicate helium-rich ejecta, and the prominent NUV \ion{Mg}{II} 
	emission suggests interaction between SN ejecta and CSM.
	\item Late-time spectra show signatures of interaction with helium-rich CSM, similar to that observed 
	in Type Ibn SNe.
	\item \name\ exploded in a compact low-metallicity 
	($Z\approx 0.005$) galaxy with small star formation rate (${\rm SFR}\approx 0.03\, M_\odot\, \rm 
	yr^{-1}$) and stellar mass ($M_\ast \approx 2.5 \times 10^{8}\, M_\odot$).
	\item The bolometric light curve of \name\ peaks at $\sim 5 \times 10^{42}\,{\rm erg\, s^{-1}}$, 
	and can be explained by a combination of two components. The first component is consistent 
	with shock cooling from an envelope of $\sim 0.1\,M_\odot$ located at $\sim 3\times 10^{12}\, \rm 
	cm$ ($40\, R_\odot$) from the progenitor. The second component is powered by 
	$\sim 0.017\, M_\odot$ of $^{56}$Ni.
	\item We estimate the ejecta mass and kinetic energy of \name\ to be $\sim 0.35\,M_\odot$ 
	and $\sim1.3\times10^{50}\,{\rm erg}$, respectively.
\end{enumerate}
We interpret \name\ as a helium-rich ultra-stripped envelope SN. 

Based on the one event, we estimate the rate density of 19dge-like ultra-stripped SNe (with $M_{\rm 
ext}\lesssim 0.15\,M_\odot$ and $R_{\rm ext}\gtrsim 2\times 10^{12}\,{\rm cm}$) to be 
1400--8200\,${\rm Gpc^{-3}\, yr^{-1}}$. This can be compared to the merger rate of DNS systems 
not formed via dynamical capture. The first detection of gravitational waves from the merging DNS 
binary GW170817 gave $R_{\rm DNS}=320$--$4740\,{\rm Gpc^{-3}\, yr^{-1}}$ \citep{GW170817}. 
Detection of GW190425 provides an update of $R_{\rm DNS}=250$--$2810\,{\rm Gpc^{-3}\, yr^{-1}}$ 
\citep{GW190425}. Based on an archival search for EM170817-like transients (known as ``kilonovae'' or 
``macronovae'') in the PTF database, \citet{Kasliwal2017} reported an upper limit on the rate of 
$800\,{\rm Gpc^{-3}\, yr^{-1}}$, which might be doubled if the typical kilonova is 50\% fainter than 
EM170817. 

It is important to compare ultra-stripped SNe rate and $R_{\rm DNS}$ constrained by future GW 
observations. If the former is smaller than the latter, it will provide evidence for the dynamical formation 
channel to be the major path for forming DNS systems. A better constraint of ultra-stripped SNe rate is 
also essential in our understanding of the final stages of helium star evolution in binary systems. As 
such, further systematic searches for ultra-stripped SNe are requried to reduce the large uncertainties 
of the current estimation. Moving forward, the discovery of ultra-stripped SNe will still rely on 
high-cadence wide-field experiments such as ZTF. In particular, the upcoming ZTF-II, with a two-day 
cadence all sky survey, coupled with higher cadence boutique experiments, is well-positioned to carry 
out this task.

\acknowledgements

We thank Takashi Moriya, Thomas Tauris, David Khatami, Dan Kasen, Sterl Phinney, and Wenbin 
Lu for valuable discussions during this work. We thank Lin Yan for sharing spectra 
of SN1993J and Gaia16apd. Y.Y.~thanks the instructors and organisers of the 
GROWTH summer school for teaching techniques in time-domain data analysis. 
This study made use of the open supernova catalog \citep{Guillochon2017}.

C.F.~gratefully acknowledges support of his research by the Heising-Simons Foundation 
(\#2018-0907).

This work was supported by the GROWTH project funded by the National Science Foundation under 
PIRE grant No.\,1545949. 

This work is based on observations obtained with the Samuel Oschin Telescope 48 inch and the 60 
inch Telescope at the Palomar Observatory as part of the Zwicky Transient Facility project. ZTF is 
supported by the National Science Foundation under grant No. AST-1440341 and a collaboration 
including Caltech, IPAC, the Weizmann Institute for Science, the Oskar Klein Center at Stockholm 
University, the University of Maryland, the University of Washington, Deutsches 
Elektronen-Synchrotron and Humboldt University, Los Alamos National Laboratories, the TANGO 
Consortium of Taiwan, the University of Wisconsin at Milwaukee, and Lawrence Berkeley National 
Laboratories. Operations are conducted by COO, IPAC, and UW. 

\software{
          \texttt{astropy} \citep{Astropy-Collaboration2013},
          \texttt{corner} \citep{Foreman-Mackey2016},
          \texttt{CIGALE} \citep{CIGALE19},
          \texttt{emcee} \citep{Foreman-Mackey2013},
          \texttt{LAMBDAR} \citep{Wright2016a},
          \texttt{Lpipe} \citep{Perley2019lpipe},
          \texttt{matplotlib} \citep{Hunter2007},
          \texttt{pandas} \citep{McKinney2010},
          \texttt{pyneb} \citep{Luridiana2013},
          \texttt{pyraf-dbsp} \citep{Bellm2016},
           \texttt{scipy} \citep{Jones2001}, 
          \texttt{simsurvey} \citep{Feindt2019},
          \texttt{sncosmo} \citep{Barbary2016}
          \texttt{ztfquery} \citep{Rigault2018}
          }

%% For this sample we use BibTeX plus aasjournals.bst to generate the
%% the bibliography. The sample63.bib file was populated from ADS. To
%% get the citations to show in the compiled file do the following:
%%
%% pdflatex sample63.tex
%% bibtext sample63
%% pdflatex sample63.tex
%% pdflatex sample63.tex

\appendix

\section{UV and Optical Data} \label{sec:appphot_data}
\startlongtable
\begin{deluxetable}{cccccc}
\tabletypesize{\scriptsize}
\tablecaption{Optical and UV photometry for SN2019dge.\label{tab:phot}}
\tablehead{
\colhead{Date (JD)}   
& \colhead{Instrument}
& \colhead{Filter}  
& \colhead{$m$} 
& \colhead{$\sigma_{m}$}
}
\startdata
58582.1544 & LT$+$IOO & $g$ & 18.590 & 0.010 \\
58582.1552 & LT$+$IOO & $r$ & 18.840 & 0.020 \\
58582.1575 & LT$+$IOO & $i$ & 19.110 & 0.020 \\
58582.1583 & LT$+$IOO & $z$ & 19.280 & 0.070 \\
58584.2341 & LT$+$IOO & $u$ & 18.570 & 0.020 \\
58580.4421 & P48$+$ZTF & $g$ & 20.828 & 0.148 \\
58580.4842 & P48$+$ZTF & $r$ & 20.891 & 0.139 \\
58582.8289 & \swift$+$UVOT & $B$ & 18.606 & 0.193 \\
58582.8280 & \swift$+$UVOT & $U$ & 18.289 & 0.113 \\
58582.8346 & \swift$+$UVOT & $UVM2$ & 18.550 & 0.068 \\
58582.8261 & \swift$+$UVOT & $UVW1$ & 18.685 & 0.108 \\
58582.8299 & \swift$+$UVOT & $UVW2$ & 18.802 & 0.103 \\
58582.8337 & \swift$+$UVOT & $V$ & 18.679 & 0.404 \\
\enddata
\tablecomments{$m$ and $\sigma_m$ are observed magnitude (without extinction correction) in AB 
system. A machine-readable table of all 117 photometric data points will be made available online.}
\end{deluxetable}

\begin{deluxetable}{llll}[htbp!]
		\tabletypesize{\scriptsize}
		\tablecaption{Photometry of the host galaxy \label{tab:host_phot}}
		\tablehead{
\colhead{Instrument/Filter}
& \colhead{$\lambda_{\rm eff}$ ($\rm \AA$)}
& \colhead{$m$}
& \colhead{$\sigma_{m}$}
	}
\startdata
% UVOT checked (lambda_eff, mag, emag)
UVOT/$UVW2$ 		& 2079.0  & 20.492 & 0.124\\
UVOT/$UVM2$		& 2255.1  &  20.471 & 0.172\\
UVOT/$UVW1$ 		& 2614.2  &  20.081  & 0.155\\
UVOT/$U$		& 3475.5  &  19.631 & 0.145\\
UVOT/$B$		& 4359.1  &  18.812 & 0.139\\
UVOT/$V$		& 5430.1  &  18.194  & 0.171\\
% SDSS checked (lambda_eff, mag, emag)
SDSS/$u'$ 		& 3561.8  &  19.636	& 0.082	\\
SDSS/$g'$ 		& 4718.9  &  18.540 & 0.015	\\
SDSS/$r'$ 		& 6185.2  & 18.056	 & 0.026 \\
SDSS/$i'$ 		& 7499.7     &17.885 & 0.028\\
SDSS/$z'$ 		& 8961.5     &17.697 & 0.089\\
% PS1 checked (lambda_eff, mag, emag)
PS1/$g_{\rm PS1}$	& 4866.5     & 18.538 & 0.042\\
PS1/$r_{\rm PS1}$	& 6214.6     & 18.029  & 0.030\\
PS1/$i_{\rm PS1}$	& 7544.6     &17.845 &0.033\\
PS1/$z_{\rm PS1}$	& 8679.5     & 17.755 & 0.050\\
PS1/$y_{\rm PS1}$	& 9633.3     &17.710  & 0.063\\
% 2MASS
2MASS/$J$ 		& 12410.5   & 17.653 & 0.215\\
2MASS/$H$ 		& 16513.7   &17.690 & 0.420\\
% WISE
WISE/$W1$ 		& 34002.6    & 18.460 & 0.069\\
WISE/$W2$   & 46520.1    & 18.953 &  0.136\\
\enddata
\tablecomments{$m$ and $\sigma_m$ are observed magnitude (without extinction correction) in the AB 
system.
}
\end{deluxetable}

The full set of photometry of \name\ is listed in Table~\ref{tab:phot}. Photometry of the host 
galaxy SDSS J173646.73+503252.3 is listed in Table~\ref{tab:host_phot}.

\section{Modeling of \name}

\begin{figure}[htbp!]
	\centering
	\includegraphics[width=\columnwidth]{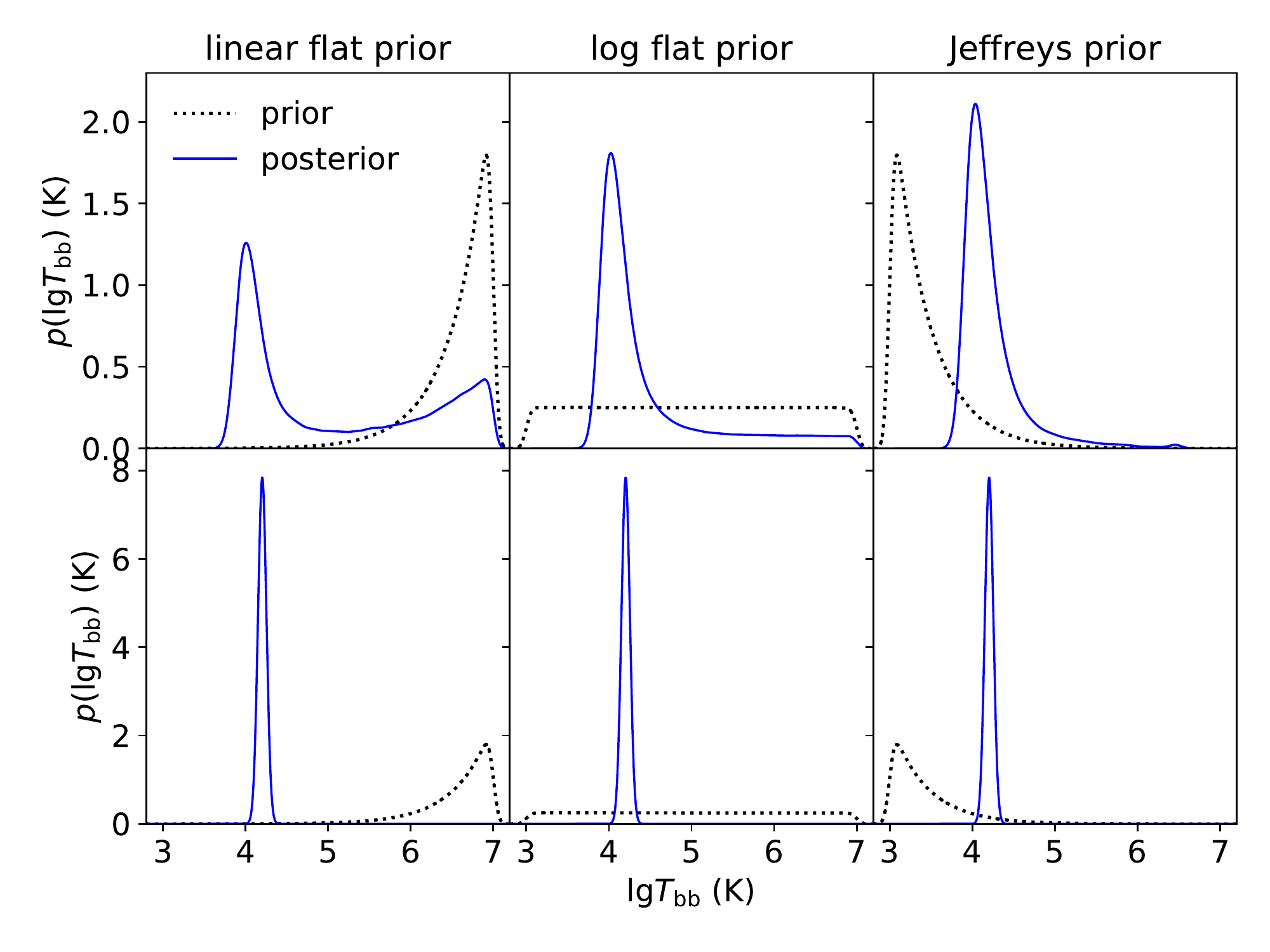}
	\caption{Posterior (solid lines) distribution of the blackbody temperature 
		$T_{\rm bb}$ on Apr 7 (upper panels) and Apr 9 (bottom panels) using three different priors 
		(dotted 
		lines).	\label{fig:bbprior}}
\end{figure}

\begin{figure*}
	\centering
	\includegraphics[width = 0.9\textwidth]{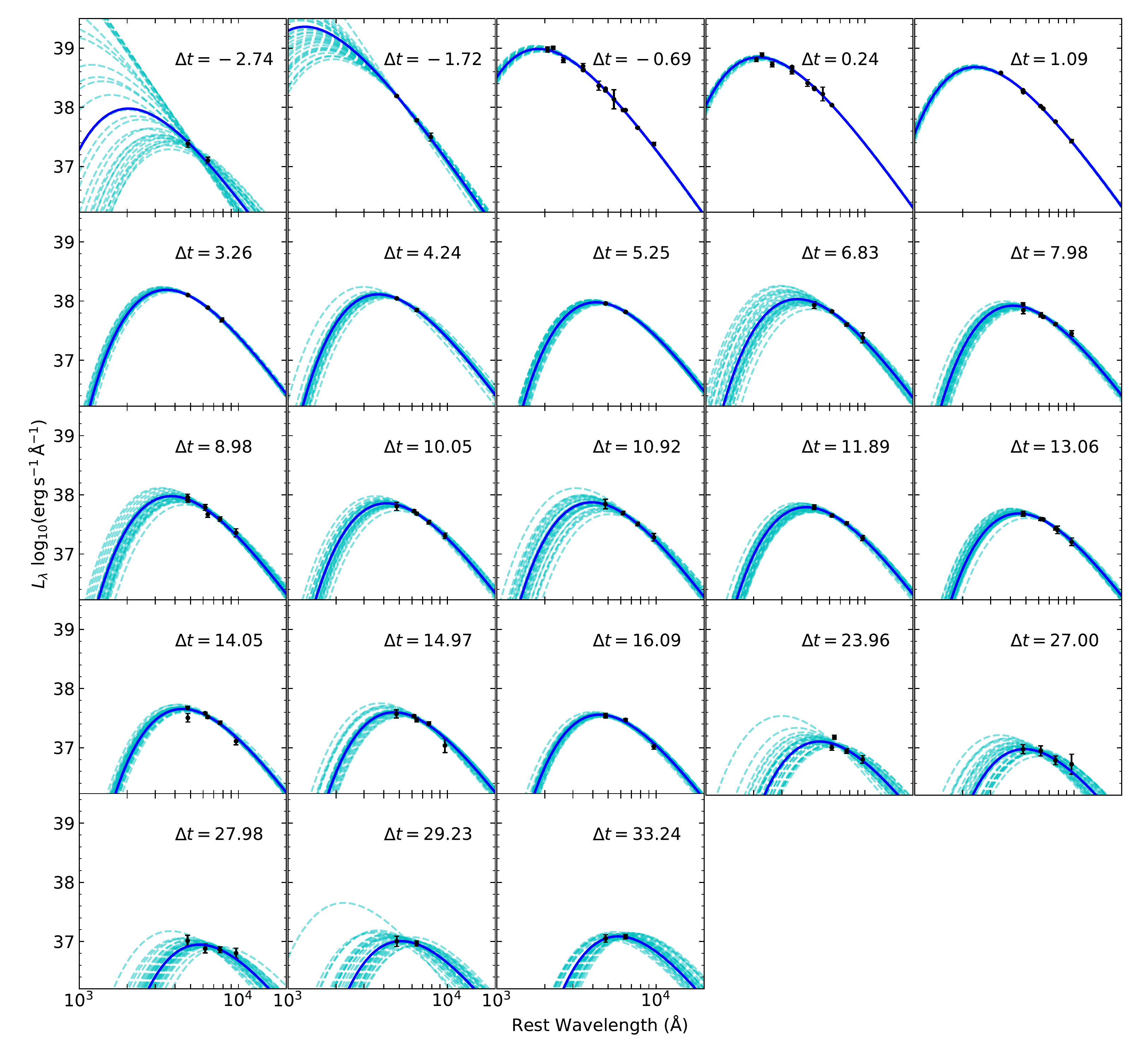}
	\caption{Black data points are \swift/UVOT and optical photometry of \name. Solid lines show 
		model fits using estimated parameters, while 30 random draws from the MCMC posterior are 
		shown with dashed lines.
		\label{fig:seds}}
\end{figure*}
\begin{table}[!htbp] 
	\centering 
	\caption{Physical evolution of SN2019dge from blackbody fits.} 
	\begin{tabular}{rrrr} 
		\hline 
		$\Delta t$ & $L (10^{41} \,{\rm erg\,s^{-1}})$ & $R$ ($10^{3}\,R_\odot$) & $T$ ($10^3$\,K) \\ 
		\hline
		$-$2.74 & $2.98^{+578.84}_{-1.41}$ &$1.45^{+1.14}_{-1.14}$ &$14.21^{+101.35}_{-5.12}$  \\
		$-$1.72 & $44.63^{+43.85}_{-15.84}$ &$2.20^{+0.44}_{-0.46}$ &$22.75^{+7.60}_{-4.14}$  \\
		$-$0.69 & $26.96^{+0.92}_{-0.87}$ &$3.50^{+0.07}_{-0.07}$ &$15.90^{+0.28}_{-0.27}$  \\
		0.24 & $22.87^{+0.62}_{-0.61}$ &$4.46^{+0.08}_{-0.08}$ &$13.51^{+0.21}_{-0.21}$  \\
		1.09 & $17.50^{+0.40}_{-0.37}$ &$4.89^{+0.10}_{-0.10}$ &$12.07^{+0.19}_{-0.18}$  \\
		3.26 & $8.34^{+0.22}_{-0.20}$ &$7.26^{+0.46}_{-0.44}$ &$8.23^{+0.31}_{-0.28}$  \\
		4.24 & $7.29^{+0.26}_{-0.22}$ &$7.38^{+0.88}_{-0.83}$ &$7.88^{+0.54}_{-0.46}$  \\
		5.25 & $6.10^{+0.16}_{-0.15}$ &$8.77^{+0.78}_{-0.73}$ &$6.92^{+0.33}_{-0.30}$  \\
		6.83 & $6.18^{+0.63}_{-0.46}$ &$7.09^{+1.01}_{-0.92}$ &$7.72^{+0.75}_{-0.62}$  \\
		7.98 & $5.29^{+0.24}_{-0.21}$ &$8.00^{+0.75}_{-0.70}$ &$6.99^{+0.39}_{-0.35}$  \\
		8.98 & $5.49^{+0.44}_{-0.36}$ &$6.79^{+0.94}_{-0.85}$ &$7.66^{+0.65}_{-0.56}$  \\
		10.05 & $4.55^{+0.37}_{-0.27}$ &$7.42^{+1.03}_{-0.96}$ &$6.99^{+0.63}_{-0.52}$  \\
		10.92 & $4.49^{+0.64}_{-0.44}$ &$6.63^{+1.25}_{-1.06}$ &$7.37^{+0.93}_{-0.75}$  \\
		11.89 & $4.04^{+0.24}_{-0.21}$ &$7.42^{+0.83}_{-0.75}$ &$6.79^{+0.45}_{-0.40}$  \\
		13.06 & $3.34^{+0.10}_{-0.10}$ &$7.52^{+0.69}_{-0.66}$ &$6.43^{+0.32}_{-0.29}$  \\
		14.05 & $3.08^{+0.10}_{-0.09}$ &$7.05^{+0.63}_{-0.59}$ &$6.51^{+0.32}_{-0.29}$  \\
		14.97 & $2.82^{+0.15}_{-0.12}$ &$7.30^{+1.21}_{-1.06}$ &$6.25^{+0.57}_{-0.49}$  \\
		16.09 & $2.45^{+0.09}_{-0.09}$ &$6.17^{+0.59}_{-0.58}$ &$6.56^{+0.33}_{-0.29}$  \\
		23.96 & $1.02^{+0.07}_{-0.06}$ &$5.47^{+1.23}_{-1.12}$ &$5.59^{+0.74}_{-0.54}$  \\
		27.00 & $0.72^{+0.08}_{-0.08}$ &$4.08^{+1.26}_{-1.01}$ &$5.93^{+0.91}_{-0.70}$  \\
		27.98 & $0.78^{+0.07}_{-0.06}$ &$5.73^{+1.68}_{-1.29}$ &$5.10^{+0.68}_{-0.57}$  \\
		29.23 & $0.83^{+0.11}_{-0.09}$ &$4.76^{+2.47}_{-1.56}$ &$5.64^{+1.23}_{-0.93}$  \\
		33.24 & $1.09^{+0.16}_{-0.11}$ &$6.99^{+2.47}_{-1.74}$ &$5.02^{+0.65}_{-0.56}$  \\
		\hline 
	\end{tabular} 
	\label{tab:bbfit} 
\end{table}

\begin{figure}[htbp!]
	\centering
	\includegraphics[width=\columnwidth]{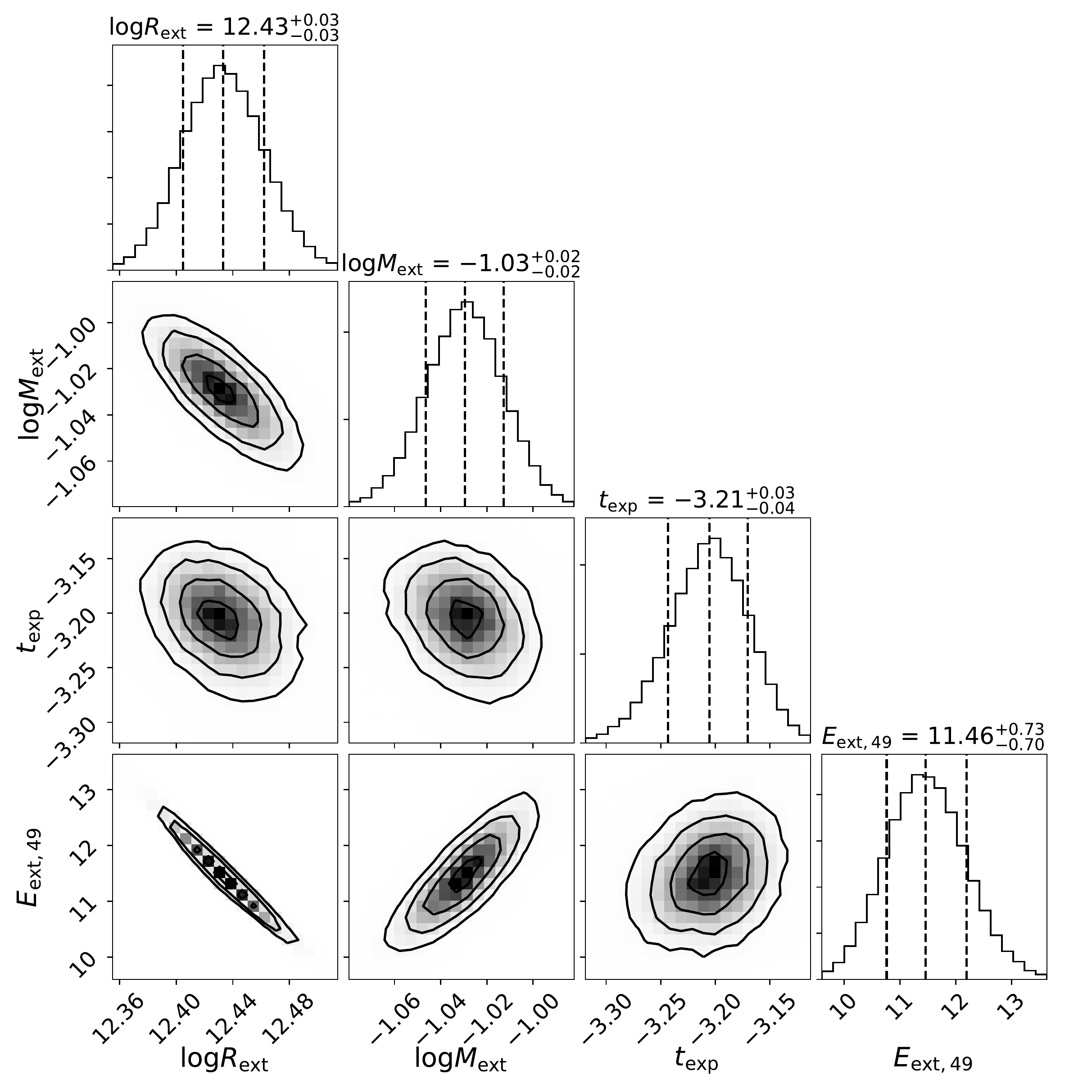}
	\caption{Corner plot showing the posterior constraints on ${\rm log}R_{\rm ext}$, ${\rm log}M_{\rm 
			ext}$, $t_\mathrm{exp}$, and $E_{\rm ext, 49}$. Marginalized one-dimensional distributions are 
		shown along the diagonal, along with the median estimate and the 68\% credible region (shown 
		with vertical 
		dashed 
		lines).	\label{fig:pirocorner}}
\end{figure}

\begin{deluxetable}{llc}[htpb!]
	\tablecaption{Shock cooling model parameters $\theta$ and their priors \label{tab:P15priors}}
	\tablehead{
		\colhead{$\theta$}
		& \colhead{Description}
		&\colhead{Prior}
	}
	\startdata
	${\rm log}R_{\rm ext}$ & log$_{10}$ of extented material radius in cm  & 
	$\mathcal{U}(-5, 25)$ \\
	${\rm log}M_{\rm ext}$ &  log$_{10}$ of extented material mass in $M_\odot$  & $\mathcal{U}(-4, 
	0)$\\
	$t_\mathrm{exp}$ & explosion epoch in MJD relative to $58583.2$& $\mathcal{U}(-8,-2.76)$ \\
	$E_{51}$ & SN energy divided by $10^{51}\,{\rm erg}$ & $\mathcal{U}(0.01, 10)$ \\
	$E_{\rm ext, 49}$ & $E_{\rm ext}$ divided by $10^{49}\,{\rm erg}$ & 
	$\mathcal{U}(0.1,100)$ \\
	\enddata
\end{deluxetable}

\begin{figure}[htbp!]
	\centering
	\includegraphics[width=\columnwidth]{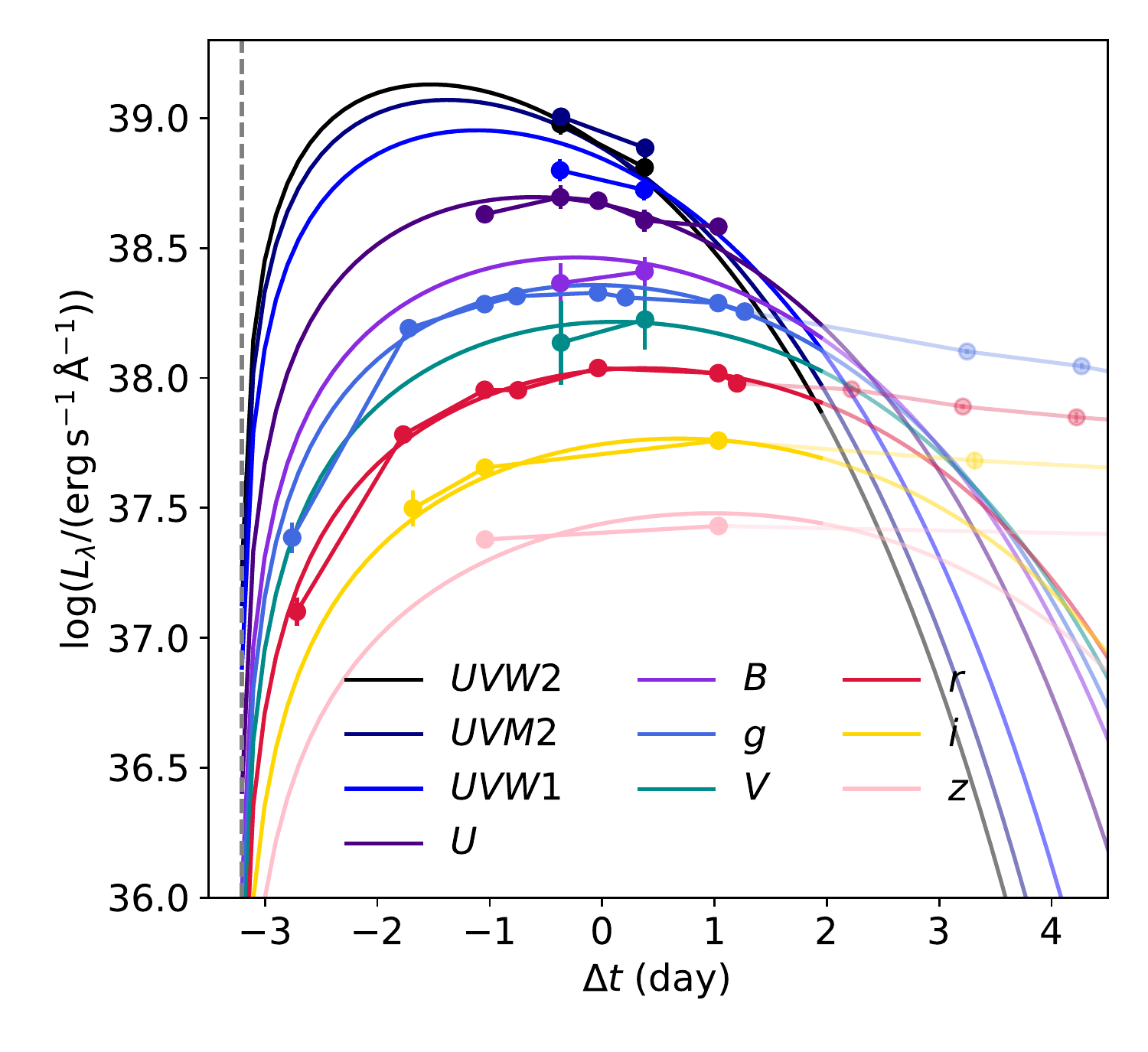}
	\caption{Cooling emission model fit to the early light curve of \name. 
		Data excluded from the fitting are shown as transparent circles. 
		The maximum a posteriori model is shown via solid lines.
		The vertical dashed line shows the median 1-D marginalized posterior value of
		$t_{\rm exp}$.
		\label{fig:piromodel}}
\end{figure}

\begin{deluxetable}{llc}[htpb!]
	\tablecaption{$^{56}$Ni decay model parameters $\theta$ and their priors \label{tab:Nidecaypriors}}
	\tablehead{
		\colhead{$\theta$}
		& \colhead{Description}
		&\colhead{Prior}
	}
	\startdata
	$\tau_{\rm m}$ &characteristic photon diffusion time in day  & $\mathcal{U}(1, 20)$ \\
	${\rm log}M_{\rm Ni}$ &  log$_{10}$ of nickel mass in $M_\odot$  & $\mathcal{U}(-4, 0)$\\
	$t_0$ & characteristic $\gamma$-ray escape time in day & $\mathcal{U}(20,  100)$  \\
	\enddata
\end{deluxetable}

\begin{figure}[htbp!]
	\centering
	\includegraphics[width=\columnwidth]{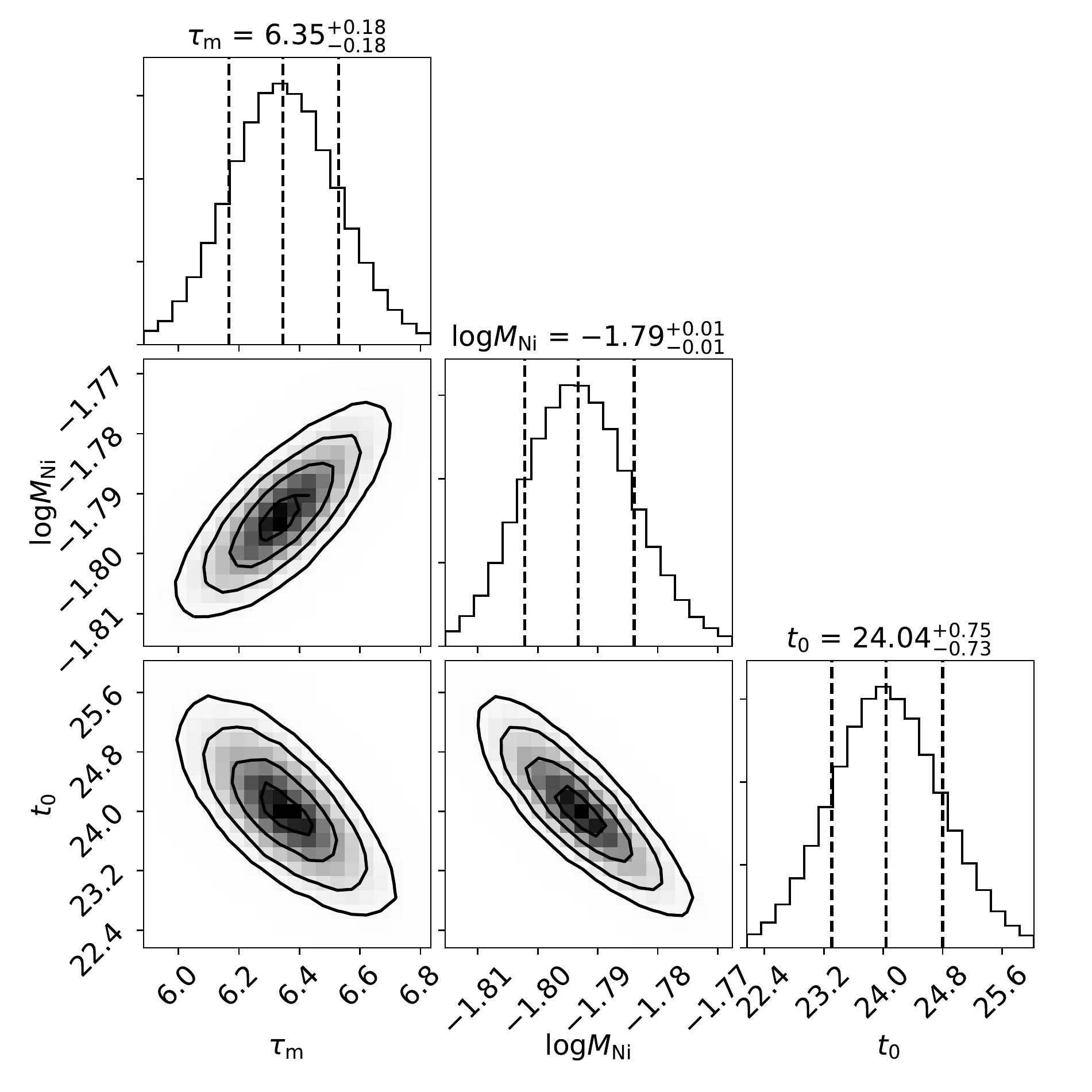}
	\caption{Corner plot showing the posterior constraints on $\tau_{\rm m}$, ${\rm log}M_{\rm 
			Ni}$, and $t_0$. Marginalized one-dimensional distributions are shown along the 
		diagonal, along with the median estimate and the 68\% credible region (shown with vertical 
		dashed 
		lines).	\label{fig:Nidecaycorner}}
\end{figure}
\subsection{Modeling the Physical Evolution} \label{subsec:bbfit}
To model the multi-band light curve with a blackbody function, we utilized the Monte Carlo Markov 
Chain (MCMC) simulations with \texttt{emcee} \citep{Foreman-Mackey2013}. We test the performance 
of three types of model priors for 
the blackbody radius ($R_{\rm bb}$) and temperature ($T_{\rm bb}$): (i) $T_{\rm bb}$ 
and $R_{\rm bb}$ are uniformly distributed in the range of [$10^3$, $10^7$]\,K and [$10$, 
$10^6$]\,$R_\odot$, respectively (ii) the two parameters are logarithmically uniformly 
distributed in the same ranges (ii) the two parameters follow Jeffreys prior \citep{jeffreys1946invariant} 
in the same ranges.

Within the ensemble, we use 100 walkers, each of which is run until convergence or 100,000 steps, 
whichever comes first. The test for convergence follows steps outlined in \citet{Yao2019} and 
\citet{Miller2020}. We adopt the 68\% credible region (i.e., $16^{\rm th}$ and $84^{\rm th}$ percentiles 
of posterior probability distributions) as the model uncertainties quoted in Table~\ref{tab:bbfit}.

We examine the fitting results under different choices of priors in Figure~\ref{fig:bbprior}, which shows 
the posterior distribution of $T_{\rm bb}$ using data obtained on Apr 7 (top panels) and Apr 9 (bottom 
panels). Early stages of SN evolution often feature extremely high temperatures. At an epoch where 
both UV and optical data are available (Apr 9), the posterior does not depend on the particular choice 
of prior, and the model parameter can thus be well constrained. However, at our first detection epoch 
where only optical data is available (Apr 7), the posterior strongly depends on the prior. For a linearly 
flat prior, high numbers receive a lot of ``weight'', making the ``multi-peaks'' shape posterior in the 
upper left panel of Figure~\ref{fig:bbprior}. Log prior and Jeffreys prior generally give similar results. 
In this study, we adopted results using log prior. However, we note that the choice of prior does not 
affect final estimates of maximum luminosity, or the model fits for shock cooling and $^{56}$Ni decay. 

In Figure \ref{fig:seds} we show the photometry interpolated onto common epochs, and fit to a 
blackbody function to derive the photospheric evolution. The resulting evolution in bolometric 
lumonosity, photospheric radius, and effective temperatures is listed in Table  \ref{tab:bbfit}.

\subsection{Modeling Early Light Curve} \label{subsec:p15fit}

We cast the P15 analytical expression for the shape of the early-time light curve in terms of $M_{\rm 
ext}$, $R_{\rm ext}$, $E_{\rm ext}$, and $E_{51}$:
\begin{subequations}
\begin{align}
 L(t) =& \frac{t_eE_{\rm ext}}{t_p^2} {\rm exp} \left[ -\frac{t (t + 2t_e)}{2t_p^2}\right] \,{\rm erg\, s^{-1}}\\
 t_e =& 10^{-9} R_{\rm ext} E_{\rm ext,49}^{-1/2} 
 \left(\frac{M_{\rm ext}}{0.01 M_\odot}\right)^{1/2} \, {\rm s}\\
 t_p =& 1.1\times 10^{5} \kappa_{0.34}^{1/2}  E_{51}^{-0.01 / 
 	1.4} \notag \\
 & \times E_{\rm ext, 49}^{-0.17 / 0.7}   \left(\frac{M_{\rm ext}}{0.01 M_\odot}\right)^{0.74} {\rm s} 
 \label{eq:tp}
\end{align}
\end{subequations}
where $t$ is time since explosion in seconds, $\kappa_{0.34} = \kappa / (0.34\,{\rm cm^2\, g^{-1}})$, 
$E_{\rm ext, 49} =  E_{\rm ext} / (10^{49}\,{\rm erg})$, $E_{51} =E /  (10^{51}\,{\rm erg})$, and $E$ is 
energy of the explosion.Following P15 we assume the emission is a blackbody at radius
\begin{align}
R(t) = R_{\rm ext} + 10^9  \left( \frac{E_{\rm ext}}{10^{49}\,{\rm 
		erg\,s^{-1}}} \right)^{1/2}  \left(\frac{M_{\rm ext}}{0.01 M_\odot}\right)^{-0.5} t
\end{align}
and temperature
\begin{align}
 T(t) = \left( \frac{L(t)}{4\pi R(t)^2 \sigma_{\rm SB}} \right)^{1/4}
\end{align}

We fix $\kappa \approx 0.2\,{\rm cm^2\, g^{-1}}$ as 
appropriate for a hydrogen-deficient ionized gas, and assign wide flat priors for all model parameters, 
as summarized in Table~\ref{tab:P15priors}. We only include observations up to $\Delta t = 2$\,d in 
the fitting. We found that this particular choice of $\Delta t$ --- 2\,d instead of 1\,d or 3\,d --- in 
general does not affect the final inference for the model parameters. Figure~\ref{fig:pirocorner} shows 
the corner plot of ${\rm log}R_{\rm ext}$, ${\rm log}M_{\rm ext}$, $t_\mathrm{exp}$, and $E_{\rm ext, 
49}$. For clarity, $E_{51}$ is excluded as it does not exhibit strong covariance with the parameters 
shown here. This can be understood by Eq.~\ref{eq:tp}, which gives $t_p \propto E_{51}^{-0.01/1.4}$, 
suggesting that the shock cooling luminosity only weakly depends on $E_{51}$. 

The maximum a posteriori model is visualized by solid lines in Figure~\ref{fig:piromodel} color-coded in 
different filters. The rising part of the model does not closely match to data due to the ignorance of 
the density structure of the stellar profile. Nevertheless, the peak of the light curve is well captured by 
this model.

\begin{figure}[htbp!]
	\centering
	\includegraphics[width=\columnwidth]{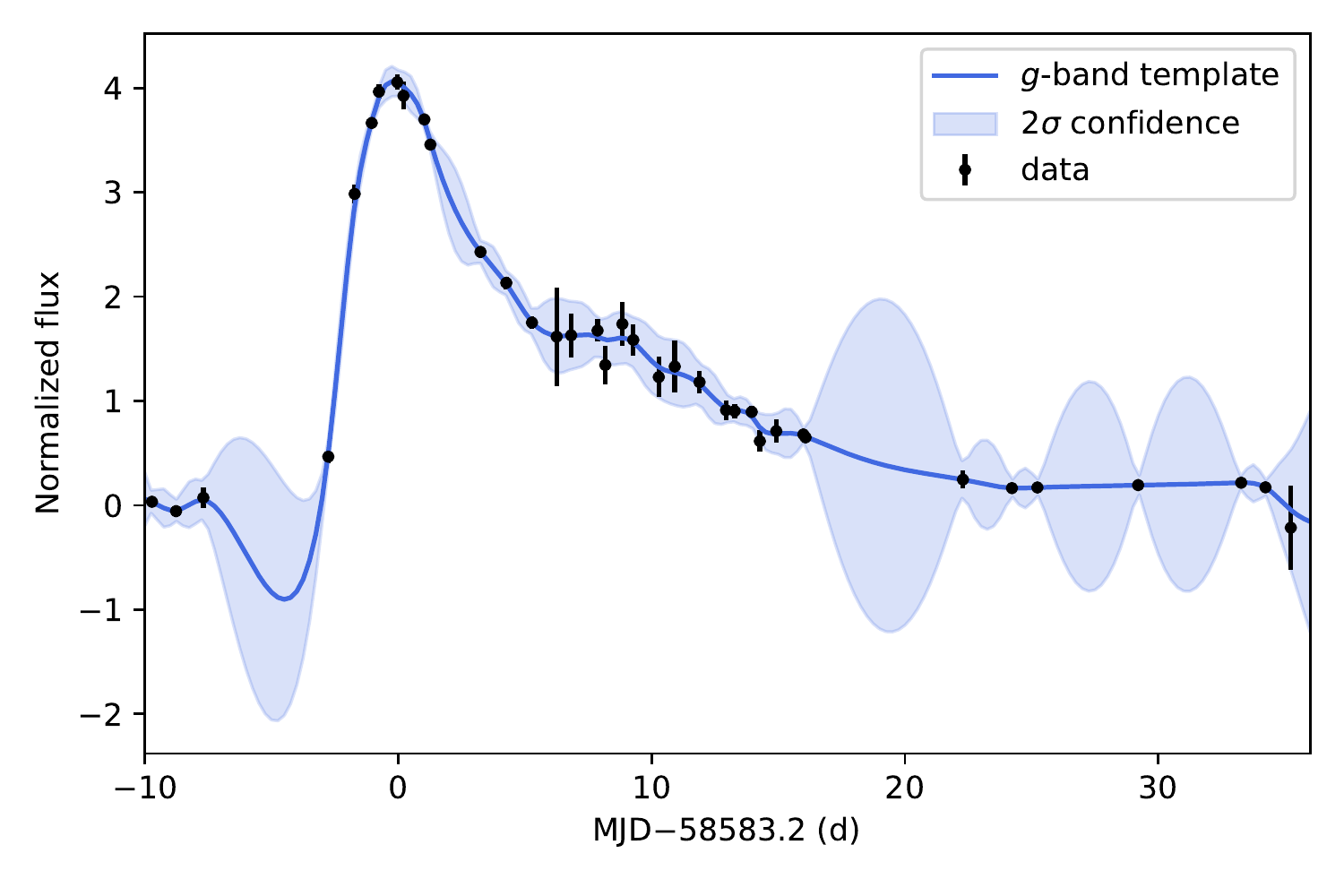}
	\includegraphics[width=\columnwidth]{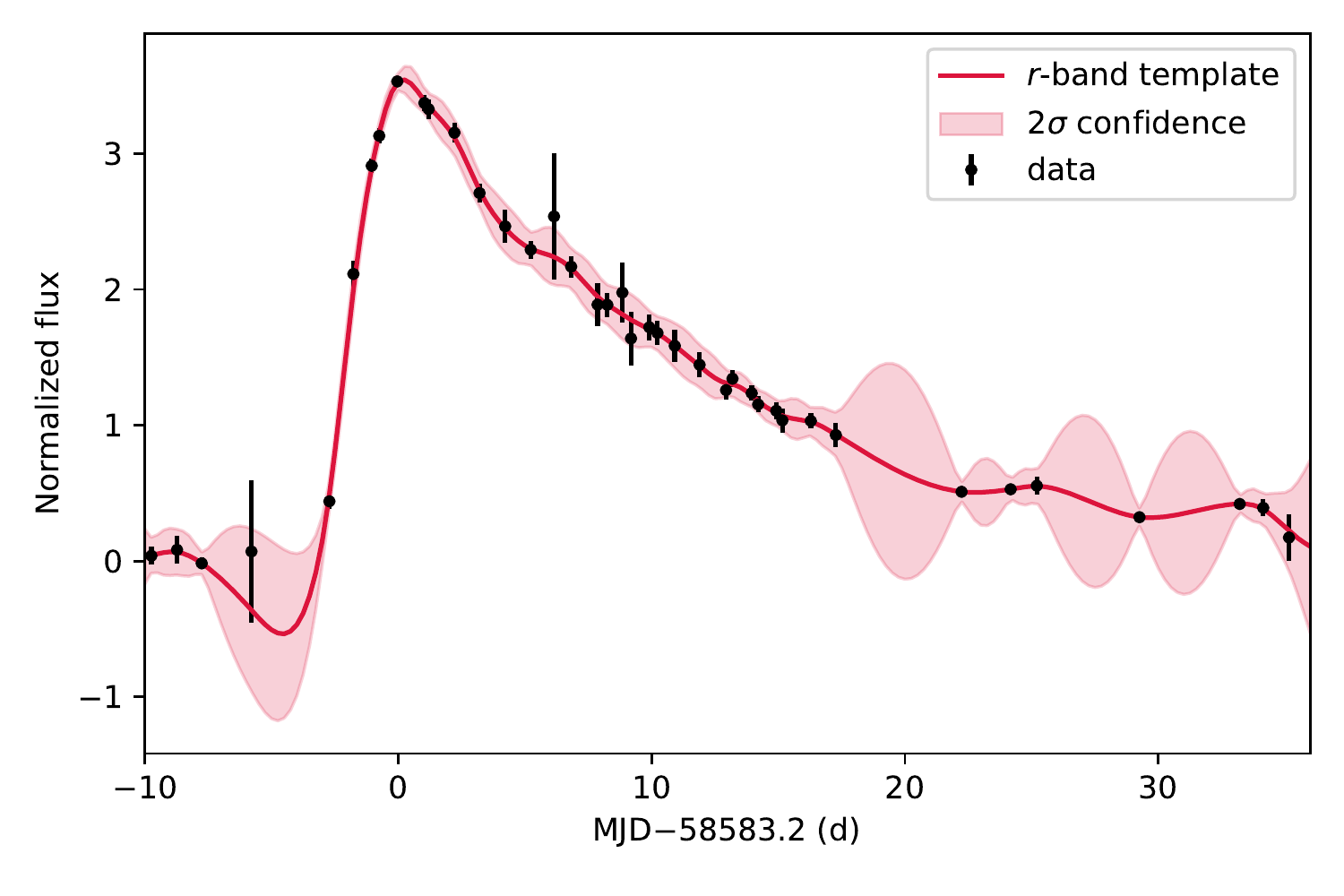}
	\caption{$g$- and $r$-band light curve templates for \name\ obtained from Gaussian process 
	fitting. 	\label{fig:template}}
\end{figure}

\subsection{Modeling the Main Peak}\label{subsec:arnettfit}

%\todo{radioactive decay give both gamma-rays and positrons}

For $^{56}$Ni$\rightarrow ^{56}$Co$\rightarrow ^{56}$Fe decay powered explosions, the energy 
deposition rate is
\begin{align}
\varepsilon_{\rm rad} =&\varepsilon_{\rm Ni, \gamma} (t) + \varepsilon_{\rm Co, \gamma} (t) 
\label{eq:heatTotal} \\
\varepsilon_{\rm Ni, \gamma} (t)   =& \epsilon_{\rm Ni}e^{-t/\tau_{\rm Ni}}  \label{eq:heatNi}\\
\varepsilon_{\rm Co, \gamma} (t)   =& \epsilon_{\rm Co} \left( e^{-t/\tau_{\rm Co}} - e^{-t/\tau_{\rm 
			Ni}} \right) \label{eq:heatCo}
\end{align}
where $\epsilon_{\rm Ni}= 3.90 \times 10^{10} \, {\rm erg\,g^{-1}\,s^{-1}}$, $\epsilon_{\rm Co}=6.78\times 
10^{9} \, {\rm erg\,g^{-1}\,s^{-1}}$, $\tau_{\rm Ni}=8.8$\,d and $\tau_{\rm Co}=111.3$\,d are the decay 
lifetimes of $^{56}\rm Ni$ 
and $^{56}\rm Co$ \citep{Nadyozhin1994}. The effective heating rate is modified by the probability of 
thermalization, and thus $\varepsilon_{\rm heat} \leq \varepsilon_{\rm rad}$.

The bolometric light curve can be generally divided into the 
photospheric phase and the nebular phase. The photospheric phase can be modelled using Equations 
given in \citet[][Appendix A]{Valenti2008}, with modifications given 
by \citet[][Eq.~3]{Lyman2016}, 
\begin{align}
 L_{\rm phot} (t) =& M_{\rm Ni} {\rm e}^{-x^2} \times \notag  \\
 & \Big[ (\epsilon_{\rm Ni} - \epsilon_{\rm Co}) \int_0^x (2z {\rm e}^{-2zy+z^2}){\rm d} z \notag \\
 & + \epsilon_{\rm Co} \int_0^x (2z {\rm e}^{-2zy+2zs + z^2}) {\rm d} z \Big]
\end{align}
where $x = t/\tau_{\rm m}$, $y = \tau_{\rm m} / (2\tau_{\rm Ni})$,
\begin{align}
s &= \frac{\tau_{\rm m} (\tau_{\rm Co} - \tau_{\rm Ni})}{2 \tau_{\rm Co} \tau_{\rm Ni}}, \\
\tau_{\rm m} &= \left( \frac{2\kappa_{\rm opt} M_{\rm ej}}{13.8 c v_{\rm phot}}\right)^{1/2}  
\label{eq:taum}
\end{align}

In the nebular phase the SN ejecta become optically thin, such that the delay between the energy 
deposition from radioactivity and the optical radiation becomes shorter. The bolometric luminosity is 
then equal to 
the rate of energy deposition: $L_{\rm neb}(t) = Q(t)$. At any given time, the energy deposition rate 
$Q(t)$ is \citep{Wheeler2015, Wygoda2019}:
\begin{align}
Q(t) \approx Q_{\gamma}(t) \left( 1 - e^{-(t_0/t)^2}\right) % Q_{\rm pos}(t),
\end{align}
where $Q_{\gamma}(t) = M_{\rm Ni}\varepsilon_{\rm rad}$ is the energy release rate of 
gamma-rays%\todo{on the spot??},
$t_0$ is the time at which the ejecta becomes optically thin to gamma rays. Here the difference 
between energy deposition rate of gamma-rays and positrons  is neglected.
%\begin{align}
%Q_{\rm pos}(t) &= M_{\rm Ni}\varepsilon_{\rm Co, pos}(t)\\
%\varepsilon_{\rm Co, pos}(t) &=  2.3\times 10^{8} \left( e^{-t/t_{s, {\rm Co}}} - e^{-t/t_{s, {\rm Ni}}}\right)
%\end{align}

%\todo{add motivation to say that this is ok}
To fit the shock cooling subtracted bolometric light curve with a simple radioactive decay model, we do 
not divide the data into photospheric phase and nebular phase, but instead adopt the following 
formula for the whole light curve:
\begin{align}
	L(t) = L_{\rm phot}(t)  \left( 1 - e^{-(t_0/t)^2}\right) 
\end{align}
Priors or the model parameters are summarized in Table~\ref{tab:Nidecaypriors}, and Figure 
\ref{fig:Nidecaycorner} shows the corner plot of $\tau_{\rm m}$, log$M_{\rm Ni}$, and $t_0$.

\subsection{Generating a Light Curve Template for SN2019dge} \label{subsec:gaussian}
To construct a template for SN2019dge in the ZTF $g$ and $r$ filters, we model the observed light 
curve by a Gaussian process. We denote the measurements as $(\mathbf{x}, \mathbf{y})$, where 
$\mathbf{x}$ is ${\rm MJD}-58583.2$, and $\textbf{y}$ is flux calculated as $10^{-0.4m}\times 10^8$ 
($m$ is magnitude). We choose a kernel in the form of Matern covariance function 
\citep[][Eq.~4.17]{rasmussen2003gaussian}:
\begin{align}
 k_{3/2}(x, x^{\prime}) = A \left( 1 + \frac{\sqrt{3}r}{l}\right) {\rm exp} \left(- 
 \frac{\sqrt{3}r}{l} \right) \label{eq:mat32}
\end{align} 
where $r = |x-x^{\prime}|$.  $A$ and $r$ in Eq.~\ref{eq:mat32} are chosen to minimize the negative log 
likelihood function (see, e.g., Eq.~2.43 of \citealt{rasmussen2003gaussian}). 

We perform the fit from $x=-10$\,d to $x=+40$\,d, and the obtained templates are shown in 
Figure~\ref{fig:template}.

\bibliography{at2019dge}{}
\bibliographystyle{aasjournal}

\end{document}